\documentclass[epjc3,final]{svjour3}  

\RequirePackage{graphicx}

\journalname{Eur. Phys. J. C}

\usepackage{amsmath}
\usepackage{graphicx}
\usepackage{pifont}
\usepackage{array}
\usepackage{rotate}
\usepackage{vmargin}
\usepackage{fancyhdr}
\usepackage{color}
\usepackage{lineno} 
\usepackage{url}
\usepackage{multirow} 
\usepackage{wasysym}

\begin{document}


\sloppy


\title{Sensitivity of LHC experiments to exotic highly ionising particles}

\author{
        A.~De~Roeck\thanksref{addr5,addr6,addr7} 
        \and
        A.~Katre\thanksref{addr1} 
        \and
        P.~Mermod\thanksref{e1,addr1,addr2}
        \and
        D.~Milstead\thanksref{addr3} 
        \and
        T.~Sloan\thanksref{addr4} 
}

\thankstext{e1}{Corresponding author, e-mail: philippe.mermod@cern.ch}

\institute{
           CERN, Geneva, Switzerland \label{addr5}
           \and
           Department of Physics, University of Antwerp, Belgium \label{addr6}
           \and
           Department of Physics, UC-Davis, USA \label{addr7}
           \and
           \textit{Current:} D\'epartement de Physique Nucl\'eaire et Corpuculaire, University of Geneva, Switzerland \label{addr1}
           \and
           Department of Physics, University of Oxford, UK \label{addr2}
           \and 
           Fysikum, Stockholm University, Sweden \label{addr3}
           \and
           Department of Physics, Lancaster University, UK \label{addr4}
}

\date{\today}

\maketitle

\begin{abstract}

The experiments at the Large Hadron Collider (LHC) are able to discover or
set limits on the production of exotic particles with TeV-scale masses possessing
values of electric and/or magnetic charge such that they behave as
highly ionising particles (HIPs). In this paper the sensitivity of the
LHC experiments to HIP production is discussed in detail. It is shown that 
a number of different detection methods are required to investigate as fully 
as possible the charge-mass range. These include direct detection as the
HIPs pass through either passive or active detectors and, in the case of magnetically charged
objects, the so-called induction method with which magnetic monopoles which stop in
accelerator and detector material could be observed. The benefit of using complementary 
approaches to HIP detection is discussed. 


\end{abstract}

\section{Introduction}
\label{intro}

Highly ionising particles (HIPs) are characterised by an ionisation energy loss which is at least dozens of times greater than minimum ionising particles while propagating through matter. Putative exotic HIPs include magnetic monopoles~\cite{Preskill1984}, particles which carry both magnetic and electric charges (so-called dyons)~\cite{Schwinger:1966nj,Schwinger1969}, and particles which can carry a high electric charge such as $Q$-balls~\cite{Kusenko:1997si} and stable micro black-hole remnants~\cite{Koch:2007}. In this work we will assume that HIPs are stable thanks to a conserved quantum number (like magnetic charge). If HIPs of accessible masses exist, they may be produced in pairs during high-energy collisions through particle-antiparticle annihilation, photon-photon fusion, or some more involved process. To date, no such particles have been found.  

Magnetic monopoles and dyons are prominent examples of HIPs envisaged in the literature. Dirac observed in 1931 that charge quantisation could be understood as a consequence of angular momentum conservation if a pole of magnetic charge existed~\cite{Dirac1931,Dirac1948}. This argument still provides compelling motivation to search for monopoles, in cosmic rays either trapped in matter~\cite{Groom86} or in flight~\cite{MACRO2002,SLIM2008,RICE2008,ANITA2011}, and at colliders at the high energy frontier~\cite{Fairbairn:2006gg}. The Dirac argument provides an approximate guideline to the value of magnetic charge which could be held by a particle. In SI units the argument yields the following relation between any electric charge $q_e$ and any magnetic charge $q_m$: $q_eq_m=nh/\mu_0$, where $h$ is Planck's constant, $\mu_0$ is the vacuum permeability, and $n$ is an integer number. With $q_e=ze$ ($e$ being the elementary charge) for electric charge and $q_m=gec$ ($c$ being the speed of light) for magnetic charge, the squares of the dimensionless quantities $z$ and $g$ represent the strengths of the corresponding electromagnetic couplings. Dirac's relation with $n=1$ and $z=1$ defines $g_D=h/(\mu_0 e^2c)=68.5$. This corresponds to an electromagnetic coupling typically several thousand times that of a particle with the elementary charge. While the Dirac argument prescribes that the minimum fundamental unit of magnetic charge is $g_Dec$, in a general picture it is allowed -- and even expected in the case of a dyon or if the fundamental electric charge is a fraction of $e$ -- to be higher~\cite{Schwinger:1966nj,Zwanziger:1969by,Preskill1984}. Predictions of monopole masses are more uncertain. Monopoles are general features of all theories where the U(1) group of electromagnetism is a subgroup of a broken gauge symmetry~\cite{Hooft1974}, in which case they typically have masses of the order of the unification scale. A large variety of models have been proposed, some of which allow for monopole masses in a range accessible to the LHC~\cite{Troost1976,Cho1997,PDG2010}.

The main aim of this paper is to provide a quantitative overview of the capabilities of the LHC experiments to detect HIPs with a range of electric and magnetic charges produced in high-energy proton-proton ($pp$) collisions. Accordingly this paper is structured as follows. Descriptions of HIP energy loss and monopole bending are given in Sections~\ref{velocity} and~\ref{bending}, respectively. HIP detection challenges and methods at colliders are described in Section~\ref{challenges}. An overview of LHC experimental setups which can potentially search for HIPs is given in Section~\ref{detectors}. A brief discussion of the existing limits on HIPs at the LHC is given in Section~\ref{limits}. Stopping positions in LHC detectors as functions of energy, direction, mass and charge (for both electric and magnetic charges) are determined in Section~\ref{stopping}. Detector acceptances as functions of HIP mass and charge and assuming a production model (with Drell-Yan as a benchmark) are given in Section~\ref{sensitivities}, and stopping acceptances of monopoles and dyons in LHC beam pipes are given in Section~\ref{trapped}. Model dependence is discussed in Section~\ref{dependence}. Finally, the sensitivities of the different LHC experiments are compared and discussed in Section~\ref{reach}, before concluding in Section~\ref{conclusions}.

\section{Ionisation energy loss in matter}
\label{velocity}

\begin{figure}[tb]
  \begin{center}
    \includegraphics[width=0.49\linewidth]{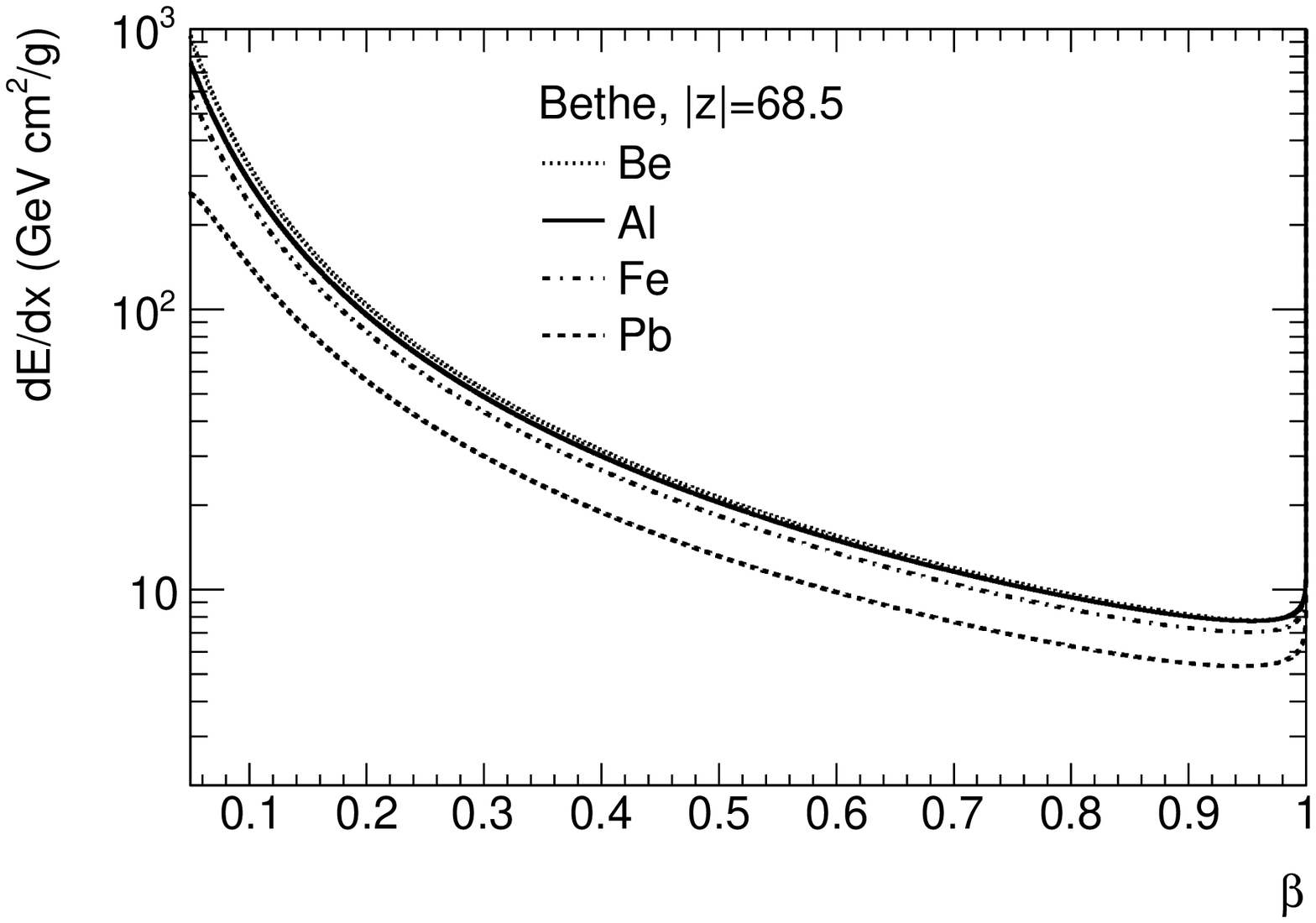}
    \includegraphics[width=0.49\linewidth]{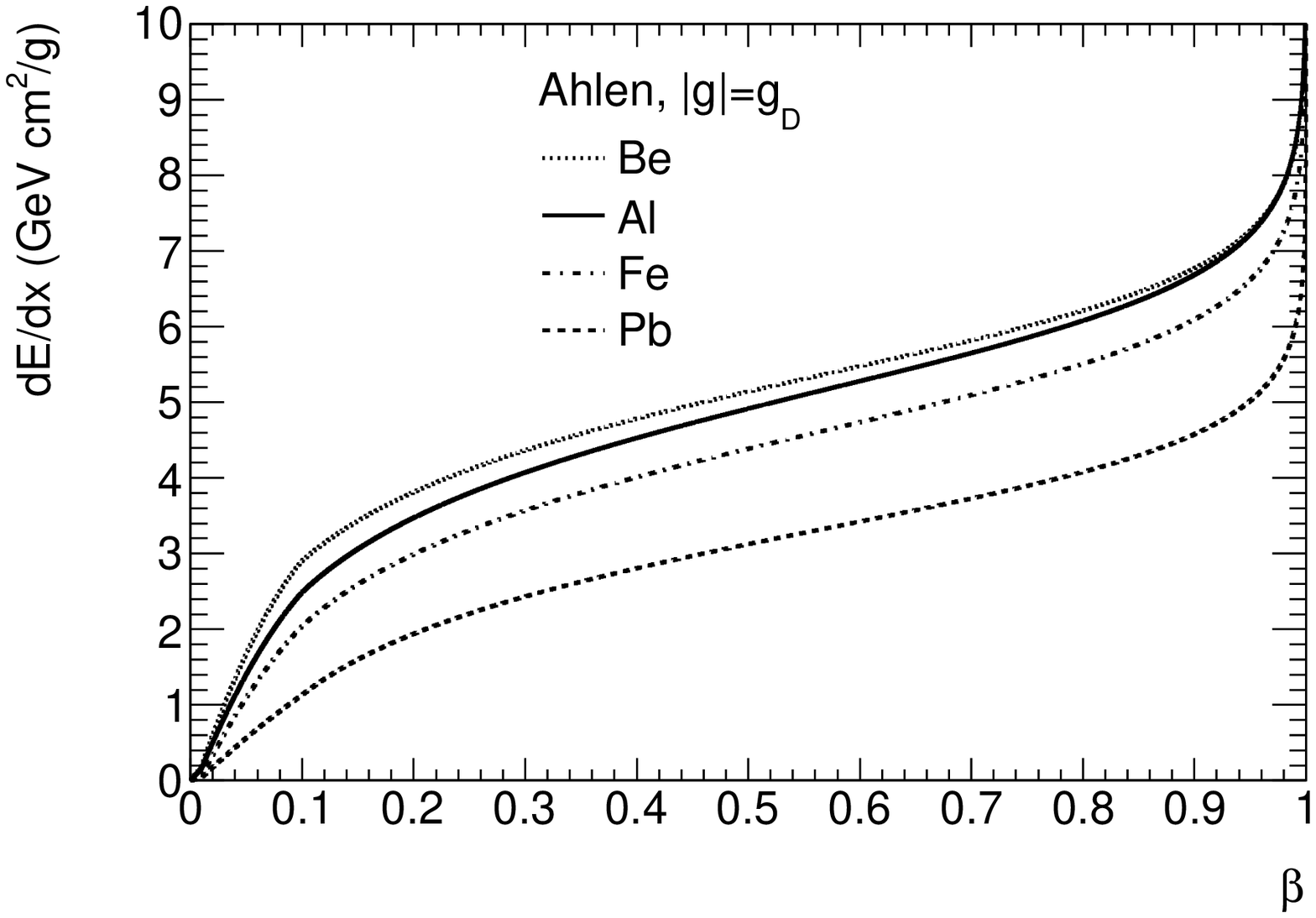}  
  \end{center}
  \caption{Calculated ionisation energy loss d$E$/d$x$ as a function of velocity $\beta$ for HIPs possessing electric charge (with $|z|=68.5$, left) and magnetic charge (with $|g|=g_D$, right) in various materials. Equations~\ref{Bethe} and~\ref{Bethe_mag} were used. Varying the correction factors by $\pm 2$ inside the brackets of the equations yields a relative uncertainty of 15\% in the energy loss. }
  \label{fig:dedx}
\end{figure}
 
Massive ($m\apprge 200$ GeV) HIPs at colliders would be produced in a velocity regime where they lose energy only through ionisation. The mean rate of energy loss per unit length d$E$/d$x$\footnote{The quantity d$x$ can be taken as the thickness of a thin layer of material, often conveniently expressed in g/cm$^2$. Dividing by the material density gives the thickness in cm.} of a massive HIP carrying an electric charge $q_e=ze$ traveling with velocity $\beta=v/c$ in a given material is modeled by the Bethe formula~\cite{PDG2010}:

\begin{equation}
-\frac{\textrm{d}E}{\textrm{d}x}=K\frac{Z}{A}\frac{z^2}{\beta^2}\left[\ln\frac{2m_ec^2\beta^2\gamma^2}{I}-\beta^2\right]
\label{Bethe}
\end{equation}

\noindent where $Z$, $A$ and $I$ are the atomic number, atomic mass and mean excitation energy of the medium, $K=0.307$ MeV~g$^{-1}$cm$^2$, $m_e$ is the electron mass and $\gamma=1/\sqrt{1-\beta^2}$. 
Higher-order terms are neglected. The uncertainty due to this approximation is estimated by performing the calculations with an additional term $\pm 2$ inside the brackets, as the sum of all corrections is expected to be significantly smaller than this for $|z|<100$~\cite{Ahlen80}. Expression \ref{Bethe} is valid only down to $\beta\sim 0.1$. Positively charged HIPs are expected to pick up electrons for $0<\beta<0.1$, which reduces their effective charge and increases their range\footnote{The range of a particle in matter is defined as the average distance before stopping.}. This generally only corresponds to the last 100~$\mu m$ before stopping, a distance much smaller than a typical detector granularity. For HIP range calculations the energy loss in this interval is approximated by a constant, using the value from Equation \ref{Bethe} at $\beta=0.1$. 

For a HIP carrying a magnetic charge $q_m=gec$, the velocity dependence of the Lorentz force causes the cancellation of the $1/\beta^2$ factor, changing the behaviour of the d$E$/d$x$ at low velocity~\cite{Ahlen78}:

\begin{equation}
-\frac{\textrm{d}E}{\textrm{d}x}=K\frac{Z}{A}g^2\left[\ln\frac{2m_ec^2\beta^2\gamma^2}{I_m}+\frac{K(|g|)}{2}-\frac{1}{2}-B(|g|)\right]
\label{Bethe_mag}
\end{equation}

\noindent where $I_m$ is approximated by the mean excitation energy for electric charges $I$. 
The Kazama, Yang and Goldhaber cross section correction and the Bloch correction are given by $K(|g|)=$ 0.406 (0.346) for $|g|=g_D$ ($2g_D$) and $B(|g|)=$ 0.248 (0.672, 1.022, 1.685) for $|g|=g_D$ ($2g_D$, $3g_D$, $6g_D$) \cite{Ahlen78}, and are interpolated linearly to intermediate values of $|g|$. The expression \ref{Bethe_mag} is valid only down to $\beta\sim 0.1$. For $\beta\leq 0.01$ we use the approximation $-$d$E$/d$x=(45$ GeV/cm$)(g/g_D)^2\beta$ \cite{Ahlen82} for all materials (units of GeV$\cdot$cm$^2$/g are obtained by dividing by the material density), and in the intermediate region $0.01<\beta<0.1$ we interpolate with a second-order polynomial. 

The $\beta$ dependencies of d$E$/d$x$ in various materials relevant for particle detectors, obtained from Equations~\ref{Bethe} and~\ref{Bethe_mag}, are shown in Fig.~\ref{fig:dedx} for an electrically charged particle with $z=68.5$ (left) and a magnetically charged particle with $g=g_D=68.5$ (right), respectively. With $g=z$, at $\beta\sim 1$ the values of d$E$/d$x$ coincide. The crucial difference between electric and magnetic charge is the behaviour at low velocity. The well-known increase of d$E$/d$x$ with decreasing $\beta$, giving rise to a large fraction of the particle's energy being deposited near the end of its trajectory (so-called Bragg peak~\cite{BraggPeak}), can be seen in the electric case. Conversely, for a magnetic monopole, d$E$/d$x$ is expected to diminish with decreasing $\beta$. 
 
\section{Monopole bending}
\label{bending}

\begin{figure}[tb]
  \begin{center}
    \includegraphics[width=0.49\linewidth]{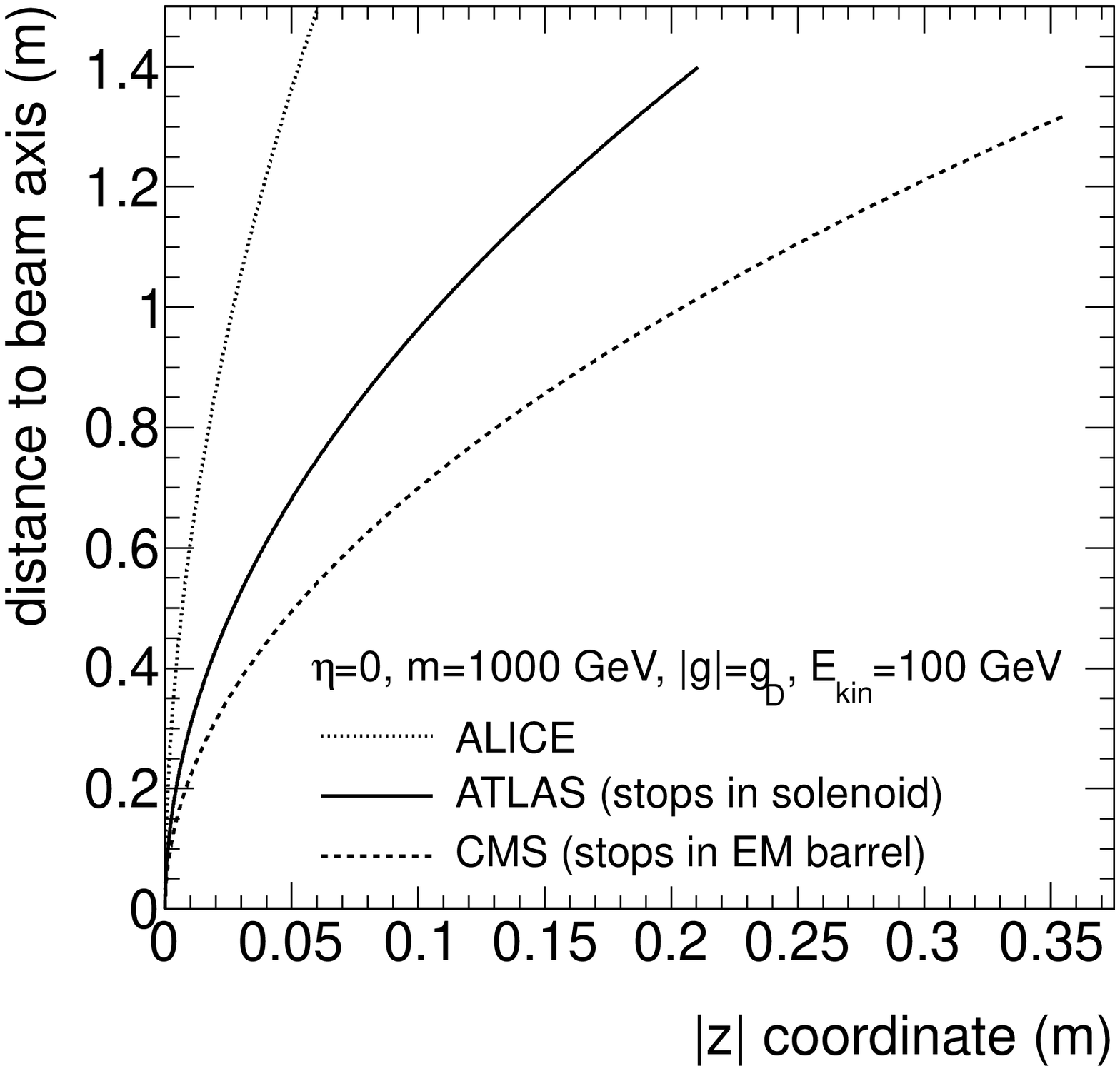}
    \includegraphics[width=0.49\linewidth]{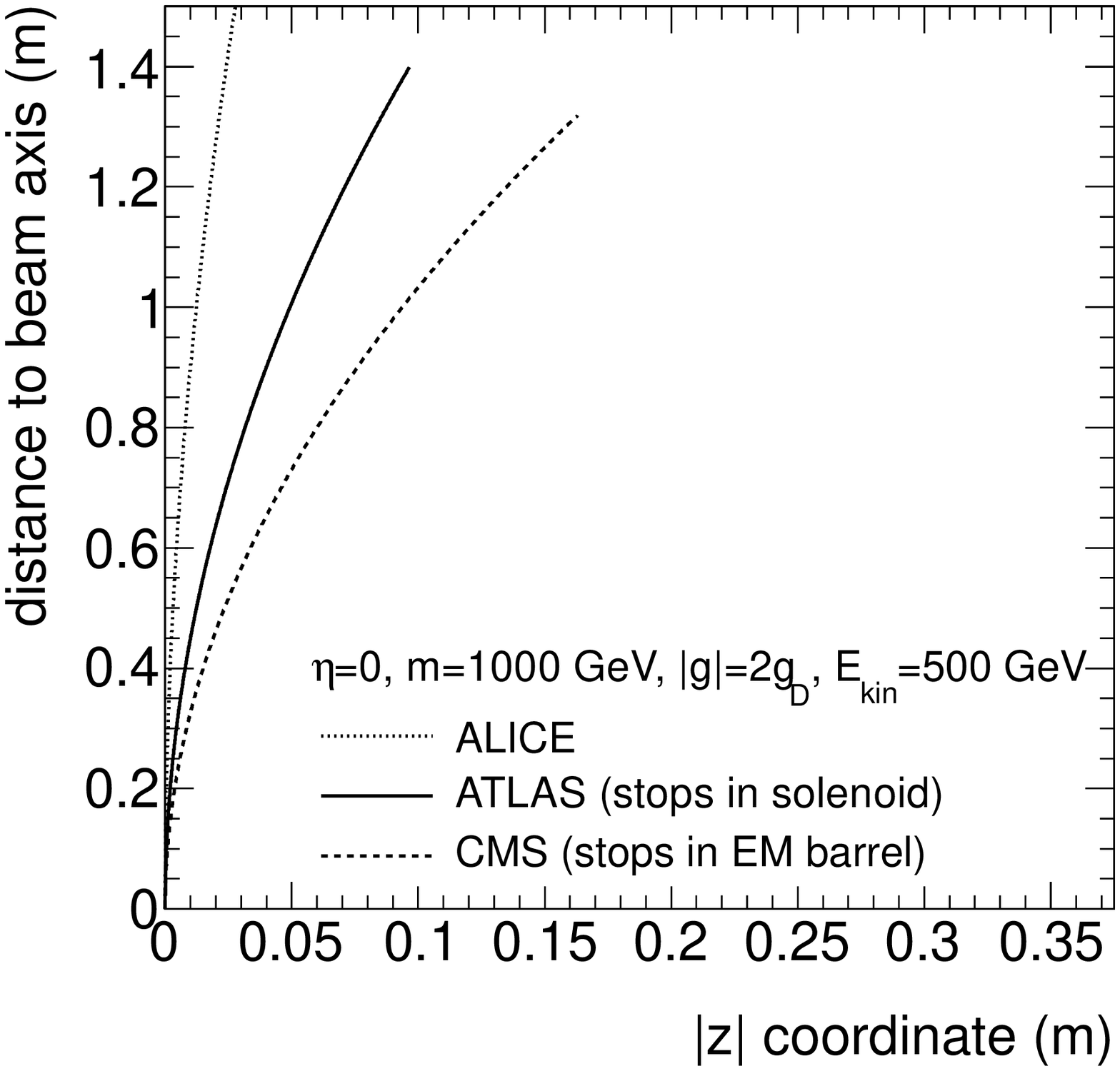}    
  \end{center}
  \caption{The approximate trajectory of a $m=1000$ GeV monopole assuming a uniform solenoid magnetic field in the ALICE ($B=0.5$~T), ATLAS ($B=2.0$~T) and CMS ($B=3.8$~T) detectors at $\eta=0$, for $|g|=g_D$ (left) and $|g|=2g_D$ (right), and corresponding to initial kinetic energies for punching through the inner detector: $E_{kin}=100$~GeV (left) and $E_{kin}=500$~GeV (right). }
  \label{fig:bending}
\end{figure}

A monopole with magnetic charge $q_m$ in a magnetic field $B$ experiences a force $q_mB$ in the field direction. In vacuum, with the field aligned along the beam axis ($z$ direction), it has a parabolic trajectory in the $r-z$ plane~\cite{HERASQUID}:

\begin{equation}
z(r)-z_v=0.5\frac{q_m|B|r^2}{p_T\beta_Tc}+\frac{r}{\tan\theta_0}
\label{traj}
\end{equation}

\noindent where $z(r)$ is the coordinate of a point on the trajectory at distance $r$ from the beam (with the particle originating at $r=0$ and $z=z_v$), $p_T$ and $\beta_T$ are the transverse momentum and transverse velocity of the monopole and $\theta_0$ is its initial angle with respect to the beam direction. Trajectories estimated using Equation~\ref{traj} for monopoles with charges $|g|=g_D$ and $|g|=2g_D$ produced at $\eta=0$ with values of initial kinetic energies chosen such that the monopoles just reach the front of the EM calorimeters are shown in Fig. \ref{fig:bending}. The largest possible displacement in the $z$ direction of a HIP exiting the inner tracker is 40 cm for a monopole with $E_{kin}=100$ GeV and $|g|=g_D$ in the CMS detector. The EM calorimeter barrel extends typically up to $|z|=3$ m. This illustrates that it is a reasonable approximation to neglect the bending of monopoles in solenoid magnetic fields when estimating monopole ranges in LHC detectors. Bending does not affect significantly the ability of a monopole with given energy to reach the EM calorimeter. It may affect monopole track reconstruction efficiency (such considerations are beyond the scope of this paper). Also, the energy gained or lost by a monopole due to the force exerted by a magnetic field in a detector is negligible compared to the ionisation energy loss.

\section{Challenges of HIP detection} 
\label{challenges}

Perturbative calculations cannot be performed in the case of a large coupling, due to divergent terms in the expansion. This is well known in soft QCD, and the same applies to the coupling to the photon in the case of large charges ($|z|$ or $|g|~\apprge 5$). Therefore, it is conservative and prudent to assume that nothing is known about HIP production mechanisms. A reliable phenomenological model of high-energy HIP interactions\footnote{On the other hand, HIP interactions at the atomic scale are fairly well understood~\cite{PDG2010}  (see Section~\ref{velocity}) and supported by measurements of heavy ions propagating in matter.} could be formulated only if the characteristics of such interactions were measured using a real signal. In the absence of such data, nothing can be certain about HIP cross sections, nor angular and energy distributions, nor their dependence upon the collision energy. Assuming a HIP production mechanism such as Drell-Yan is useful as a benchmark but does not provide a reliable prediction. Likewise, searches at different collision energies or in different types of collisions (such as $e^+e^-$, $ep$, $p\bar{p}$ and $pp$) cannot be directly compared. What motivates searching for HIPs at the LHC is that a new energy regime is opened. There is always the possibility that HIP production would be made possible only in the new energy regime, and certainly there is a new possibility for probing higher HIP masses than before. For a given mass and charge, a model-independent way to present HIP search results is to provide an estimate of the detector efficiency as a function of HIP production angle and energy~\cite{QballATLAS10}. It can also be useful to set cross section limits for a well-defined presumed pair production model. While HIP mass limits may also be quoted accompanied with a careful description of the chosen cross section, they do not provide additional information and can be misleading since this choice is arbitrary. Search results are often interpreted only for a monopole with the Dirac charge. Considering various charges within the bounds of the detector capabilities is more informative and allows for interpolations to intermediate values~\cite{TASSO88}. It is important that experiments try to test as many mass and charge combinations as possible, in addition to trying to cover as broad a range as possible in particle angle and energy.

HIPs can be sought by directly tracking their passage through an active general-purpose detector and looking specifically for a highly-ionising signature. Searches were made recently with the OPAL detector at LEP~\cite{OPALdirect} and with the CDF detector at the Tevatron~\cite{Abulencia:2005hb}, interpreted for monopoles with Dirac magnetic charge ($|g|=g_D$); and with the ATLAS detector at the LHC~\cite{QballATLAS10}, interpreted for electrically charged particles in the range $6\leq |z|\leq 17$. The difficulties in these searches come from the fact that the subdetectors of general-purpose experiments are designed to detect minimum-ionising particles and not HIPs. Effects arising from the particle's low velocity and from the high density of the energy deposition, such as electronics saturation, light quenching in scintillators, and adjacent hits from delta electrons, require a non-trivial treatment. The response of each subdetector to high ionisation cannot be easily calibrated, and signal efficiency estimates tend to rely heavily on simulations. To illustrate this point, in the ATLAS search, the dominant source of uncertainty comes from the modeling of the effect of electron-ion recombination in the liquid-argon calorimeter in the case of a high energy loss~\cite{QballATLAS10,Birks11}.

An approach for dedicated HIP searches which was used at LEP~\cite{Kinoshita:1992wd,Pinfold:1993mq} and the Tevatron~\cite{Price:1987py,Price:1990in,Bertani:1990tq} to set limits in ranges of both electric and magnetic charges (up to $\sim 4g_D$), is the so-called track-etch technique. In such experiments, which are run in a passive mode (without electronics), the interaction point is surrounded by thin plastic sheets. A HIP would cause permanent damage along its trajectory in the plastic. After etching the surface of the plastic in a controlled manner, this would be revealed in the form of etch-pit cones of micrometer dimensions. The advantages of such a device include the lack of triggering constraints and the possibility to easily test its response using ion beams~\cite{Pinfold:2009zz}. This technique will be applied at the LHC with the MoEDAL\footnote{MoEDAL stands for Monopole and Exotics Detector at the LHC.} experiment, which consists of an array of plastic track-etch detectors to be deployed around the LHCb interaction point for 14 TeV collision runs~\cite{Moedal}. MoEDAL is described in more detail along with the other relevant LHC detectors in Section~\ref{detectors}.

Another complementary approach for monopole searches is the induction method, in which parts of accelerator and detector material surrounding the collisions (such as the beam pipe) are analysed by passing them through a sensing coil coupled to a superconducting quantum interference device (SQUID). The binding energies of monopoles in nuclei with finite magnetic dipole moments are estimated to be hundreds of keV -- much greater than the binding energies of electrons to atoms and of atoms in solids~\cite{Milton2006}. Somewhat more speculatively, as the magnetic field in the vicinity of the monopole is so strong that it could disrupt the nucleus, it is thought that binding would even occur to individual protons and neutrons, even in spin zero nuclei without magnetic moment. In any case, both $^9$Be and $^{27}$Al, the principal materials used for beam pipes, have strong nuclear magnetic moments and should form bound states with monopoles~\cite{Milton2006}. Hence, in searches with the induction method, it is assumed to be difficult to dislodge a monopole from the sample once it has stopped. If a trapped magnetic charge is present in the sample, a measurable current will appear in the superconducting coil by induction and remain after the sample has passed through. This technique was used at HERA~\cite{HERASQUID} and the Tevatron~\cite{TEVATRONSQUID2000,TEVATRONSQUID2004} and allows the extraction of limits on monopole production in a specific kinematic regime (low angle and/or low energy) and for magnetic charges up to very high values. Monopoles that traverse very short distances (of the order of a few mm) before being trapped can be detected with the induction method. This is complementary to searches that involve tracking the particle through the entire detector. The possible application of the induction method to search for monopoles in the ATLAS and CMS beam pipes is discussed in Section~\ref{trapped}. 

An indirect way to search for HIPs is through photon signatures. With their large electromagnetic coupling, HIPs may be expected to radiate photons, and also to annihilate into photons shortly after their production. However, as discussed above, the non-perturbative processes involved are impossible to quantify in a reliable way. That does not prevent experiments from looking for such signatures, as in a search with the D0 detector~\cite{D0indirect}, but it makes interpretations of the results difficult~\cite{photoncritic}.

\section{Description of LHC detectors}
\label{detectors}

Detailed descriptions of the ATLAS, CMS, LHCb, ALICE and MoEDAL detectors are given in Refs.~\cite{ATLAS08,CMS08,LHCb08,ALICE08,Moedal}, respectively. An overview of their relevant features is given below. 

ATLAS~\cite{ATLAS08} is a multipurpose particle physics detector with a forward-backward symmetric cylindrical geometry. Inner detector (ID) tracking covers the pseudorapidity\footnote{The pseudorapidity is defined as a function of the polar angle (relative to the beam axis) $\theta$ as: $\eta=-\ln\left(\tan(\theta/2)\right)$.} range $|\eta|<2.5$ and is performed by silicon-based detectors and a transition radiation tracker. A uniform magnetic field of 2~T along the beam axis is produced by a thin superconducting solenoid surrounding the ID. A liquid-argon sampling electromagnetic (EM) calorimeter with accordion-shaped electrodes and lead absorbers surrounds the ID. It comprises barrel ($|\eta|<1.475$) and endcap ($1.375<|\eta|<3.2$) components and has two main layers (EM1 and EM2). Hadronic calorimetry (Had) is provided by a scintillating tile detector with iron absorbers for $|\eta|<1.7$, and liquid-argon detectors for $1.5<|\eta|<4.9$. Beyond the calorimeters, muons are detected with a multi-system muon spectrometer. In the forward region ($2.5<|\eta|<4.9$), the coarse granularity of the calorimeters and the lack of tracking detectors would make HIP identification difficult. In this work, it is assumed that ATLAS would be able to efficiently separate HIPs which penetrate the EM calorimeter from backgrounds for $|\eta|<2.5$. 

CMS~\cite{CMS08} is a multipurpose particle physics detector with a forward-backward symmetric cylindrical geometry. Inner tracking is performed by silicon pixel and strip detectors in the pseudorapidity range $|\eta|<2.5$. These are followed by a lead tungstate crystal EM calorimeter and a brass-scintillator hadronic calorimeter, which cover the range $|\eta|<3.0$. Surrounding the calorimeters, a superconducting solenoid provides a uniform magnetic field of 3.8~T along the beam axis. A steel/quartz-fiber Cherenkov calorimeter extends the calorimeter coverage to $|\eta|<5.0$. The steel return yoke outside the solenoid is instrumented with gas detectors used to identify muons. As in ATLAS, HIP identification by the inner tracker and EM calorimeter in CMS is assumed to be efficient in the range $|\eta|<2.5$.

LHCb~\cite{LHCb08} is a single-arm spectrometer specialised in the identification of heavy flavoured hadrons. It provides reconstruction of charged particles in the forward pseudorapidity range $2.0<\eta<4.9$. The detector elements are placed along the LHC beam line starting with the vertex detector (VELO), a silicon strip device that surrounds the interaction region. Track momenta are measured with the VELO along with a large area silicon strip detector located upstream of a dipole magnet and a combination of silicon strip detectors and straw drift-tubes placed downstream. The magnet has a bending power of 4 Tm. Two ring imaging Cherenkov detectors (RICH1 and RICH2) are used to identify charged hadrons. Beyond RICH2, at a distance of about 12 m from the interaction point, the first muon station is followed by an EM calorimeter, a hadron calorimeter, and the rest of the muon system. In this work, it is assumed that LHCb would be able to efficiently separate HIPs which penetrate the EM calorimeter from backgrounds for $2.0<\eta<4.9$. 

The first level calorimeter trigger systems~\cite{LHCtriggers,ATLASCaloTrig} of ATLAS, CMS and LHCb base their decision on the sum of the energy deposition in calorimeter towers of fixed size, with transverse energy thresholds of the order of 30 GeV for the 2012 LHC runs. Every 25 ns, they must deliver a decision and identify detector regions and a bunch crossing for further processing by high level triggers. A collision event where a calorimeter signal arrives later than 25 ns after the default arrival time of a particle traveling at the speed of light would be either triggered in the next collision and thus assigned to the wrong bunch crossing, or rejected if there are no colliding bunches in the next 25 ns time slot. Recovering such events in a search is very difficult or impossible. This limits the range in $\beta$ for which one can trigger directly on HIPs and stable massive particles in general~\cite{Fairbairn:2006gg,CMSslepton10,ATLASslepton11}. Further higher level trigger requirements are designed by default to identify objects like jets, electrons and photons. Although none of the experiments have documented such developments yet, based on the ATLAS experience~\cite{QballATLAS10} it can be conjectured that specific high level triggers can potentially be implemented for a HIP selection based on the extremely high density of the energy deposition.  

ALICE~\cite{ALICE08} is a general-purpose detector which specialises in the reconstruction of hadrons, electrons, muons, and photons produced in the collisions of heavy ions, up to high multiplicities and down to low momenta. It is also capable of collecting $pp$ collision events at low instantaneous luminosity. It comprises a central barrel in the range $|\eta|<0.9$ and a forward muon spectrometer. The central barrel is interesting for HIP searches thanks to its low material budget. Its detectors are located inside a 0.5~T solenoidal magnet. The Inner Tracking System (ITS) consists of silicon detectors surrounding the beam pipe at radial positions between 3.9 cm and 43.0 cm. The time projection chamber (TPC) is a cylindrical drift volume of 250 cm outer radius, used for tracking as well as particle identification via a measurement of the ionisation energy loss (d$E$/d$x$) in the detector gas. Beyond the TPC, further detection is performed by the transition radiation detector (TRD), the time-of-flight detector (TOF), and finally two EM calorimeters which do not cover the full angular acceptance. In $pp$ collision mode, two VZERO detectors are used for triggering. They consist of scintillator arrays and cover the pseudorapidity ranges $-3.7<\eta<-1.7$ and $2.8<\eta<5.1$. The minimum bias (MB) trigger combines information from the VZERO detectors and the ITS to trigger on inelastic collisions with an efficiency of 86.4\%~\cite{ALICE12}. In this work, it is assumed that ALICE has the capability to efficiently trigger on (using the MB trigger) and identify events where at least one HIP traverses the TPC. 

\begin{table}[tb]
  \begin{center}
    \begin{tabular}{l|c|cc|cc|cc|}
      \multicolumn{2}{c|}{} & \multicolumn{2}{c|}{inner tracker} & \multicolumn{2}{c|}{services/other} & \multicolumn{2}{c|}{EM calorimeter}\\ 
      detector & pseudorapidity range & $d$ (mm) & d$x$ (g$\cdot$cm$^{-2}$) & $d$ (mm) & d$x$ (g$\cdot$cm$^{-2}$) & $d$ (mm) & d$x$ (g$\cdot$cm$^{-2}$) \\ 
      \hline \hline    
      ATLAS  & $|\eta|<0.9$   & 1080 & 11.0  & 1300 & 20.8 & 1970 & 264.5 \\ 
      CMS    & $|\eta|<0.9$   & 1290 &  9.6  & --   & --   & 1520 & 190.1 \\ 
      ALICE  & $|\eta|<0.9$ & 2500 &  2.8  & 4000 & 10.5 & 4800 & 127.4 \\ 
      \hline
      ATLAS  & $0.9<|\eta|<1.5$ & 2330 & 28.8 & 2800 & 38.8 & 4240 & 287.5 \\     
      CMS    & $0.9<|\eta|<1.5$ & 2780 & 40.8 & --   & --   & 3270 & 190.1 \\     
      \hline 
      ATLAS  & $1.5<|\eta|<2.5$ & 3700 & 24.0 & 3800  & 20.8 & 4350  & 431.3 \\     
      CMS    & $1.5<|\eta|<2.5$ & 2900 & 28.8 & 3300  & 19.1 & 3700  & 182.0 \\     
      LHCb   & $2.0<\eta<3.9$ & 1000 &  4.2 & 12230 & 36.7 & 12400 & 159.3 \\       
      \hline
    \end{tabular}
  \caption{Typical depth $d$ (from the interaction point to the exit of the detector) and thickness d$x$ (in g$\cdot$cm$^{-2}$) for components of the ATLAS, CMS, LHCb and ALICE detectors, in pseudorapidity ranges for which they have near 100\% geometrical acceptance. The values are obtained from Refs.~\cite{ATLAS08,CMS08,LHCb08,ALICE08}.}
  \label{tab:detectors}
  \end{center}
\end{table}

\begin{table}[tb]
  \begin{center}
    \begin{tabular}{c|c|c|c|}
      pseudorapidity range & \multicolumn{2}{c|}{geometrical acceptance} & d$x$ (g$\cdot$cm$^{-2}$) \\ 
       & standard MoEDAL ($z/\beta\geq 5$) & VHCC ($z/\beta\geq 50$) &  \\ 
      \hline \hline
      $1.5<\eta<5.0$   & 0\%  & 95\% & 4.2 \\ 
      \hline
      $0.8<\eta<1.5$   & 50\% & 70\% & 72  \\ 
      \hline
      $-3.2<\eta<0.8$  & 60\% & 40\% & 14  \\ 
      \hline
      $-4.2<\eta<-3.2$ & 90\% & 60\% & 96  \\ 
      \hline
      $-5.0<\eta<-4.2$ & 90\% & 90\% & 7.2   \\ 
      \hline
    \end{tabular}
  \caption{Typical geometrical acceptance and thickness d$x$ of materials in front of the MoEDAL detectors for various pseudorapidity ranges. The values are obtained from Refs.~\cite{Moedal,VHCC}.}
  \label{tab:moedal}
  \end{center}
\end{table}

Material budgets of the various detector components are provided in Refs.~\cite{ATLAS08,CMS08,LHCb08,ALICE08} and reported in Table~\ref{tab:detectors} for inner, intermediate, and EM parts of the ATLAS, CMS, LHCb and ALICE detectors. This information is used in this work to compute HIP ranges and fractions of HIPs which reach the sensitive parts of the detectors (EM calorimeter for ATLAS, CMS and LHCb, end of TPC for ALICE). Uncertainties due to approximations made for thicknesses (given in radiation lengths in the original references) and exact types of material in complex detector parts were estimated. For instance, the ATLAS and CMS inner tracking systems comprise various elements such as silicon, aluminium, copper and carbon, and an average was used. The resulting relative uncertainty in the detector thicknesses given in Tables~\ref{tab:detectors} and~\ref{tab:moedal} is 20\%. 

The MoEDAL detector is made of thin sheets of two types of polymers: type one (CR39) is sensitive to particles with d$E$/d$x$ corresponding to $z/\beta\geq 5$ (see Equation~\ref{Bethe}), and type 2 (lexan and makrofol) is sensitive to $z/\beta\geq 50$~\cite{Moedal,VHCC}. The standard MoEDAL detector array described in the experiment's technical design report~\cite{Moedal} consists of stacks of three 0.5 mm thick type-one sheets, three 0.5 mm thick type-two sheets, and three 0.2 mm thick type-two sheets, with 1 mm thick aluminum front and end plates. This array will be deployed on the walls of the LHCb VELO cavern, covering the pseudorapidity range where it does not overlap with LHCb ($-5.0<\eta<1.5$). As indicated in Table~\ref{tab:moedal}, in this range, the LHCb VELO detector and its housing (in front of MoEDAL) are responsible for an amount of material typically comparable with or larger than that of the ATLAS and CMS inner trackers (shown in Table~\ref{tab:detectors}). There is a recent proposal~\cite{VHCC} for adding a thinner detector array called very high charge catcher (VHCC). The VHCC consists of three 0.1 mm thick type-two sheets encased in a 0.025 mm thick aluminium foil envelope. It will be deployed around the VELO vacuum chamber, and also on the front and rear faces of the LHCb RICH1 detector. Thus, it will cover the area in front of the VELO tracking system ($1.5<\eta<4.9$) which amounts for an average thickness of only 4.2 g$\cdot$cm$^{-2}$. Being so thin, the VHCC will not disturb the normal LHCb operation. On the other hand, being entirely made of type-two sheets, it will only be sensitive to high ionisation energy loss, corresponding to $z/\beta\geq 50$. 

\begin{table}[tb]
  \begin{center}
    \begin{tabular}{l|c|c|}
      interaction point & $L$ (cm$^{-2}$s$^{-1}$) & $\int{L dt}$ (fb$^{-1}$) \\
      \hline
      ATLAS  & $5\cdot 10^{33}$ & 20 \\
      CMS    & $5\cdot 10^{33}$ & 20 \\           
      LHCb/MoEDAL   & $5\cdot 10^{32}$ & 2 \\           
      ALICE  & $10^{30}$        & 0.004 \\          
      \hline  
    \end{tabular}
  \caption{Average and integrated luminosities at the ATLAS, CMS, LHCb and ALICE interaction points corresponding to what can roughly be expected from 7-8 TeV $pp$ collisions by the end of 2012~\cite{lumiplans2012}.}
  \label{tab:lumi}
  \end{center}
\end{table}

Finally, Table~\ref{tab:lumi} gives estimates of LHC experiment luminosities for $pp$ collision runs. In 2011, the ATLAS and CMS experiments each accumulated around 5 fb$^{-1}$ of integrated luminosity. In 2012, the LHC is expected to collide protons at 8 TeV and deliver 15 fb$^{-1}$ more to ATLAS and CMS~\cite{lumiplans2012}. The integrated luminosities at the LHCb and ALICE interaction points are typically 10 times and 5000 times lower, respectively~\cite{ALICE08,LHCb08,lumiplans2012}. After the planned shutdown in 2013, the LHC is expected to start running with 14 TeV $pp$  collisions in 2014, with similar relations between the luminosities at the different interaction points, and where the LHCb luminosity will also apply to MoEDAL.

\section{Current limits on HIPs at the LHC}
\label{limits}

To date, the only bounds on HIPs at LHC collision energies (7 TeV $pp$ collisions) were obtained at the ATLAS experiment in an early search for direct production of particles in the electric charge range $6\leq |z|\leq 17$ and mass range $200\leq m\leq 1000$ GeV~\cite{QballATLAS10}. This search makes use of an electron trigger which requires an inner detector track matched to an EM energy deposition with a relatively low (10~GeV) transverse energy threshold, relying on the signature of a HIP stopping inside the EM2 calorimeter layer. Using the equations described in Section~\ref{velocity} for energy loss calculations (see Fig.~\ref{fig:dedx}), the possibility of an interpretation of this search within a monopole scenario can be investigated. Due to the lack of a Bragg peak, monopoles are less likely than electrically charged particles to stop and deposit large amounts of energy in a given detector layer. For instance, while the energy loss of a monopole with $|g|=g_D/4$ is equivalent to that of a $|z|=17$ particle at high $\beta$, such a monopole would not deposit enough energy to induce a trigger while slowing down and stopping in the ATLAS EM calorimeter. It would also need to possess a low ($<100$ GeV) initial kinetic energy to avoid punching through the EM calorimeter, which means that the signal would also likely arrive too late for a reliable first level trigger. A monopole with $|g|=g_D/2$ or higher would satisfy the trigger energy and time criteria upon reaching the EM calorimeter. However, such a monopole would be expected to produce more delta-electrons in the tracker than a $|z|=17$ particle, thus affecting tracking and track-matching related efficiencies -- which is the cause of the $|z|\leq 17$ bound in the ATLAS search~\cite{QballATLAS10}. The conclusion of this study is that, even without considering possible tracking inefficiencies due to monopole bending in the solenoid field, the initial ATLAS search results cannot be extrapolated to magnetic charges. Hence, at the time of writing this paper, magnetic monopoles remain unconstrained by direct LHC searches for highly ionising particles.

\section{The stopping of HIPs at LHC experiments}
\label{stopping}

\begin{figure}[tbh]
  \begin{center}
    \includegraphics[width=0.49\linewidth]{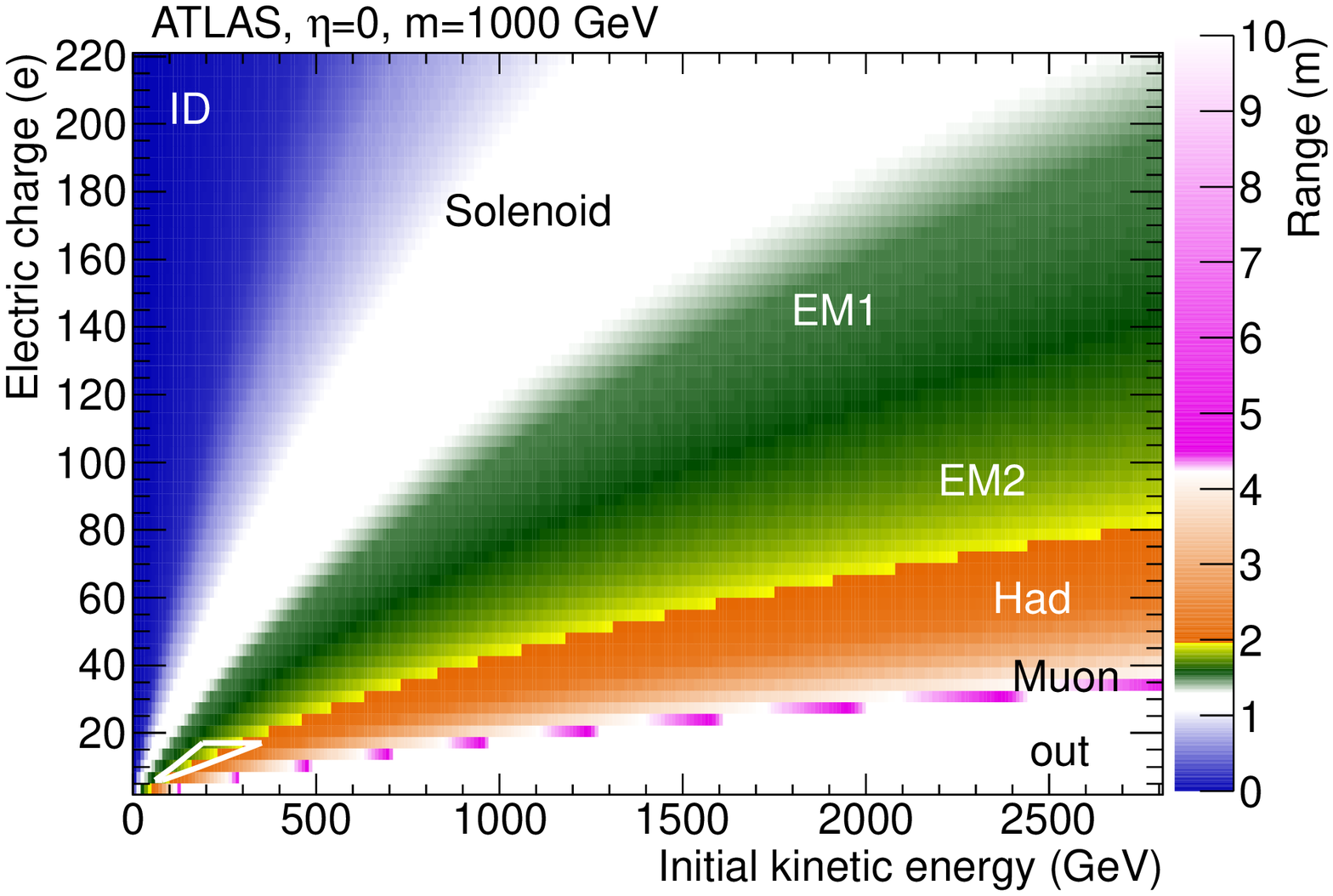}
    \includegraphics[width=0.49\linewidth]{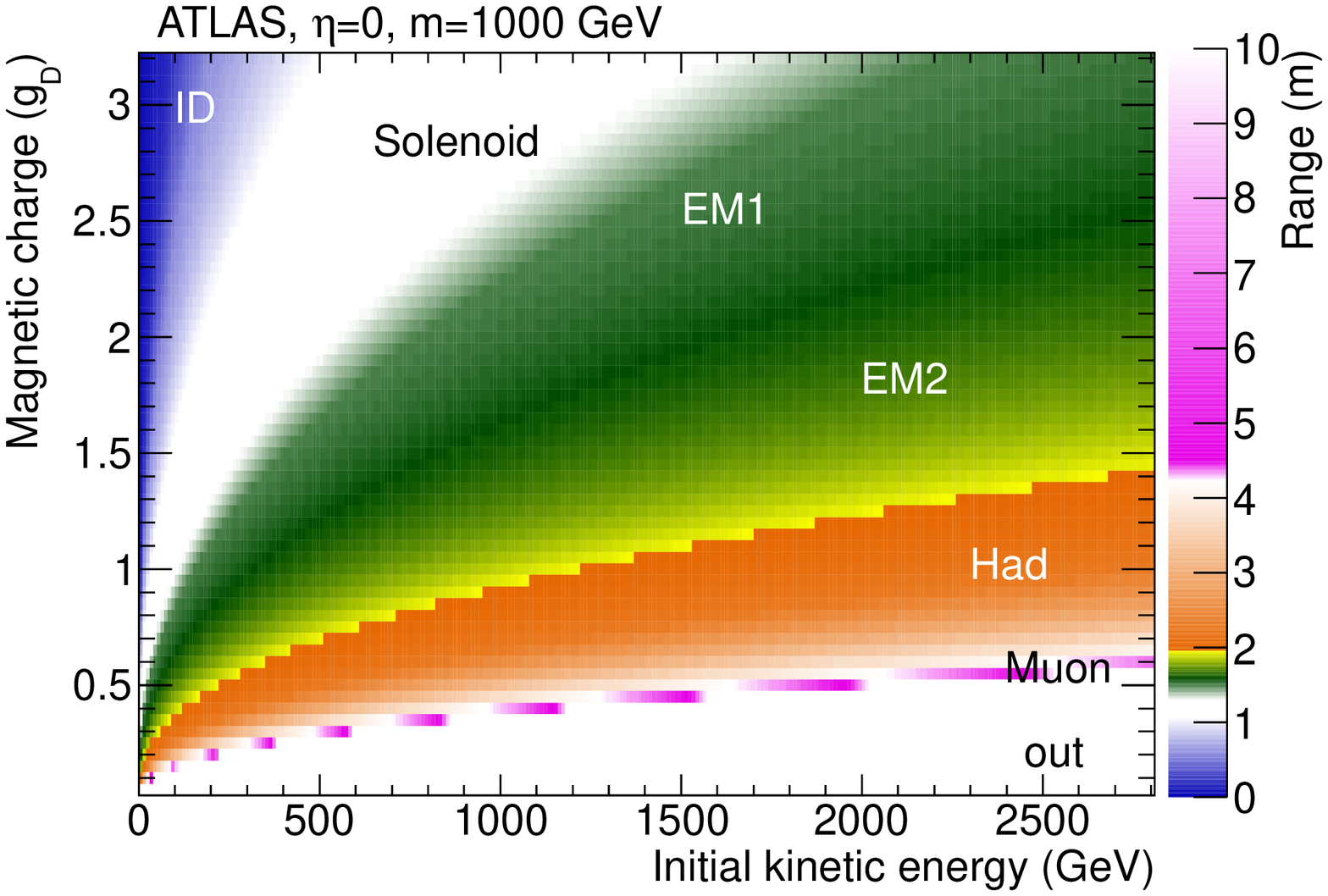}  
    \includegraphics[width=0.49\linewidth]{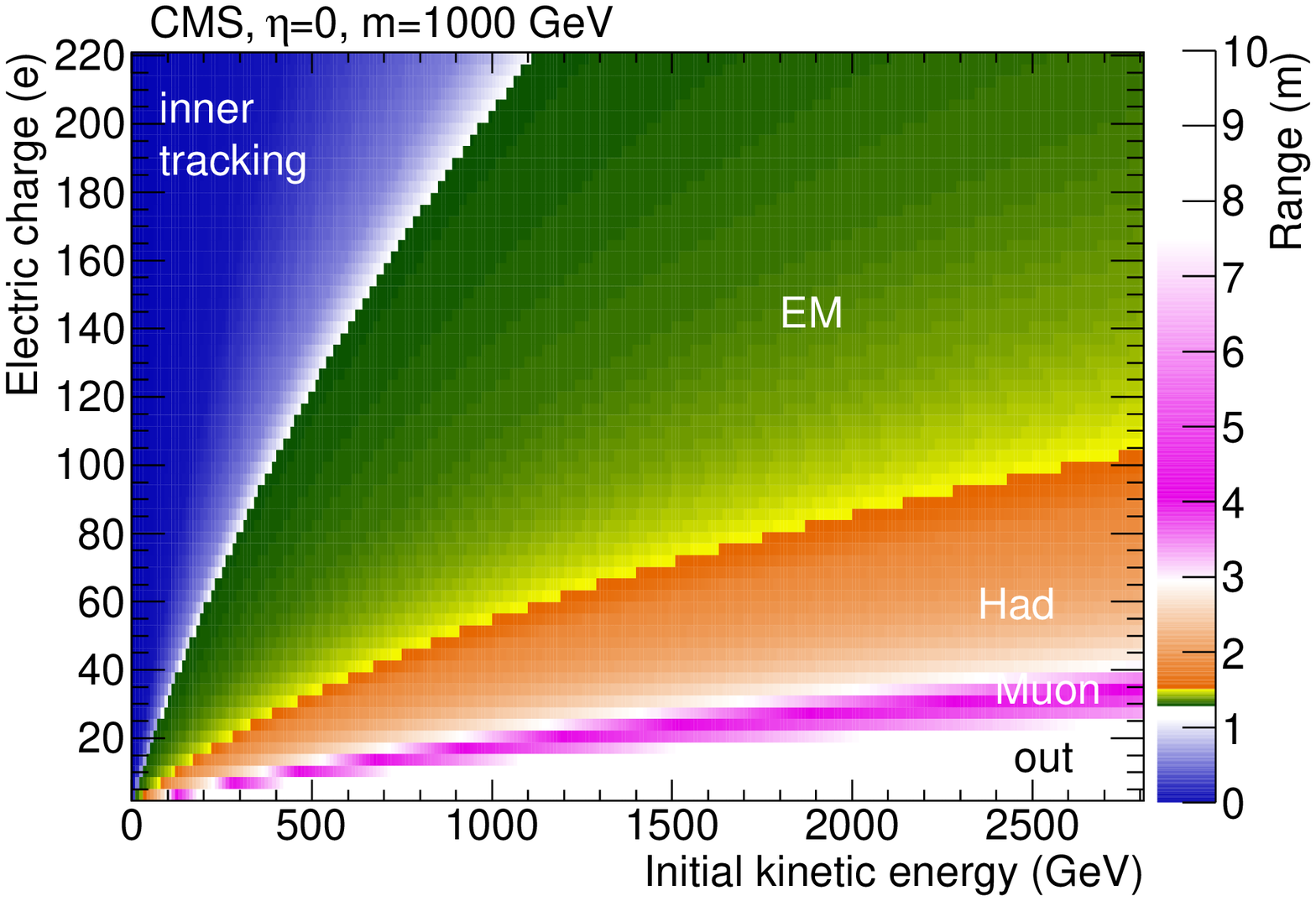}
    \includegraphics[width=0.49\linewidth]{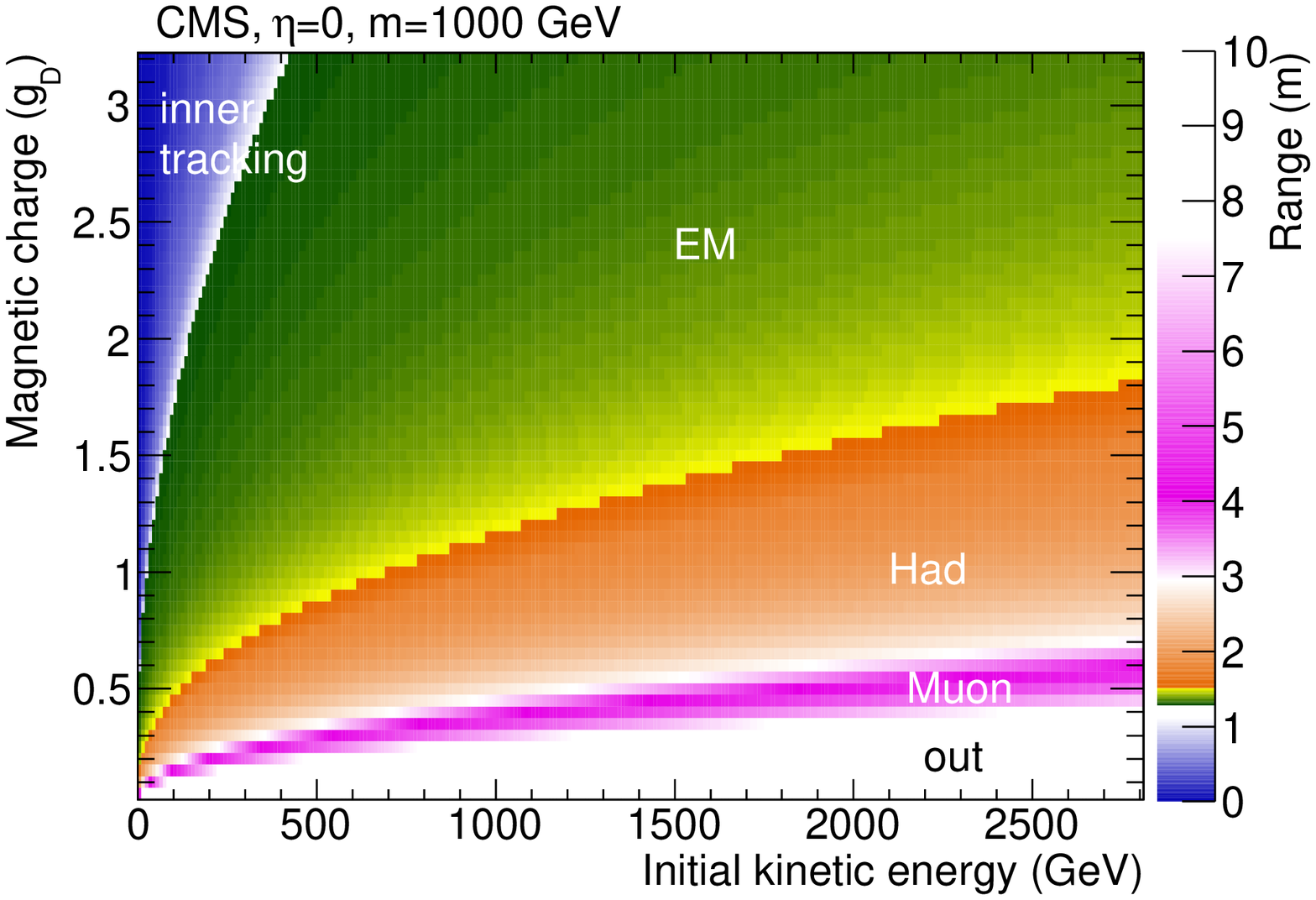}  
    \includegraphics[width=0.49\linewidth]{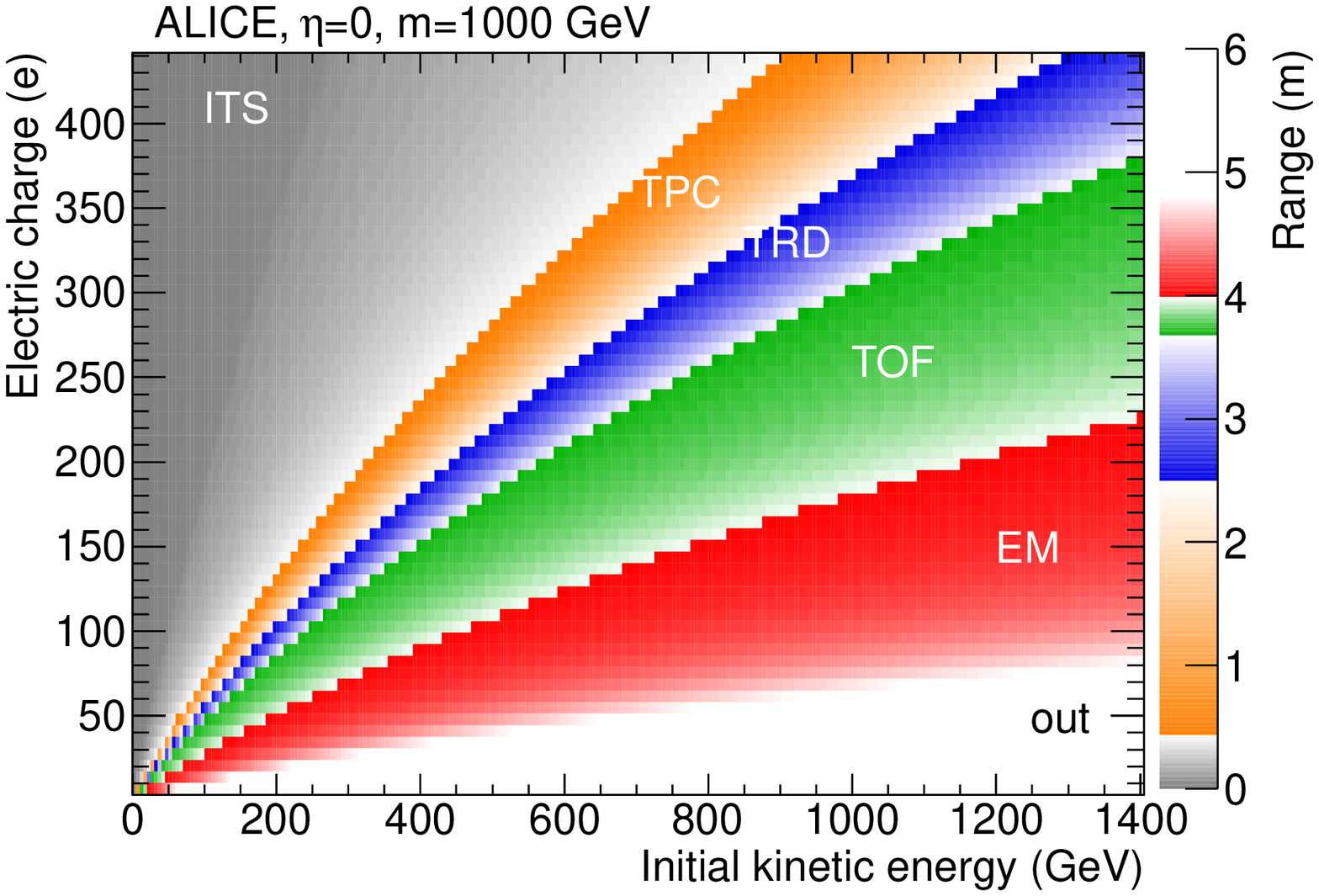}
    \includegraphics[width=0.49\linewidth]{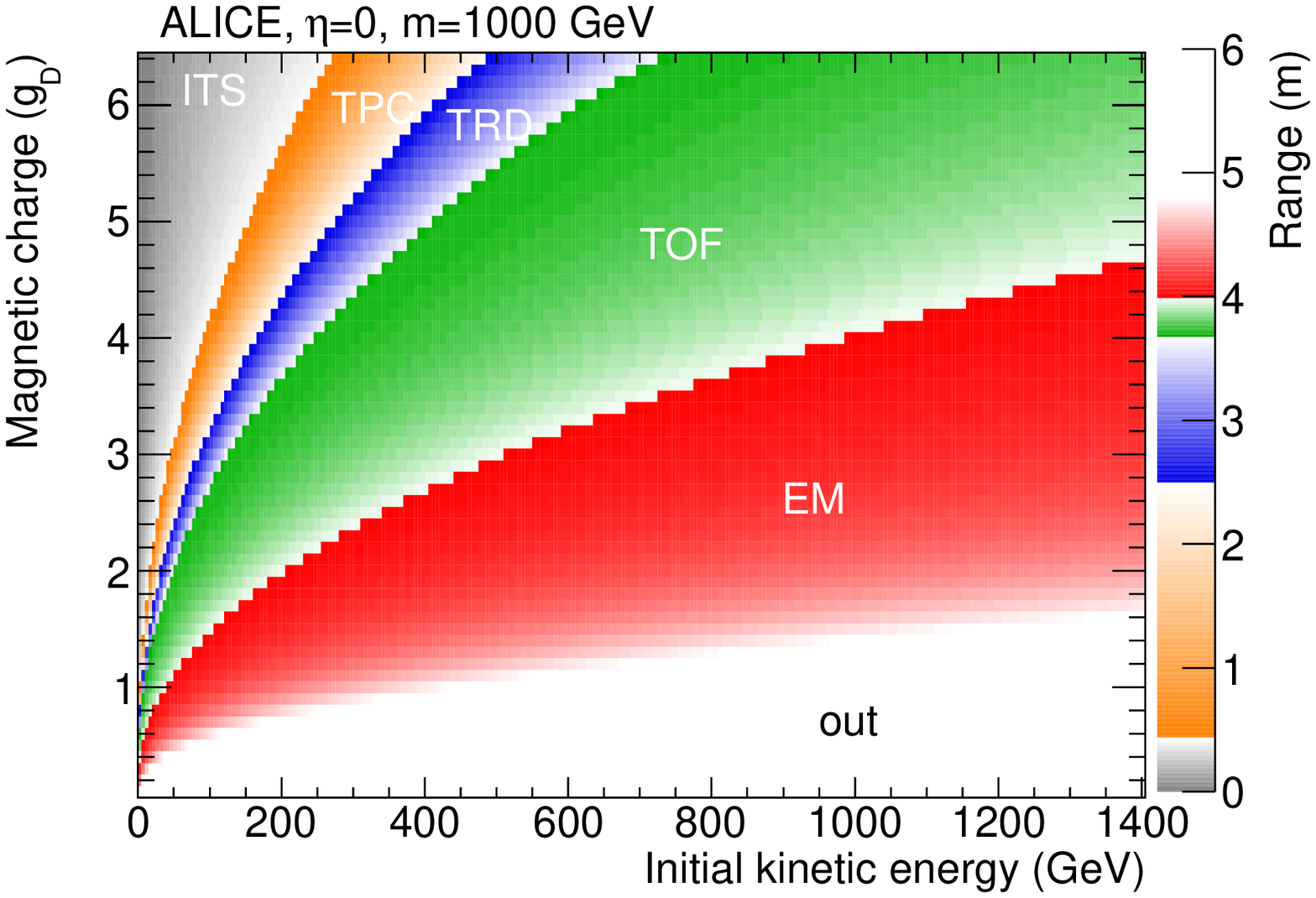}  
  \end{center}
  \caption{Two-dimensional histograms showing the range in meters (with scale indicated by the stripes on the right of the plots) as a function of HIP initial kinetic energy (horizontal axis) and charge (vertical axis), for HIPs with $m=1000$ GeV at $\eta\sim 0$, for electric (left) and magnetic (expressed in units of $g_D=68.5$, right) charges. The area for which the early ATLAS search is sensitive, corresponding to particles stopping in EM2, is indicated by a white line on the top left plot, using the $|q_e|$ and $E_{kin}$ ranges quoted in Ref.~\cite{QballATLAS10}.}
  \label{fig:HIPrange_eta0}
\end{figure}

\begin{figure}[tbh]
  \begin{center}
    \includegraphics[width=0.49\linewidth]{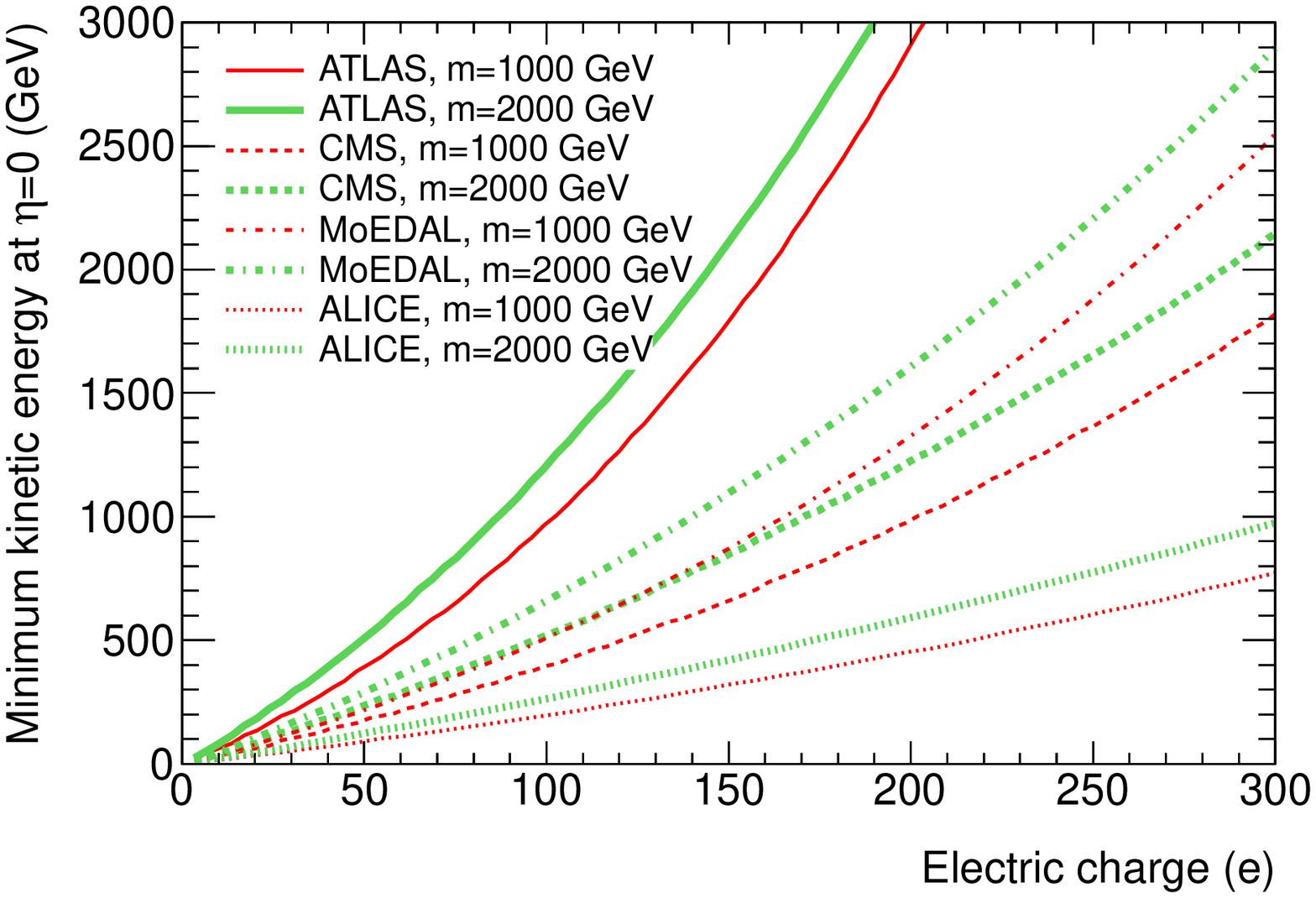}
    \includegraphics[width=0.49\linewidth]{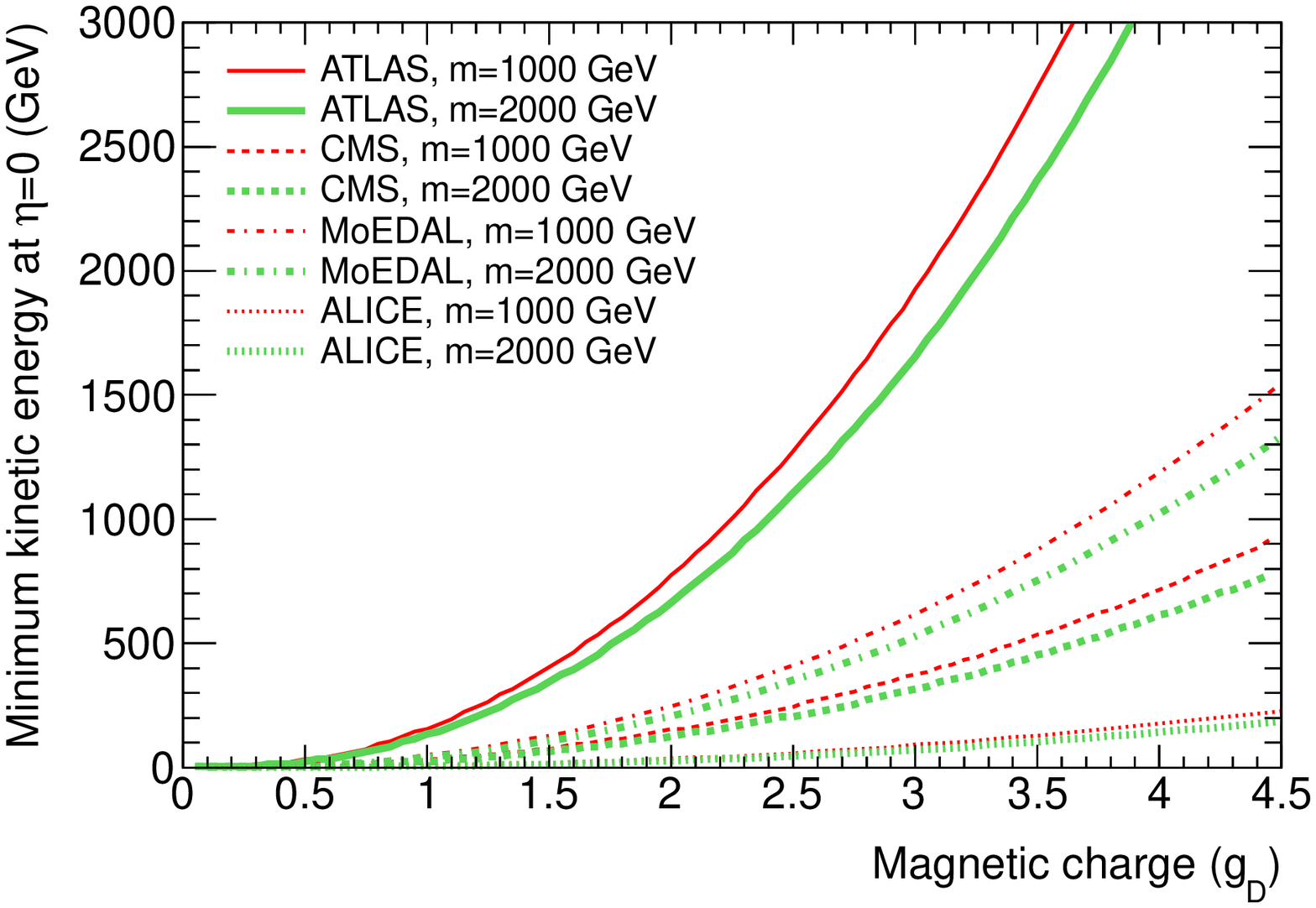}
    \includegraphics[width=0.49\linewidth]{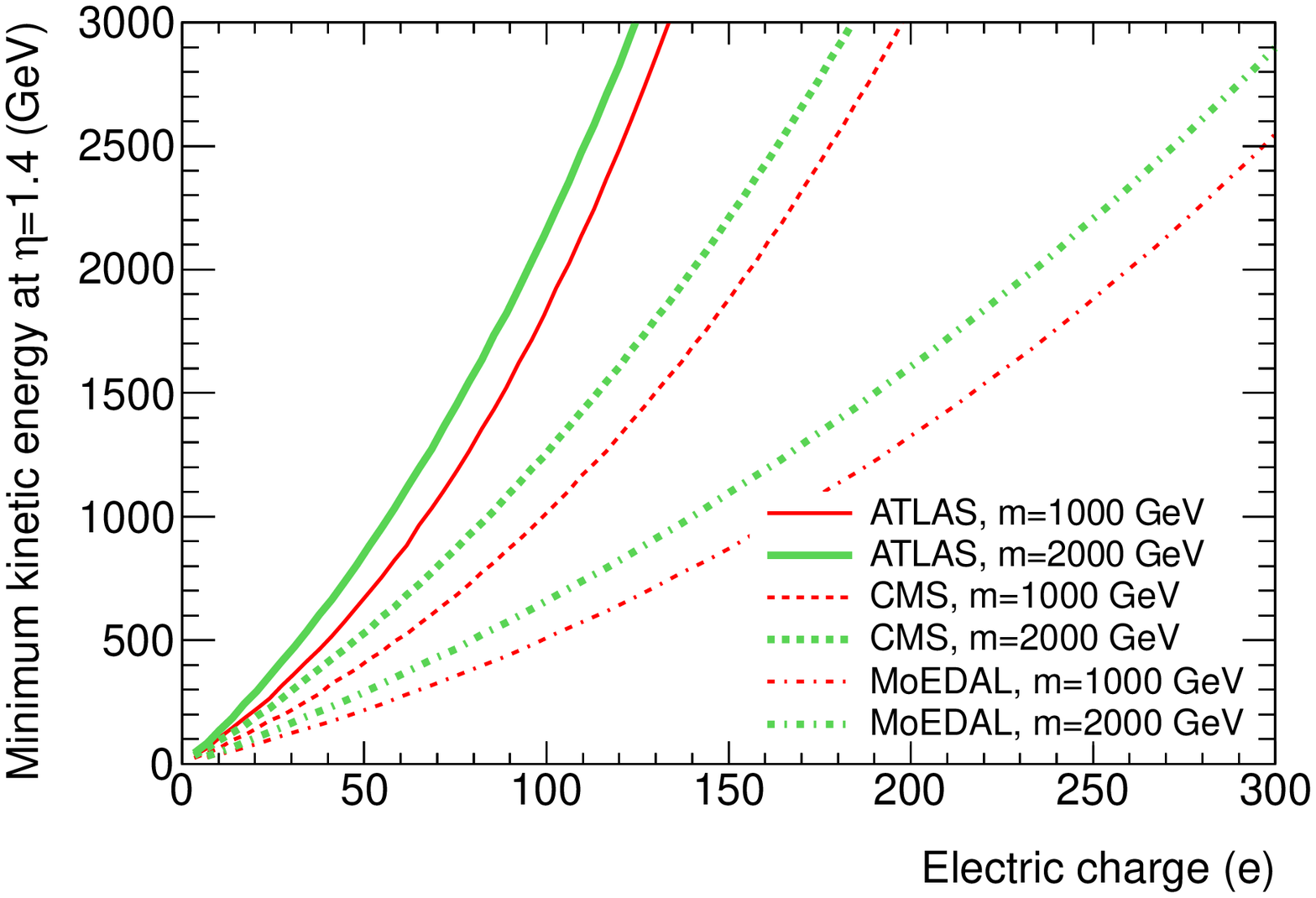}
    \includegraphics[width=0.49\linewidth]{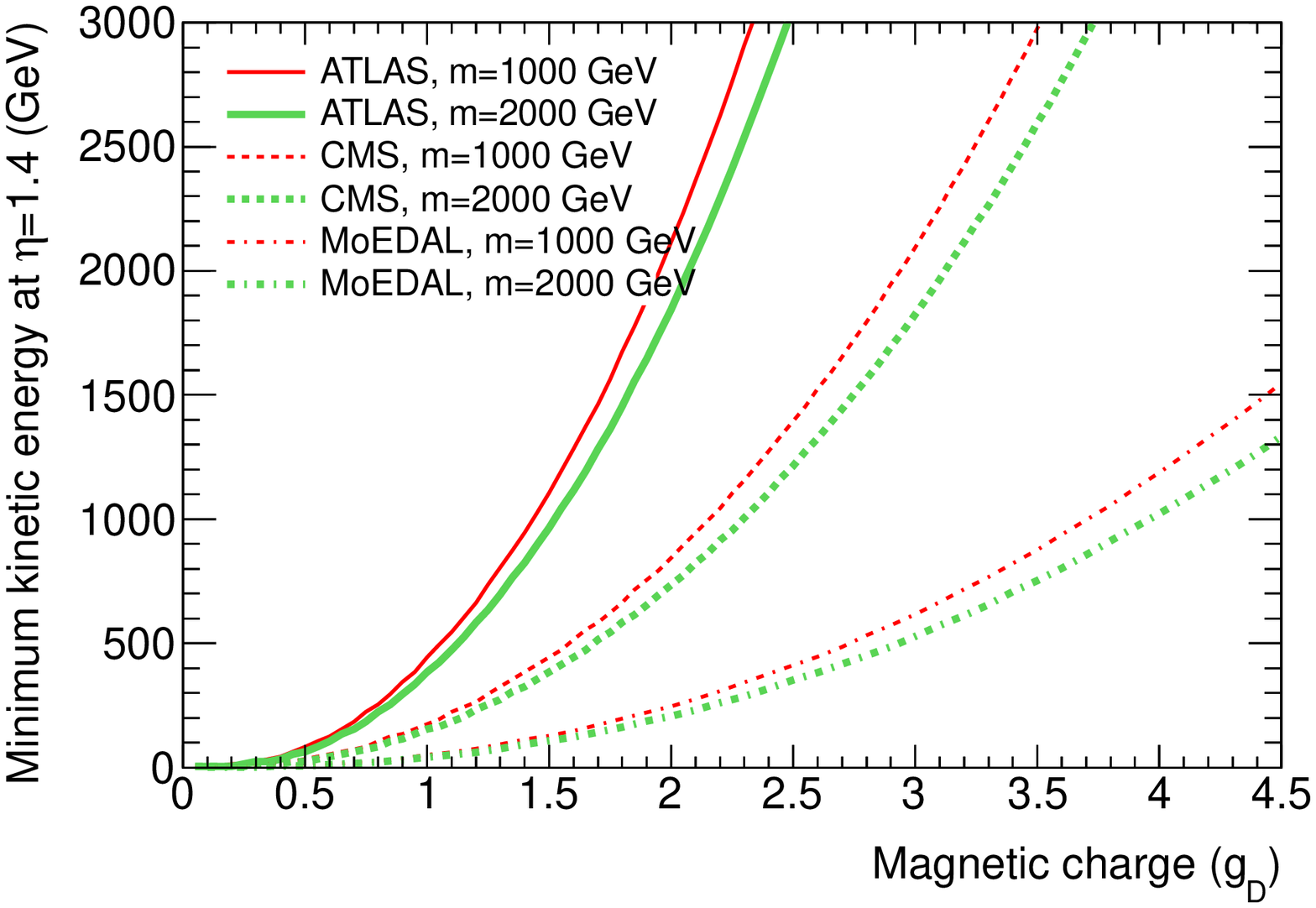}
    \includegraphics[width=0.49\linewidth]{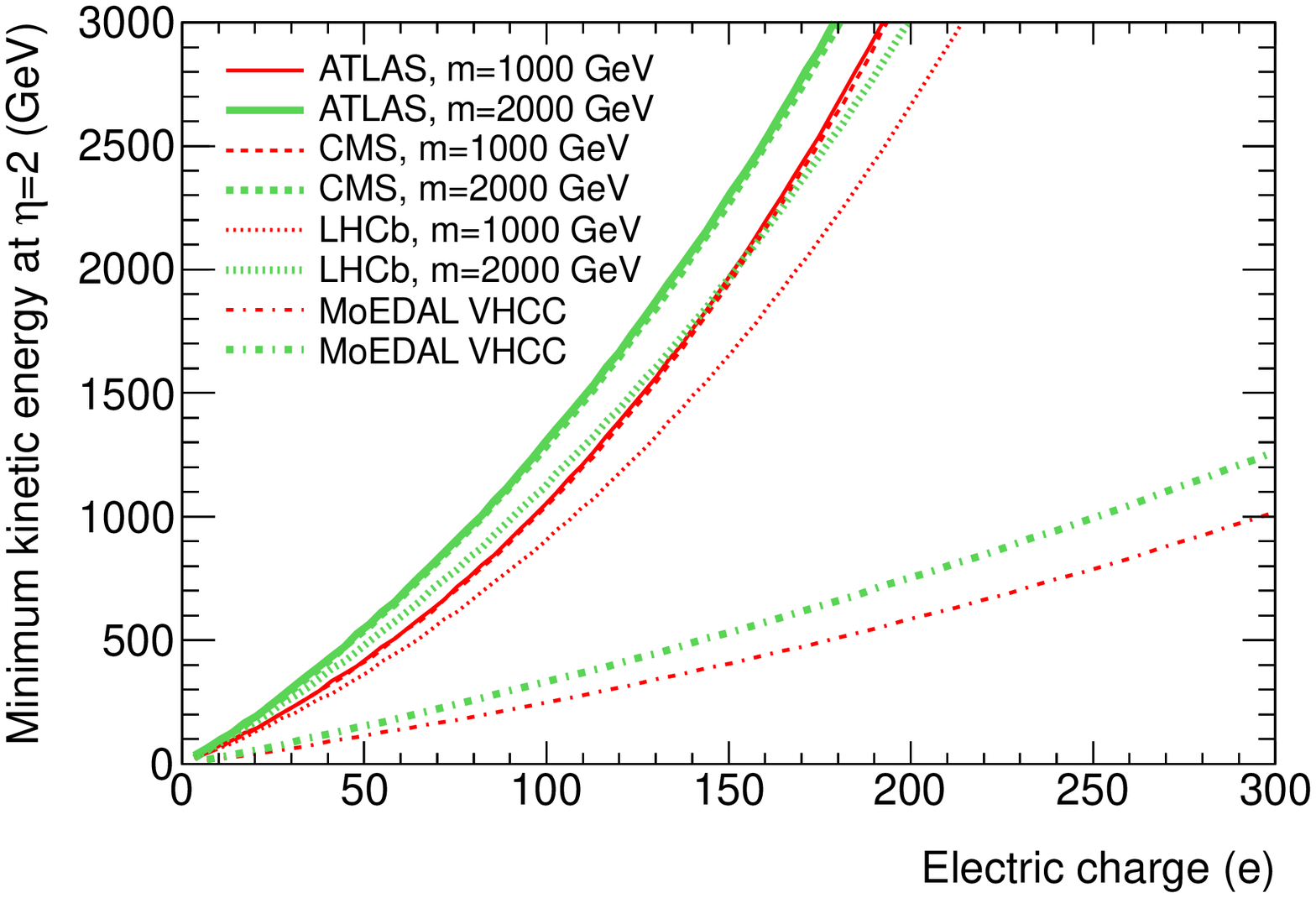}
    \includegraphics[width=0.49\linewidth]{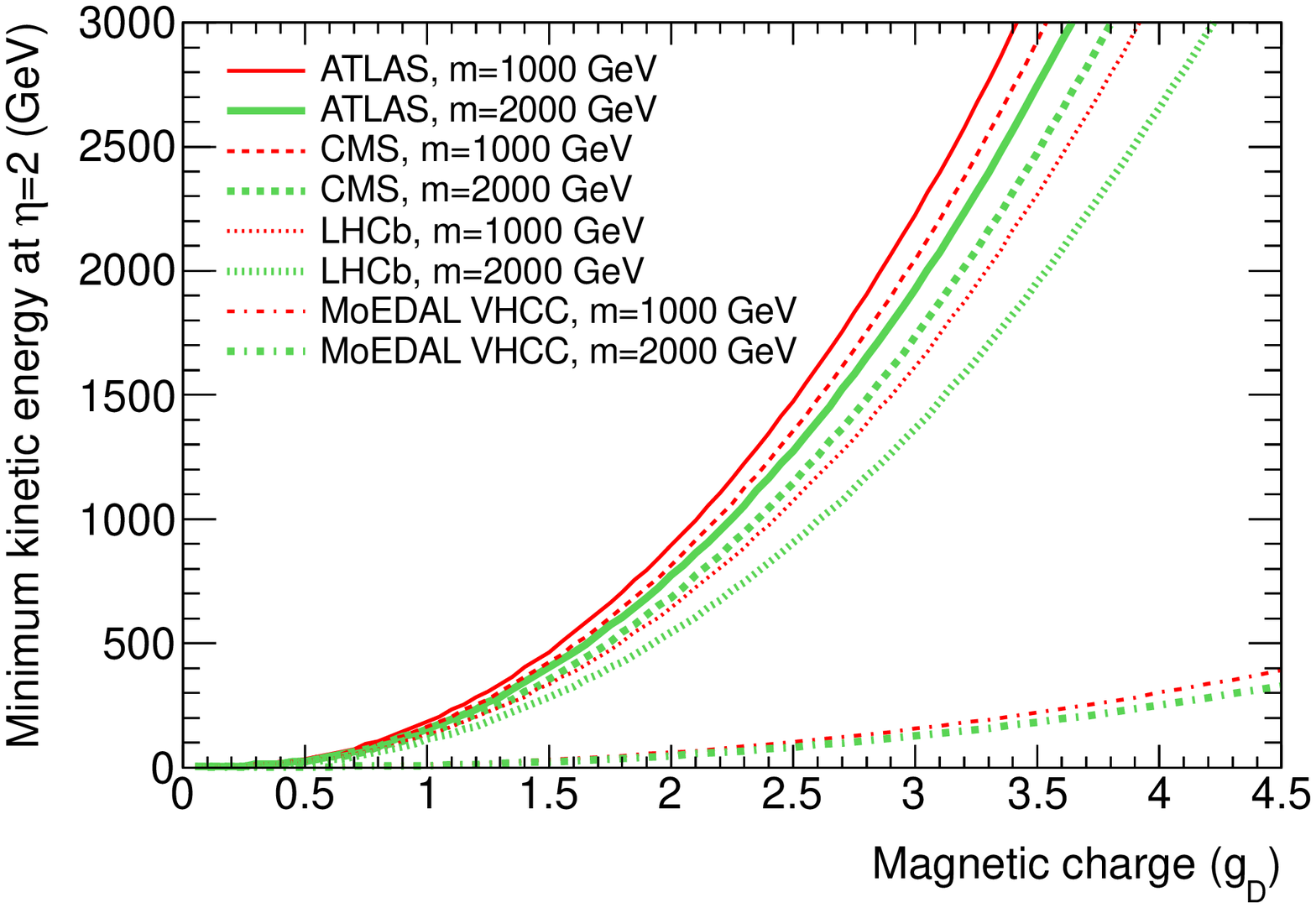}
  \end{center}
  \caption{Minimum initial kinetic energies required for HIPs to reach the sensitive parts of the detectors (front of the EM calorimeters for ATLAS, CMS and LHCb, end of TPC for ALICE, plastic sheets for MoEDAL), as functions of electric charge (left) and magnetic charge (right). The curves correspond to ATLAS (solid lines), CMS (dashed lines), MoEDAL (dashed dotted lines), ALICE (dotted lines in the top plot), and LHCb (dotted lines in the bottom plot), for 1000 GeV (thin) and 2000 GeV (thick) masses, at angles corresponding to $\eta\sim 0$ (top), $\eta\sim 1.4$ (middle) and $\eta\sim 2.0$ (bottom). }
  \label{fig:thresholds}
\end{figure}

Using the tools described in Section~\ref{velocity}, the loss of energy d$E$ at a given step d$x$ along the HIP trajectory is estimated and the position in the detector at which the energy is zero is recorded. This range is computed as a function of the pseudorapidity $\eta$, charge $z$ (or $g$), mass $m$, and initial kinetic energy $E_{kin}$. For ATLAS, CMS and ALICE, the results of these calculations for a HIP with $m=1000$~GeV traversing the central region of the detector (around $\eta=0$) are shown in Fig.~\ref{fig:HIPrange_eta0} for electrically (left) and magnetically (right) charged HIPs. The calculations are also performed -- but not shown in this figure -- for the LHCb and MoEDAL setups, and for a broad interval of masses and various relevant $\eta$ values in all detectors. With the same initial conditions, the range of a magnetic monopole is systematically longer than that of an electrically charged particle with charge $|z|=|g|$. This is due to the lower energy loss of monopoles at low $\beta$ in Equation \ref{Bethe_mag} (see Fig.~\ref{fig:dedx}). 

One notable difference between ATLAS and CMS, which affects significantly the range for high charges (see top and middle plots in Fig.~\ref{fig:HIPrange_eta0}), is the location of the solenoid magnet: beyond the hadronic calorimeter in CMS and just beyond the inner detector in the case of ATLAS. The ATLAS solenoid and its cryogenic system add 20.8 g$\cdot$cm$^{-2}$ of extra material (almost twice as much as the whole inner detector) before the HIP reaches the calorimeters. This feature gives an advantage to CMS since a HIP which stops prior to the EM calorimeter would not be able to trigger the event. Indeed, as mentioned in Section~\ref{detectors}, first level triggers used currently in both ATLAS and CMS (LHCb as well) are based on either an energy deposition in the calorimeters or a track in the muon spectrometer~\cite{LHCtriggers}. 

Thanks to the lower material budget of ALICE, the charge scale extends further in the bottom plots in Fig.~\ref{fig:HIPrange_eta0}, for comparatively lower $E_{kin}$ values. This allows ALICE to cover more phase space than the other experiments: for instance, a monopole with charge $3g_D$ and $E_{kin}=100$ GeV at $\eta=0$, which would always be stopped and lost before it can reach the sensitive parts of the other detectors, would still traverse the ALICE TPC. The disadvantages of the ALICE detector are the low luminosity in $pp$ collisions (see Table~\ref{tab:lumi}) and the limited pseudorapidity coverage ($|\eta|<0.9$). 

To present a direct comparison between the potential performances of the various experiments, and to illustrate the mass and $\eta$ dependencies of the HIP ranges, punch-through energies of HIPs of various masses for reaching the sensitive parts of the detectors are plotted in Fig.~\ref{fig:thresholds} as functions of electric (left) and magnetic (right) charges, for $\eta\sim 0$ (top), $\eta\sim 1.4$ (middle) and $\eta\sim 2.0$ (bottom). In the barrel region ($|\eta|<1.5$), for a given charge, the energy required to reach the CMS EM calorimeter is typically half the one needed in ATLAS. The difference is much less pronounced in the endcaps ($1.5<|\eta|<2.5$), where the ATLAS solenoid does not stand in the way. ALICE has the lowest thresholds in the region it covers ($|\eta|<0.9$). In the more forward region, MoEDAL has the lowest thresholds, especially for $\eta\geq 1.5$ thanks to the VHCC. LHCb has thresholds comparable to those of ATLAS and CMS in the region it covers ($2.0<\eta<4.9$). As expected, for electric charges the energy loss is higher (shorter range) for higher masses (lower $\beta$), while it is the opposite for magnetic charges.

\section{LHC detector acceptances}
\label{sensitivities}

\begin{figure}[tb]
  \begin{center}
    \includegraphics[width=0.49\linewidth]{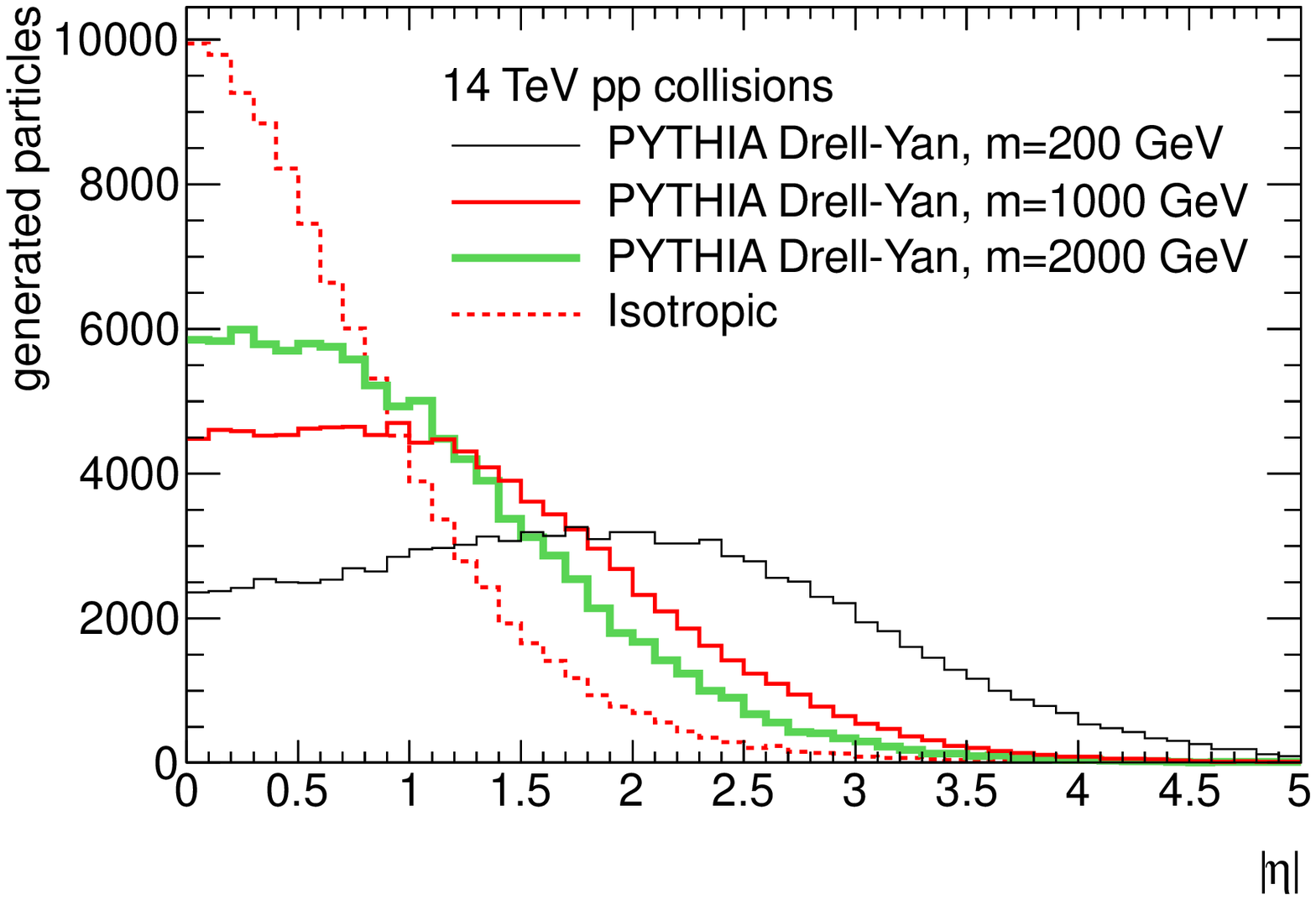}
    \includegraphics[width=0.49\linewidth]{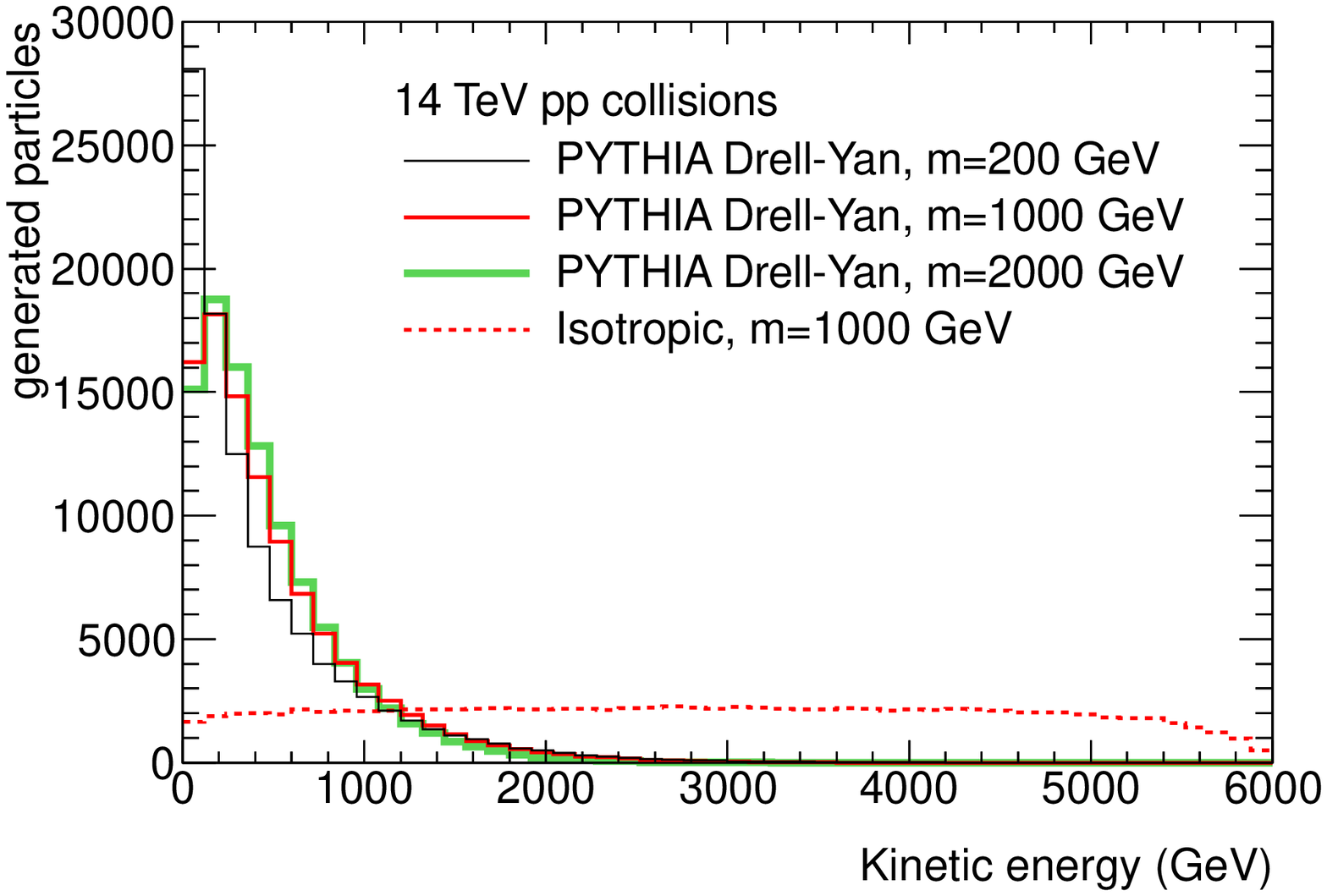}    
    \includegraphics[width=0.49\linewidth]{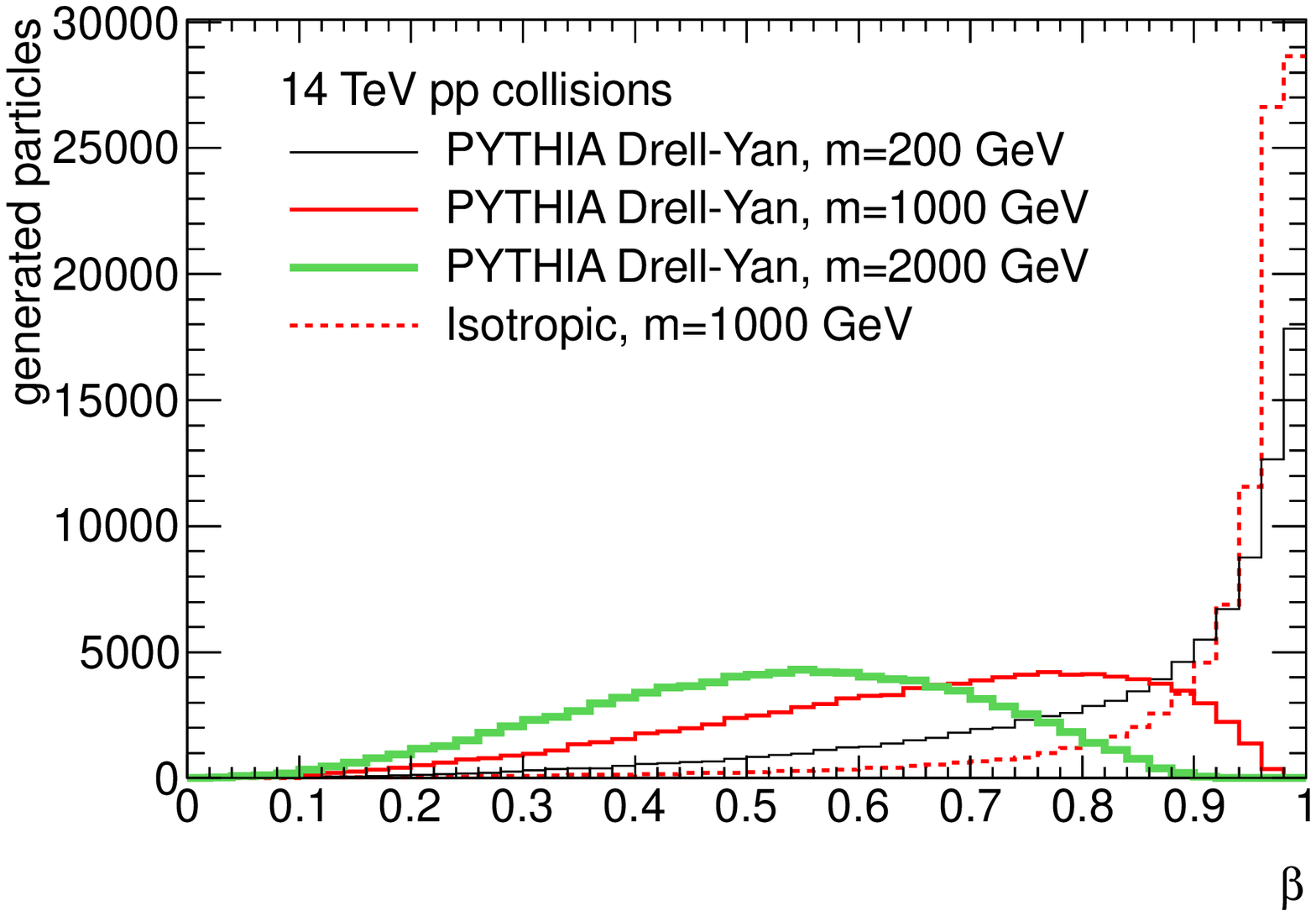}        
    \includegraphics[width=0.49\linewidth]{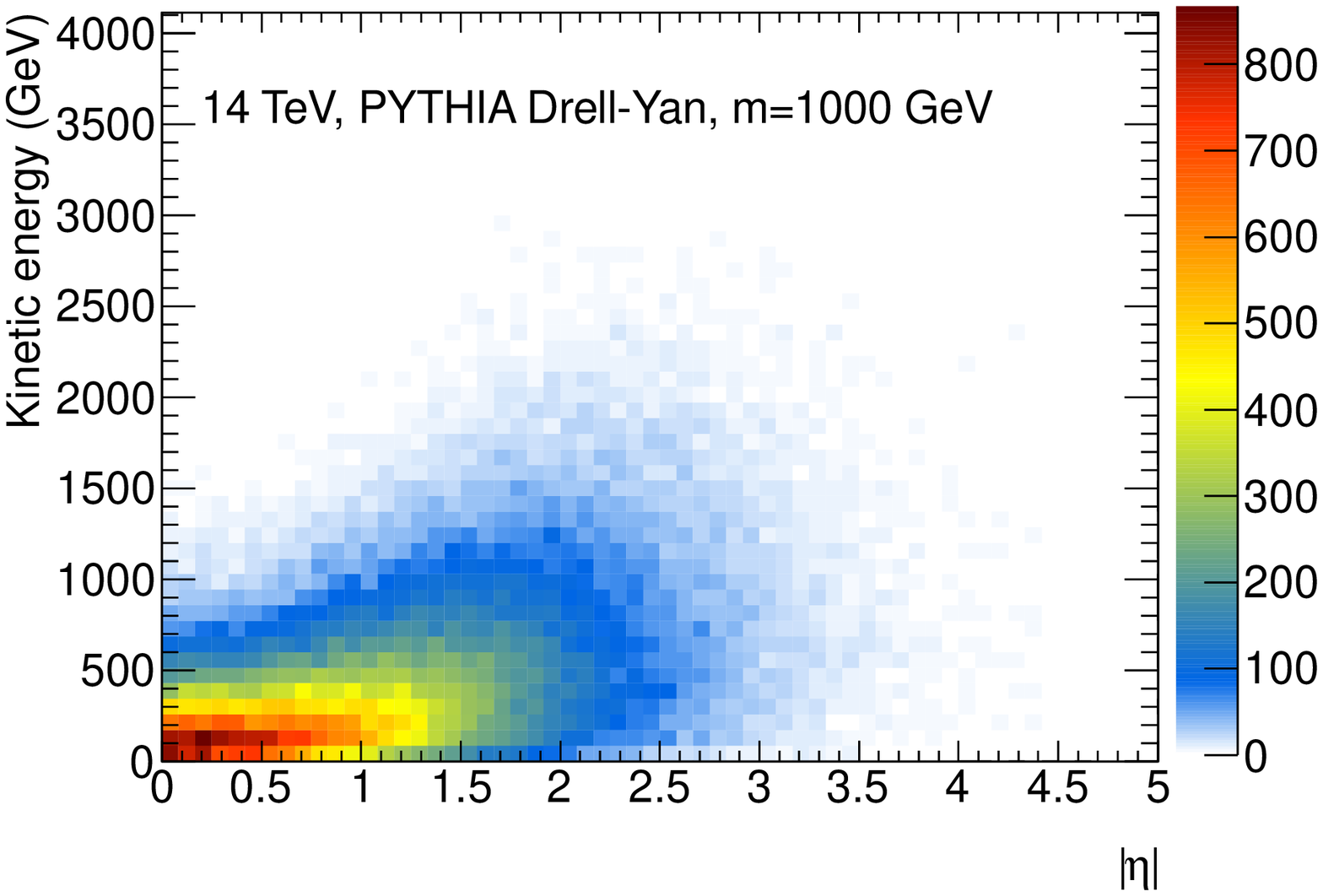}    
  \end{center}
  \caption{Drell-Yan (solid lines) and isotropic (dashed lines) pair production kinematics for $pp$ collisions at 14 TeV center-of-mass energy. Distributions of absolute value of pseudorapidity (top, left), kinetic energy (top, right) and velocity (bottom, left) are shown for 200 (thin), 1000 (medium) and 2000 (thick) GeV masses. A scatter plot of kinetic energy versus pseudorapidity is shown for the 1000 GeV case for Drell-Yan (bottom, right). These distributions are based on 50000 generated events.}
  \label{fig:DYkin}
\end{figure}

\begin{figure}[[tbh]
  \begin{center}
    \includegraphics[width=0.49\linewidth]{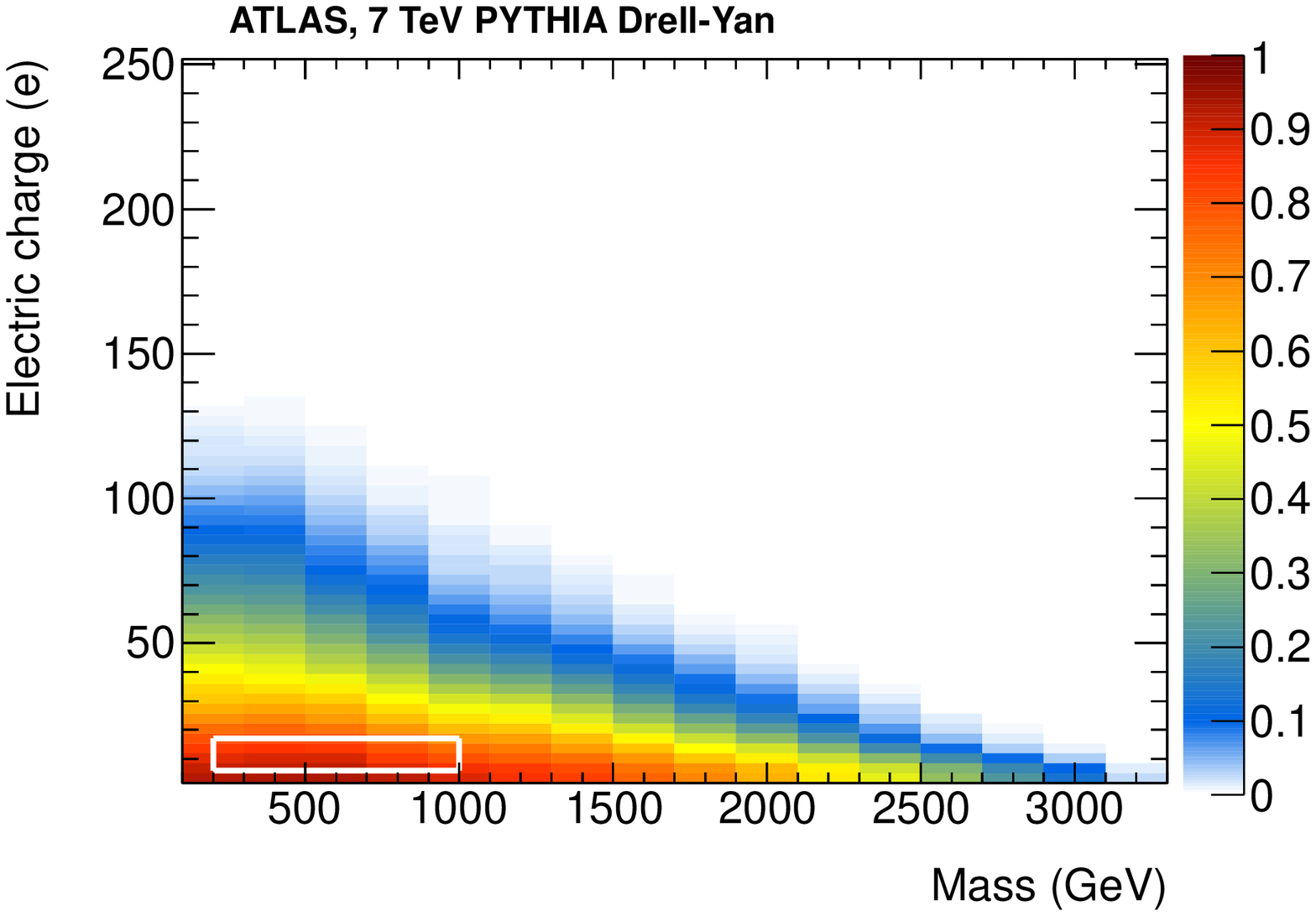}
    \includegraphics[width=0.49\linewidth]{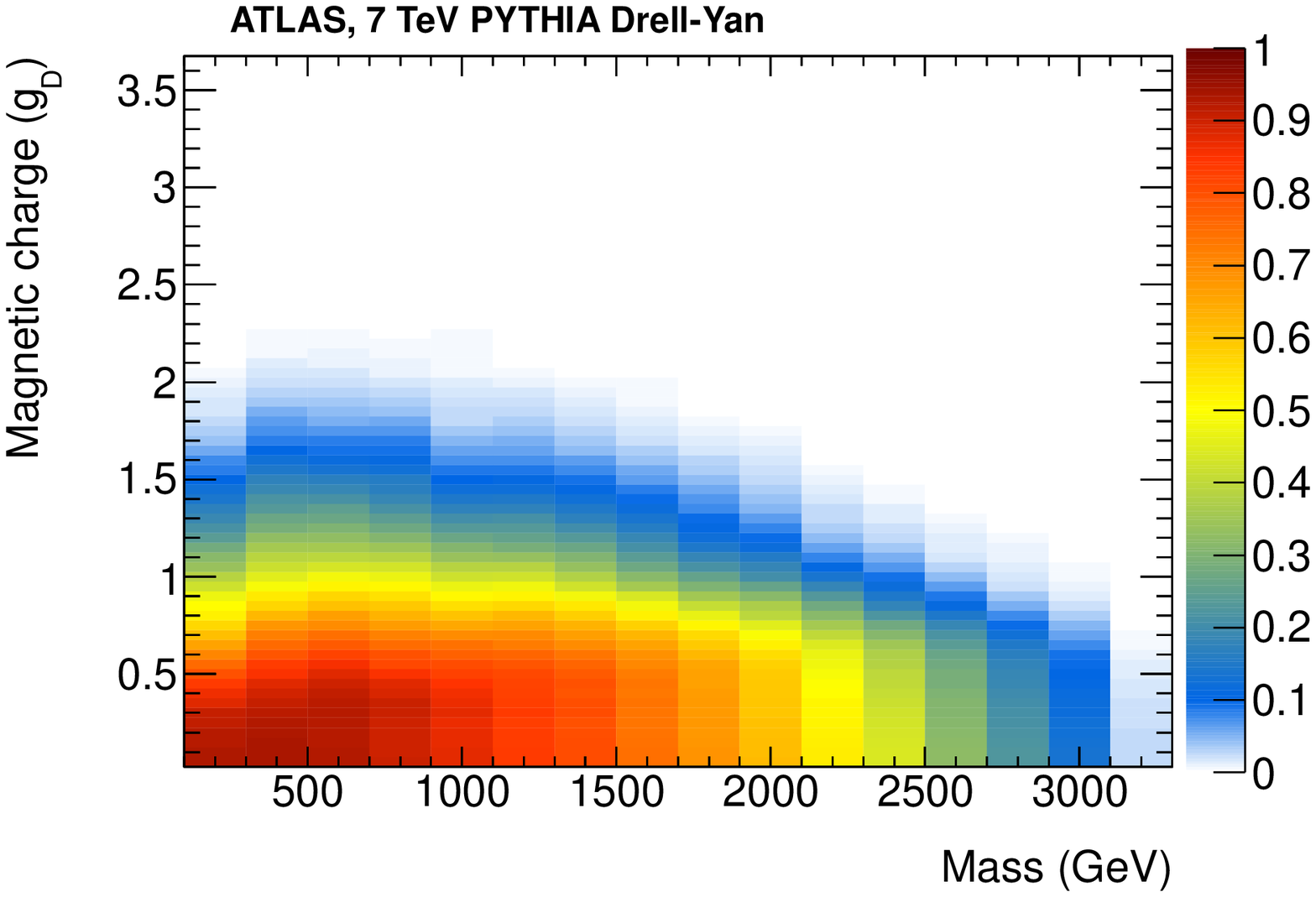}  
    \includegraphics[width=0.49\linewidth]{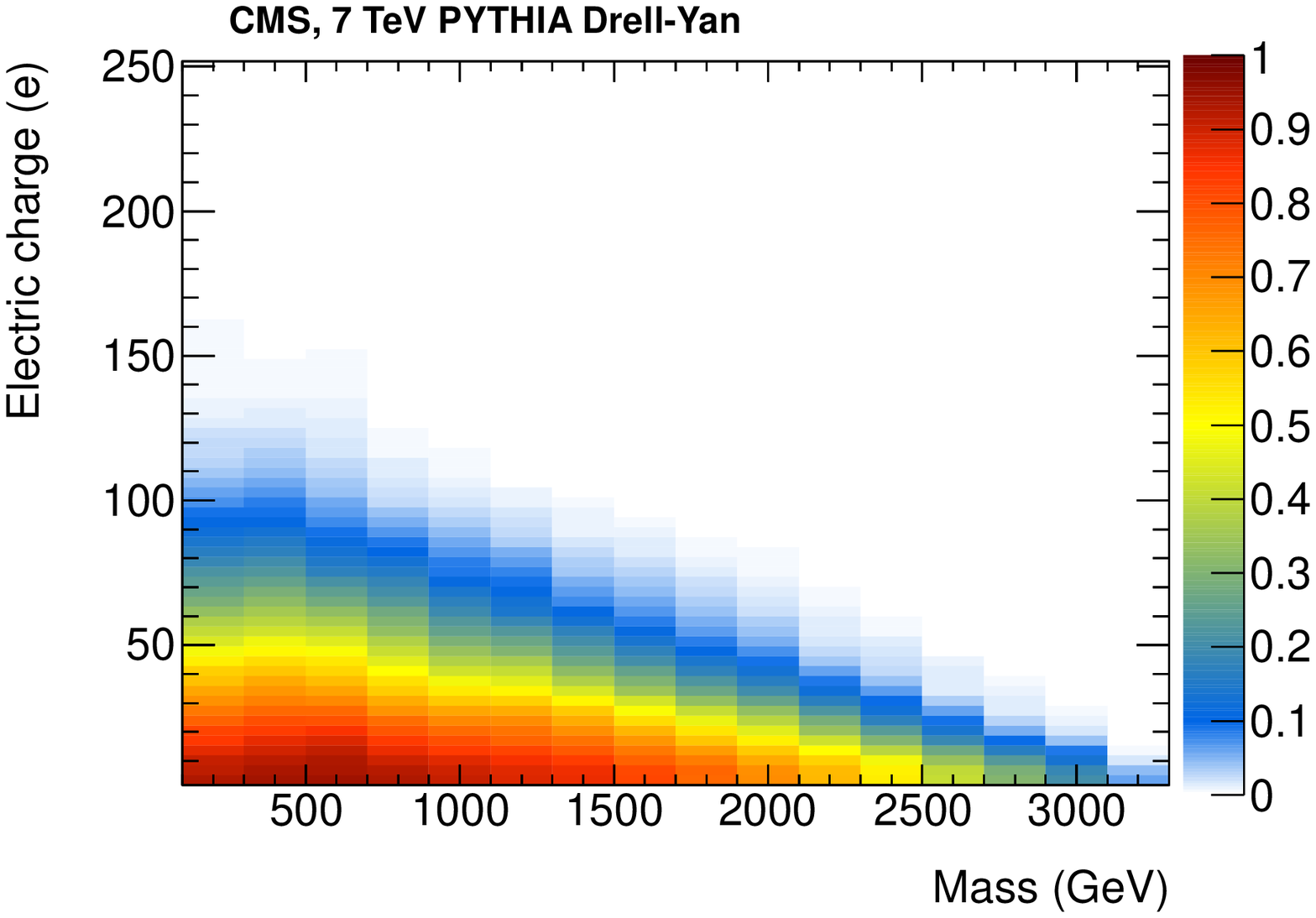}
    \includegraphics[width=0.49\linewidth]{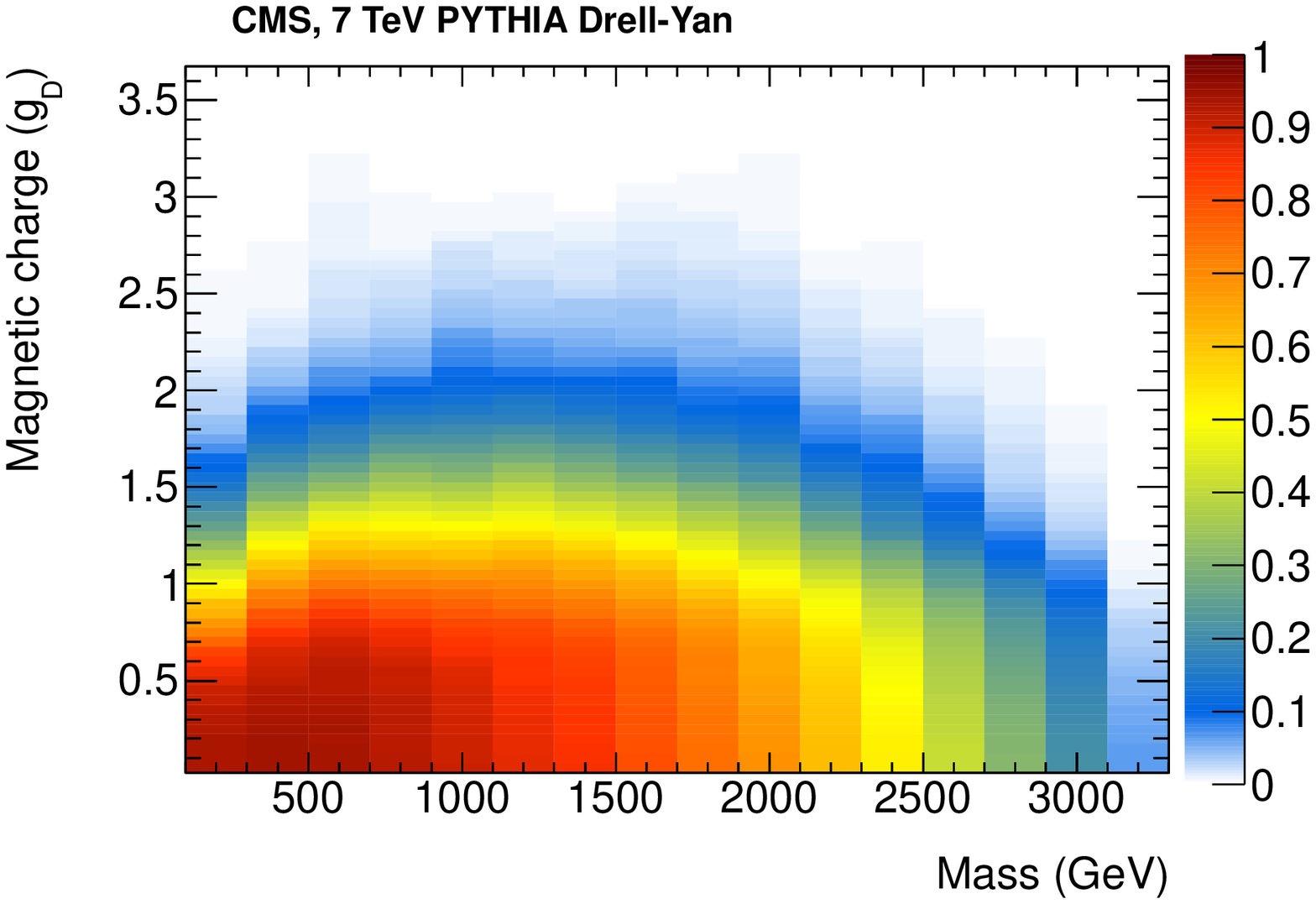}  
    \includegraphics[width=0.49\linewidth]{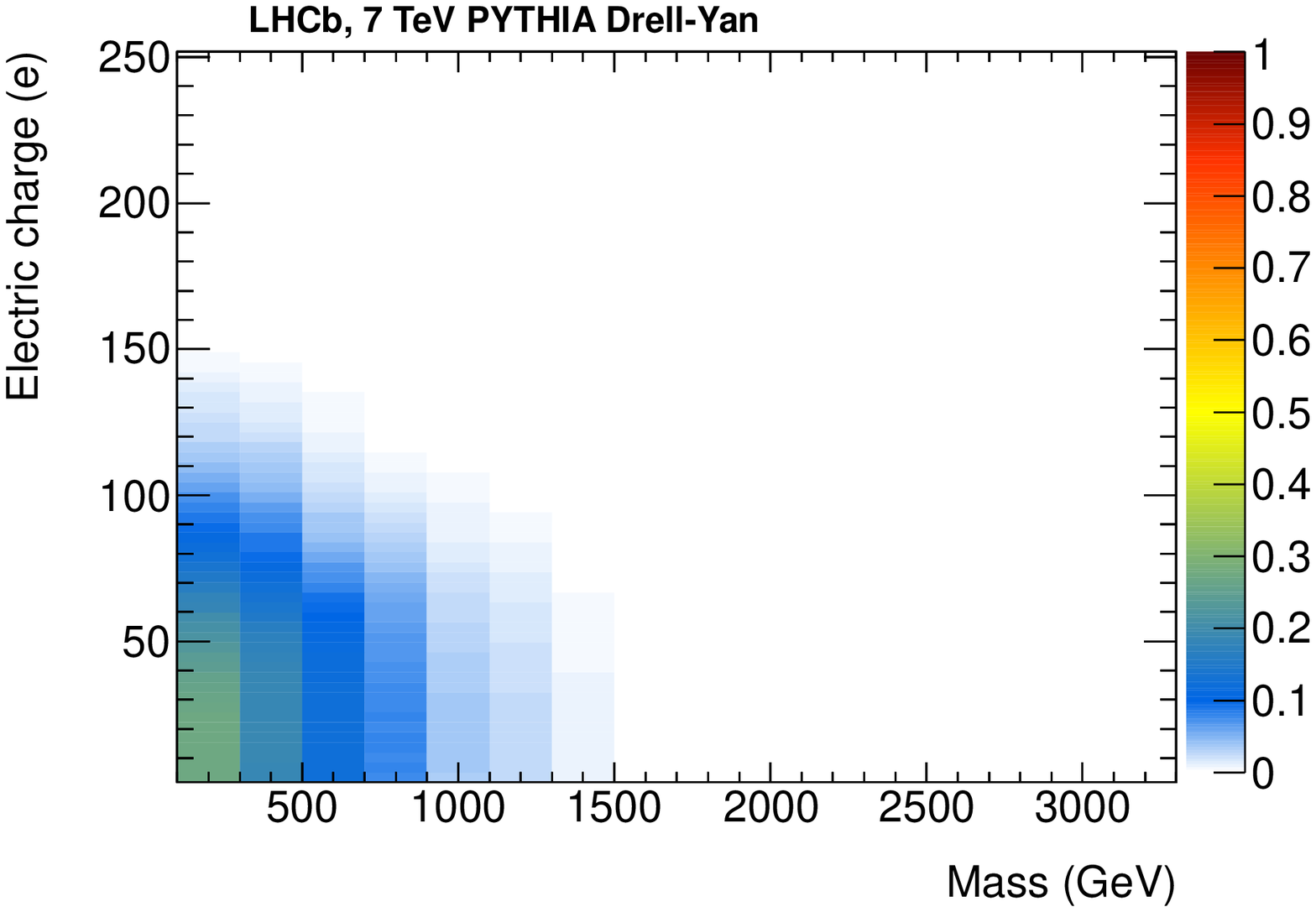}  
    \includegraphics[width=0.49\linewidth]{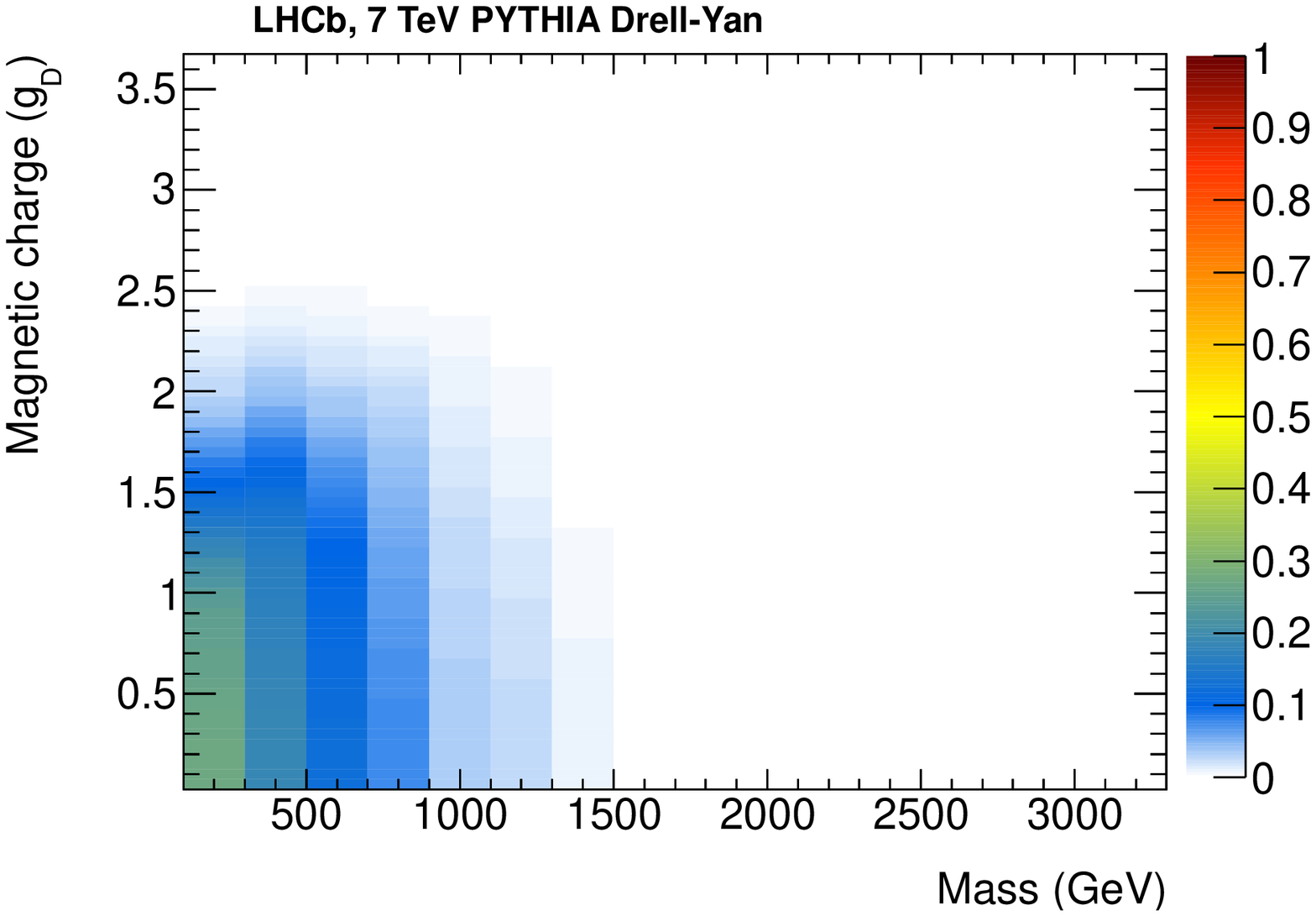}  
    \includegraphics[width=0.49\linewidth]{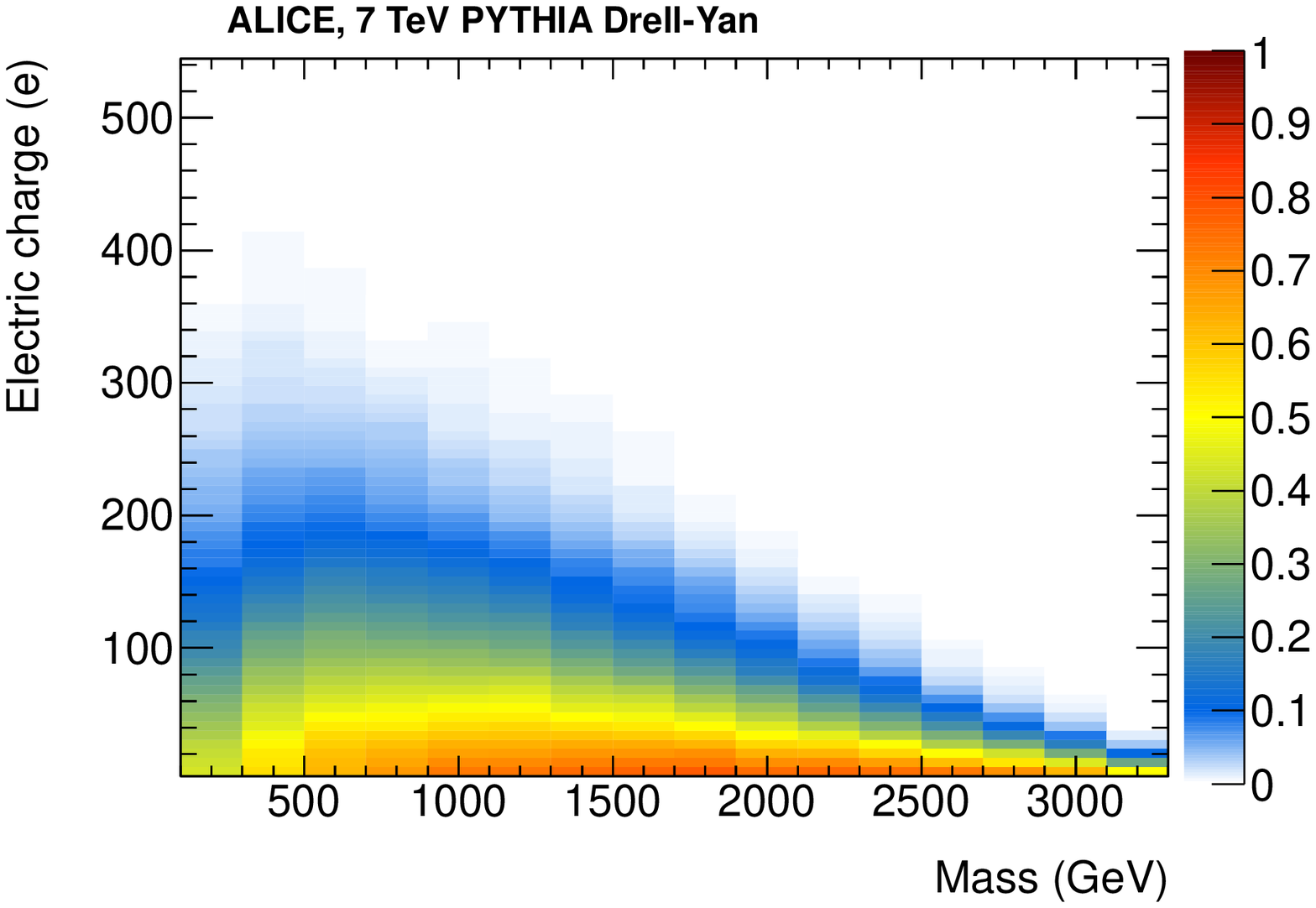}
    \includegraphics[width=0.49\linewidth]{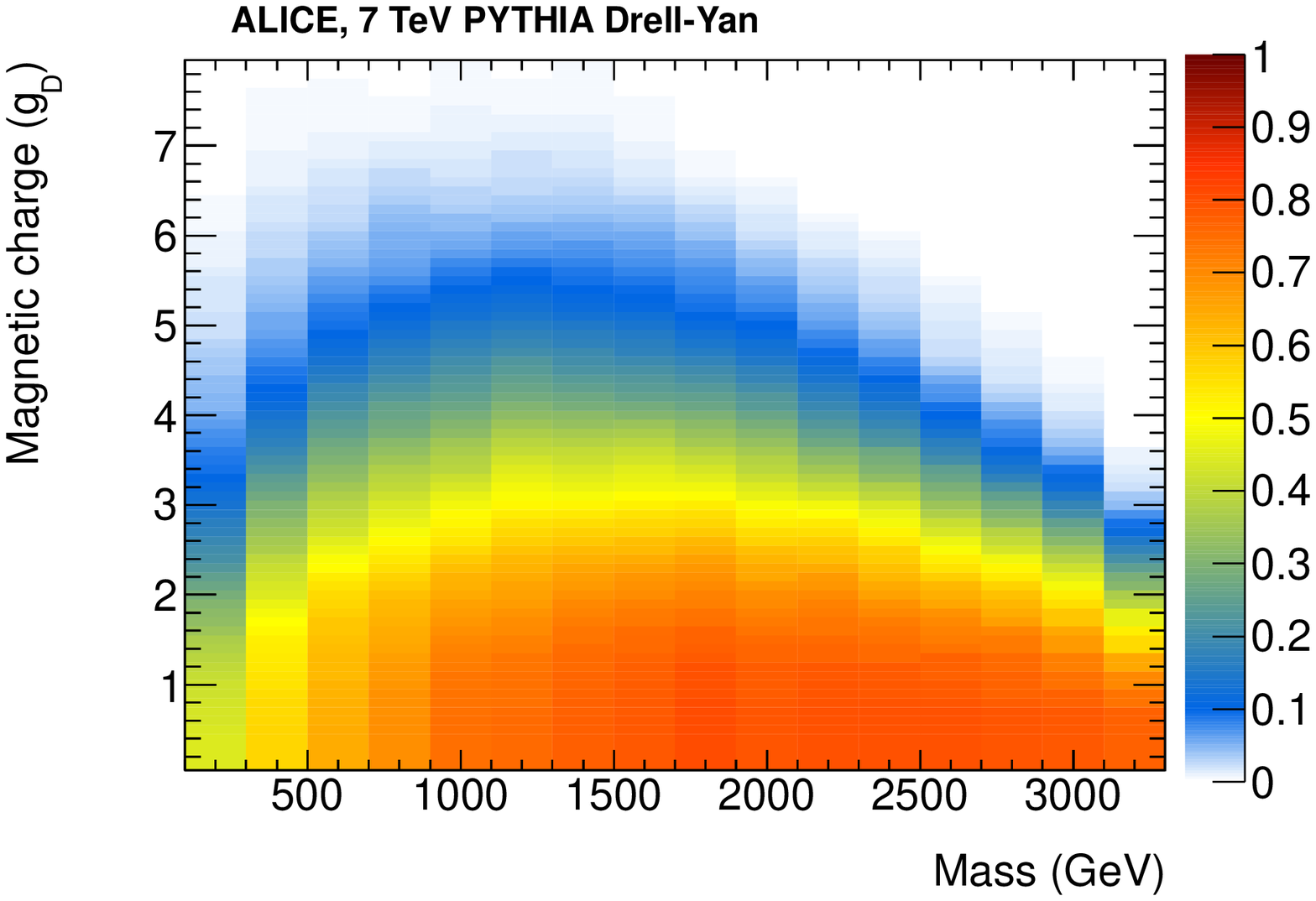}  
  \end{center}
  \caption{Acceptances as functions of HIP mass and charge, for electric (left) and magnetic (right) charges, for the ATLAS (top), CMS (middle, top), LHCb (middle, bottom) and ALICE (bottom) detectors, assuming a Drell-Yan pair production mechanism with 7 TeV $pp$ collisions. The area for which limits were set in the ATLAS search~\cite{QballATLAS10} is indicated by a white line in the top left plot. Note that the vertical axis scale is different for ALICE. The relative binwise uncertainties in the acceptance $a$ are 15\% (systematics) and $(1000\cdot a)^{-1/2}$ (statistics).}
  \label{fig:HIPsensitivity_7TeV}
\end{figure}

\begin{figure}[tbh]
  \begin{center}
    \includegraphics[width=0.49\linewidth]{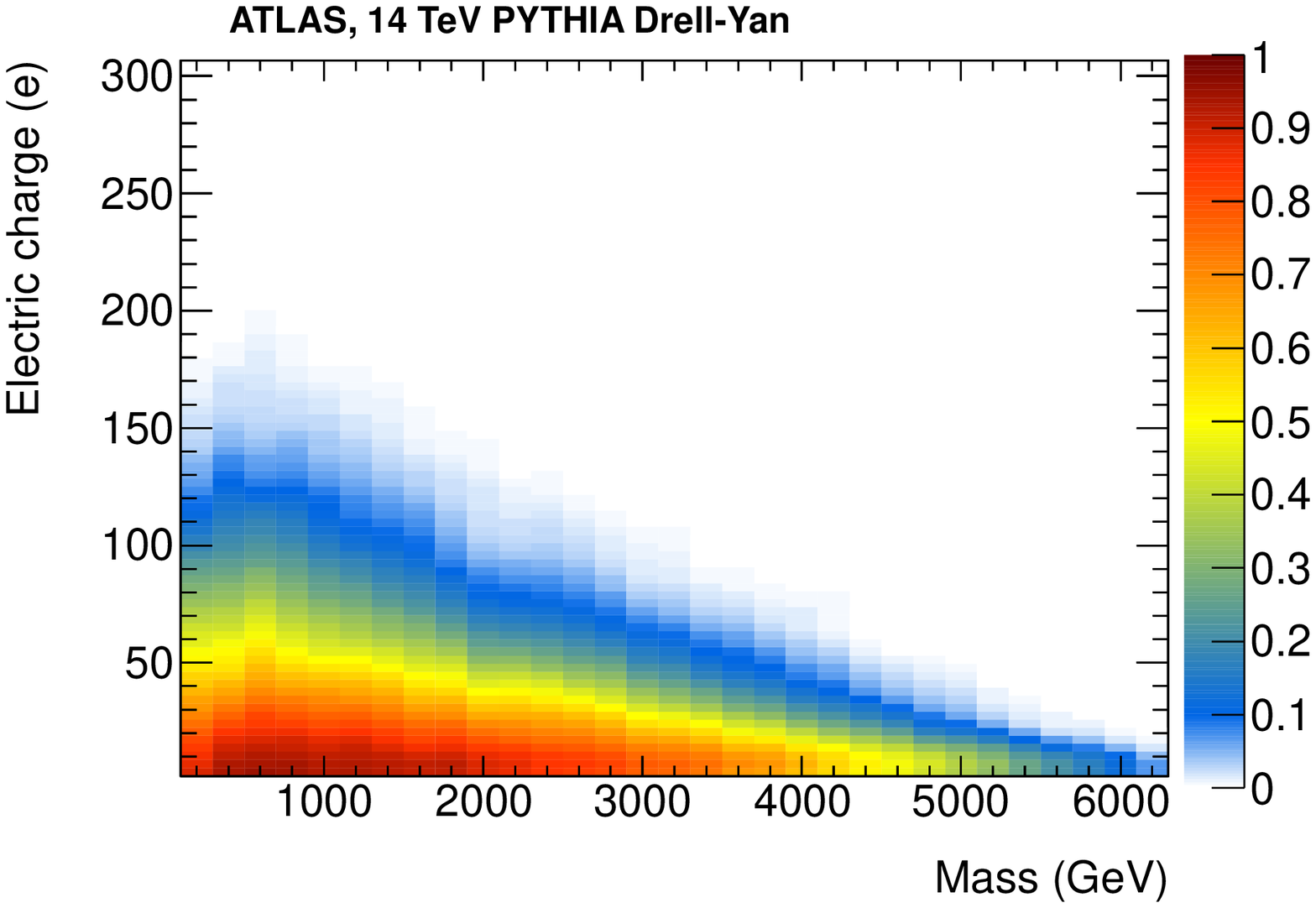}
    \includegraphics[width=0.49\linewidth]{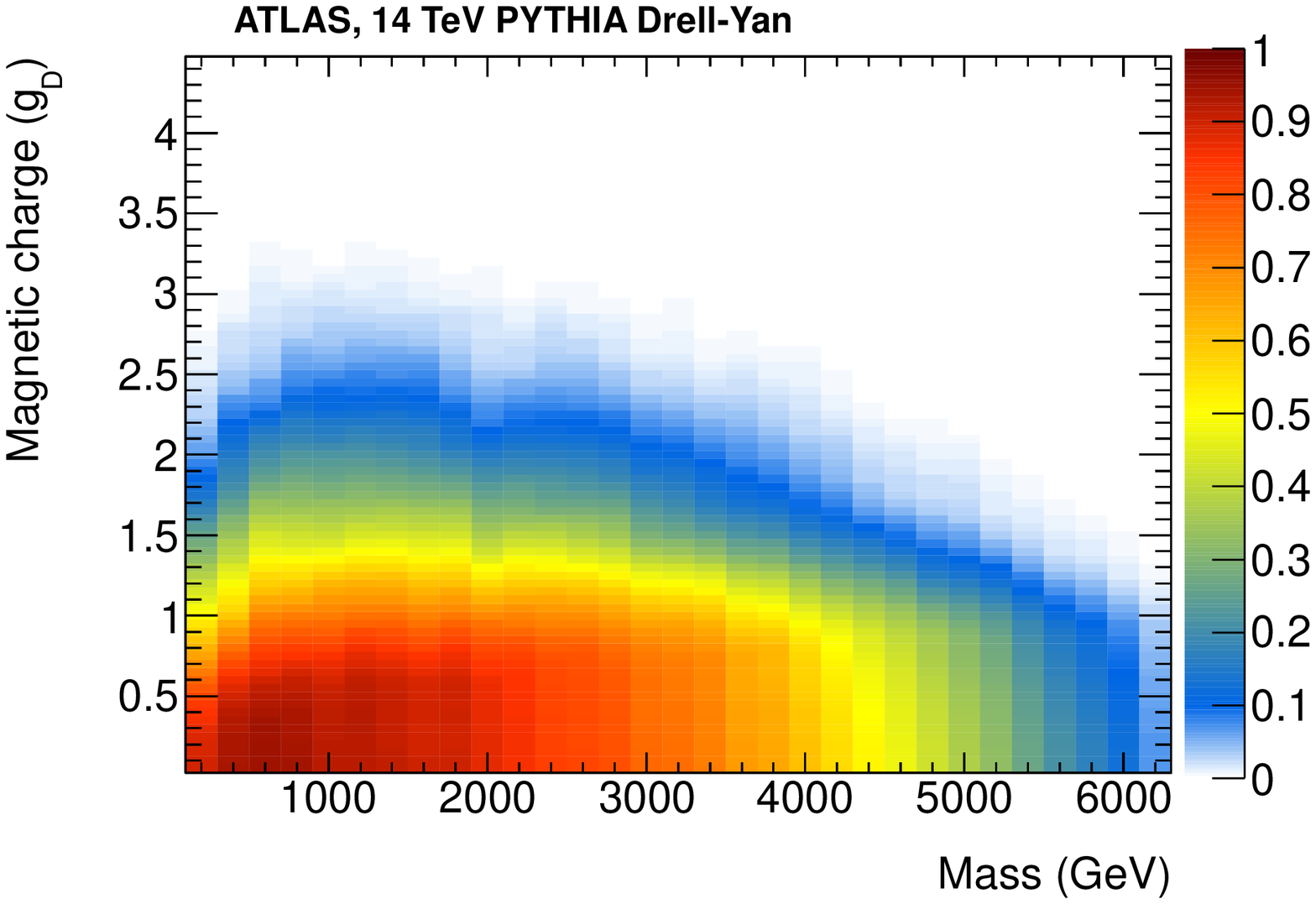}  
    \includegraphics[width=0.49\linewidth]{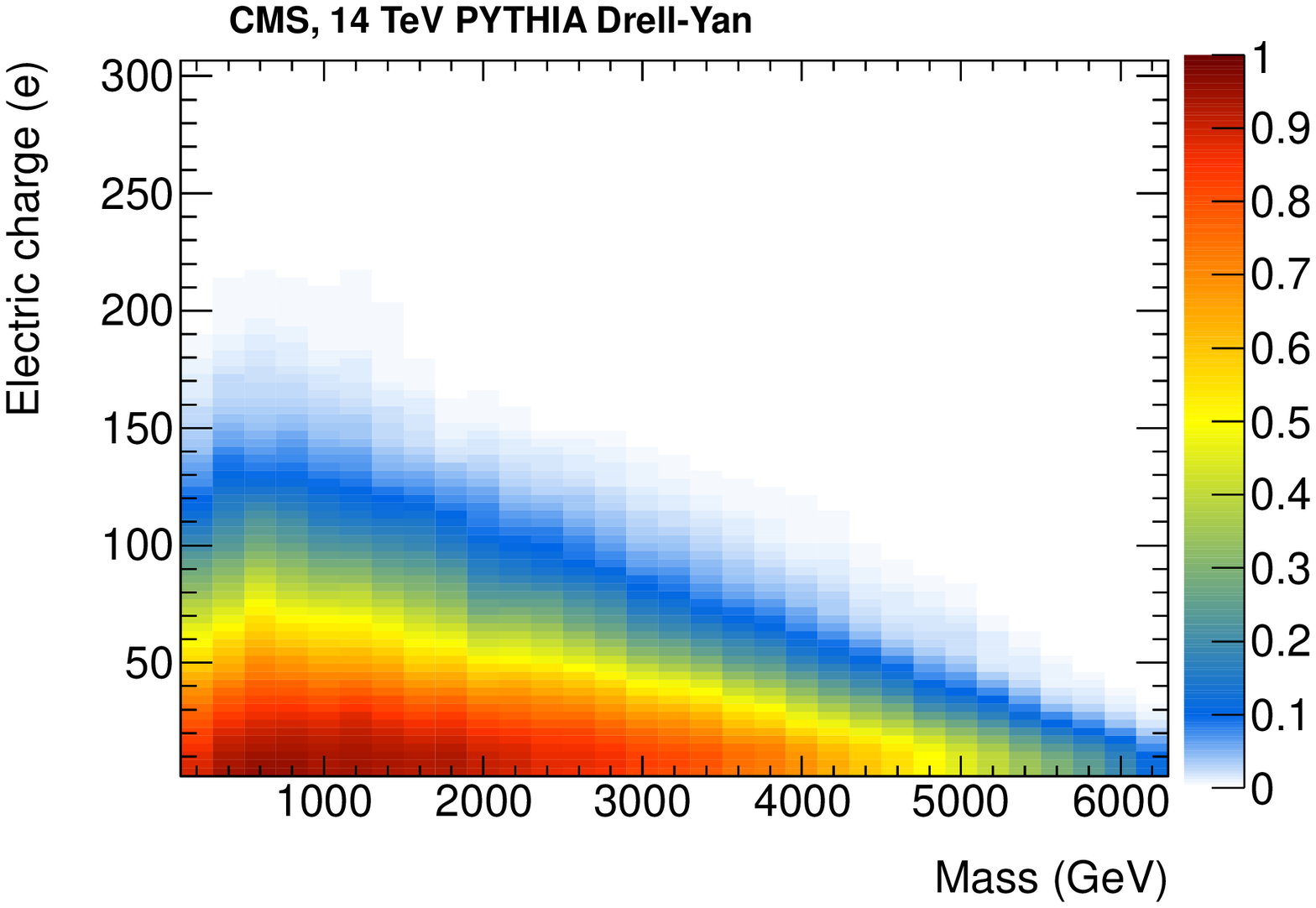}
    \includegraphics[width=0.49\linewidth]{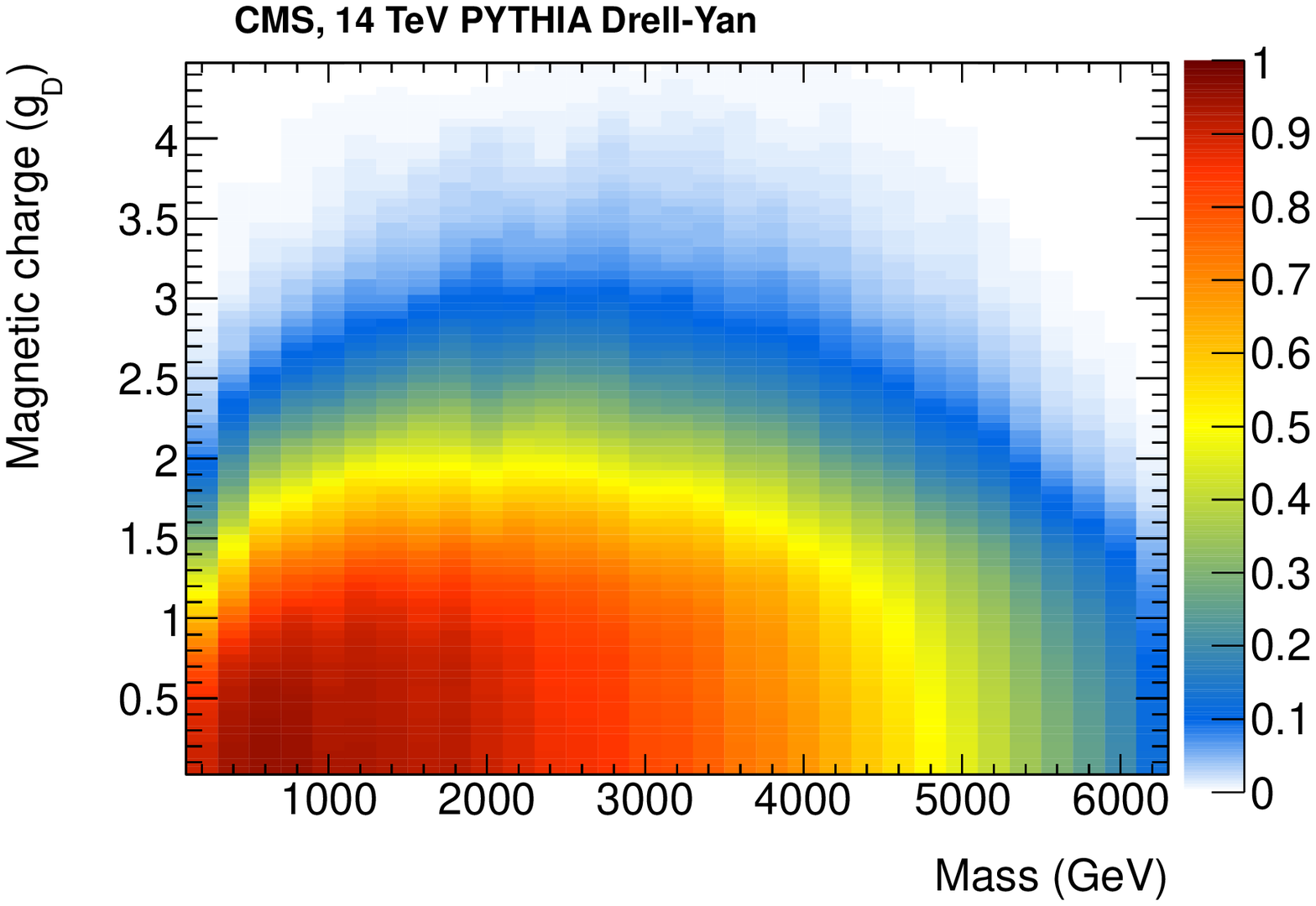}  
    \includegraphics[width=0.49\linewidth]{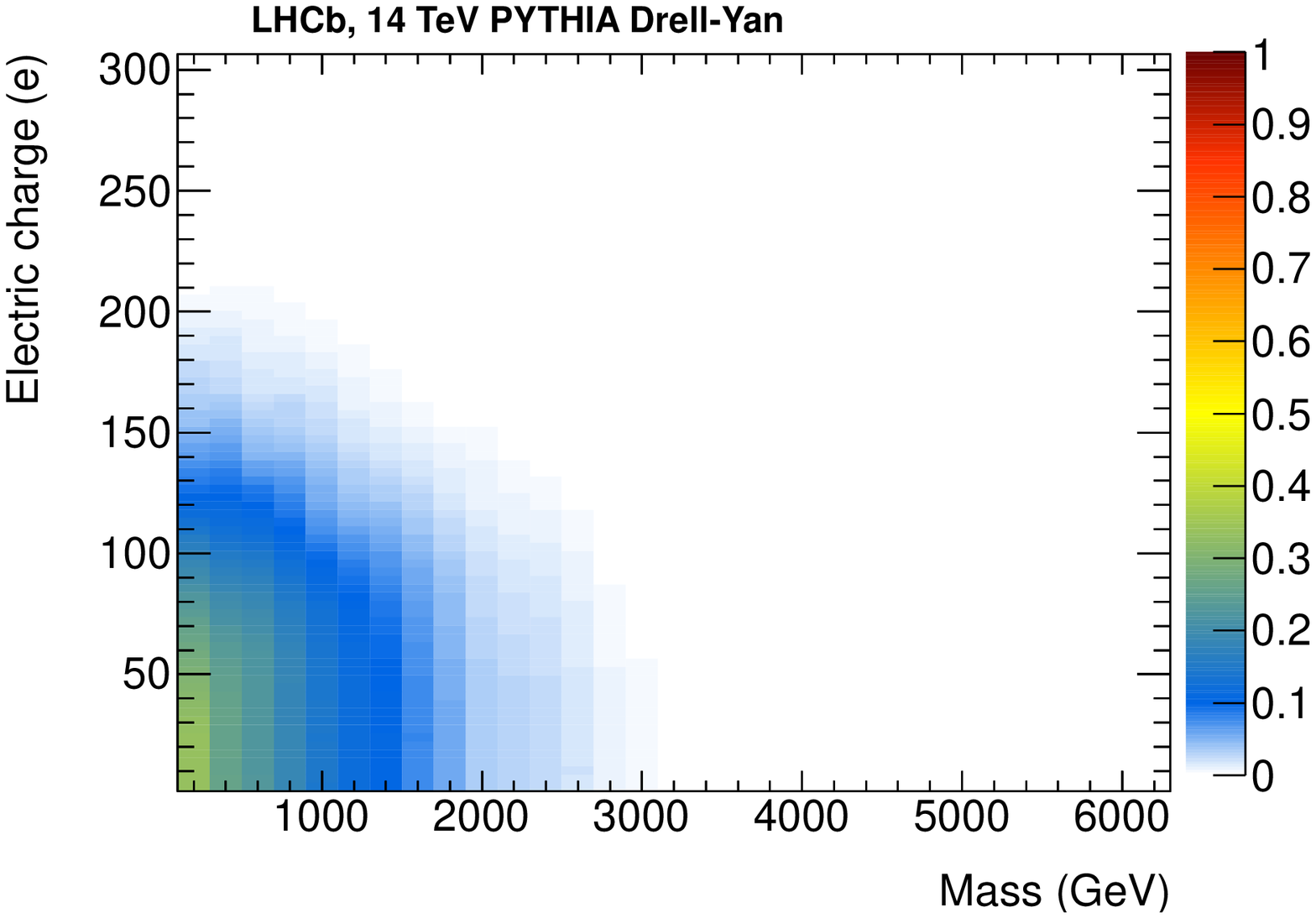}  
    \includegraphics[width=0.49\linewidth]{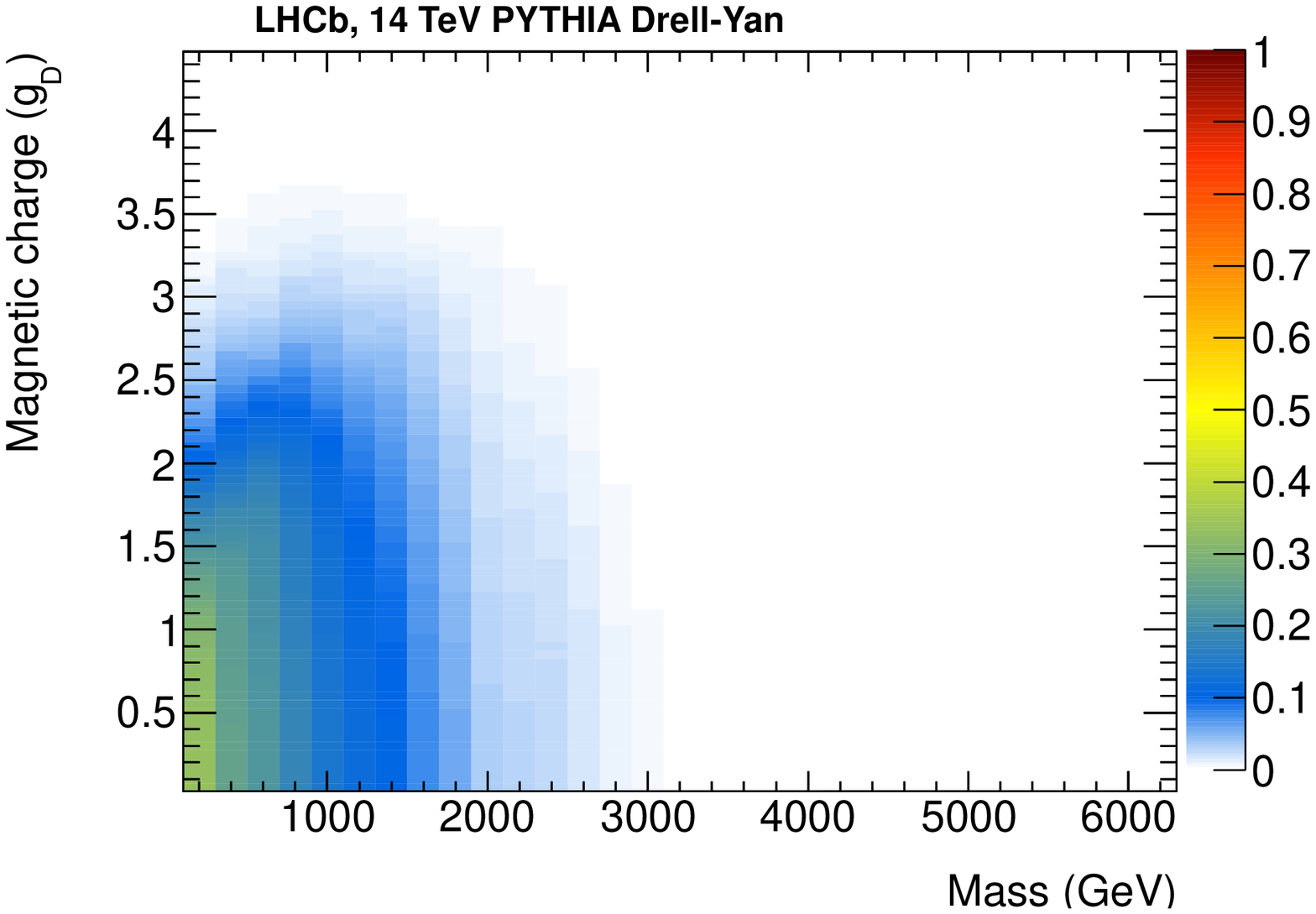}  
    \includegraphics[width=0.49\linewidth]{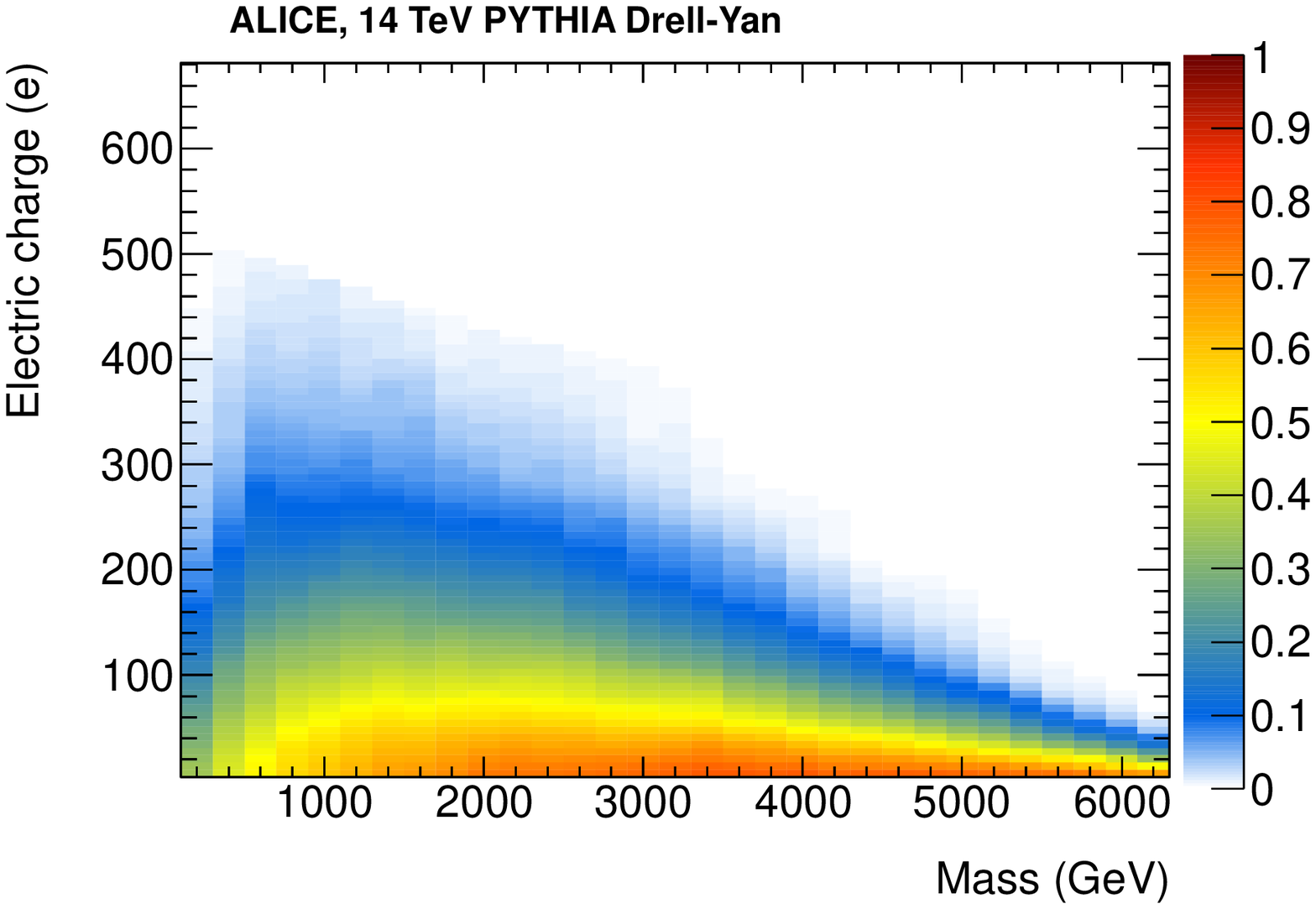}
    \includegraphics[width=0.49\linewidth]{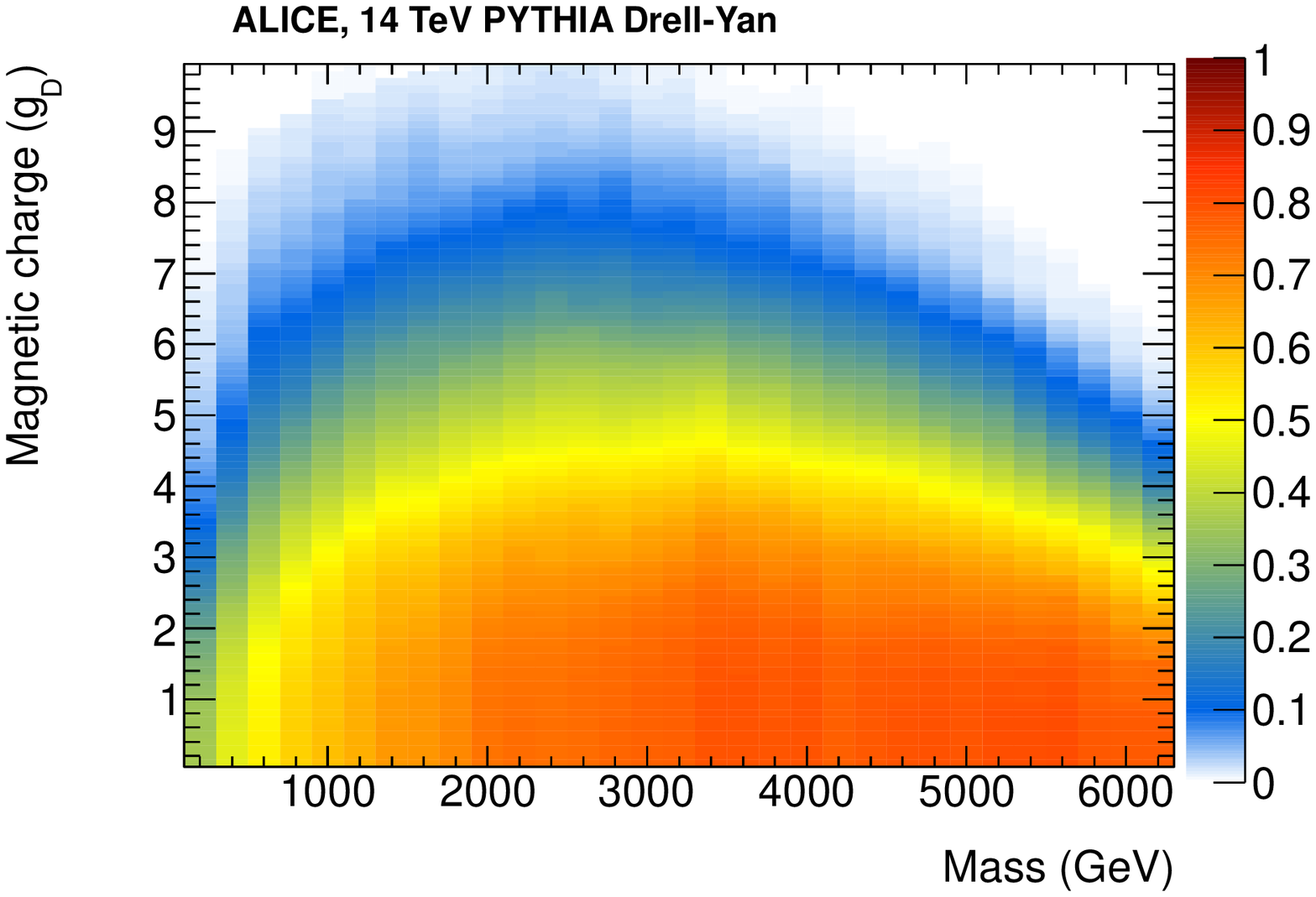}  
  \end{center}
  \caption{Acceptances as functions of HIP mass and charge, for electric (left) and magnetic (right) charges, for the ATLAS (top), CMS (middle, top), LHCb (middle, bottom) and ALICE (bottom) detectors, assuming a Drell-Yan pair production mechanism with 14 TeV $pp$ collisions. Note that the vertical axis scale is different for ALICE. The relative binwise uncertainties in the acceptance $a$ are 15\% (systematics) and $(1000\cdot a)^{-1/2}$ (statistics).}
  \label{fig:HIPsensitivity_14TeV}
\end{figure}

\begin{figure}[tb]
  \begin{center}
    \includegraphics[width=0.49\linewidth]{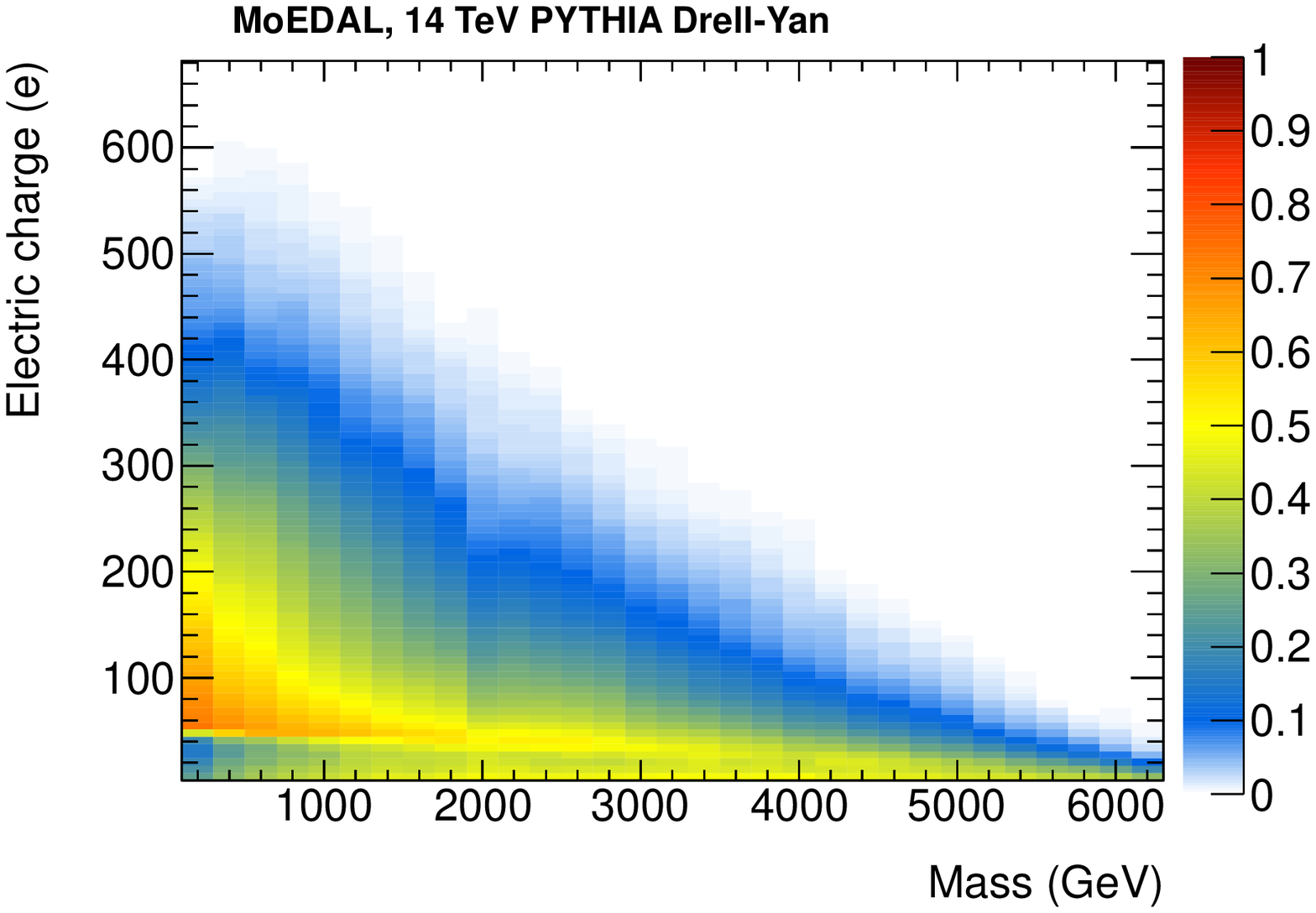}
    \includegraphics[width=0.49\linewidth]{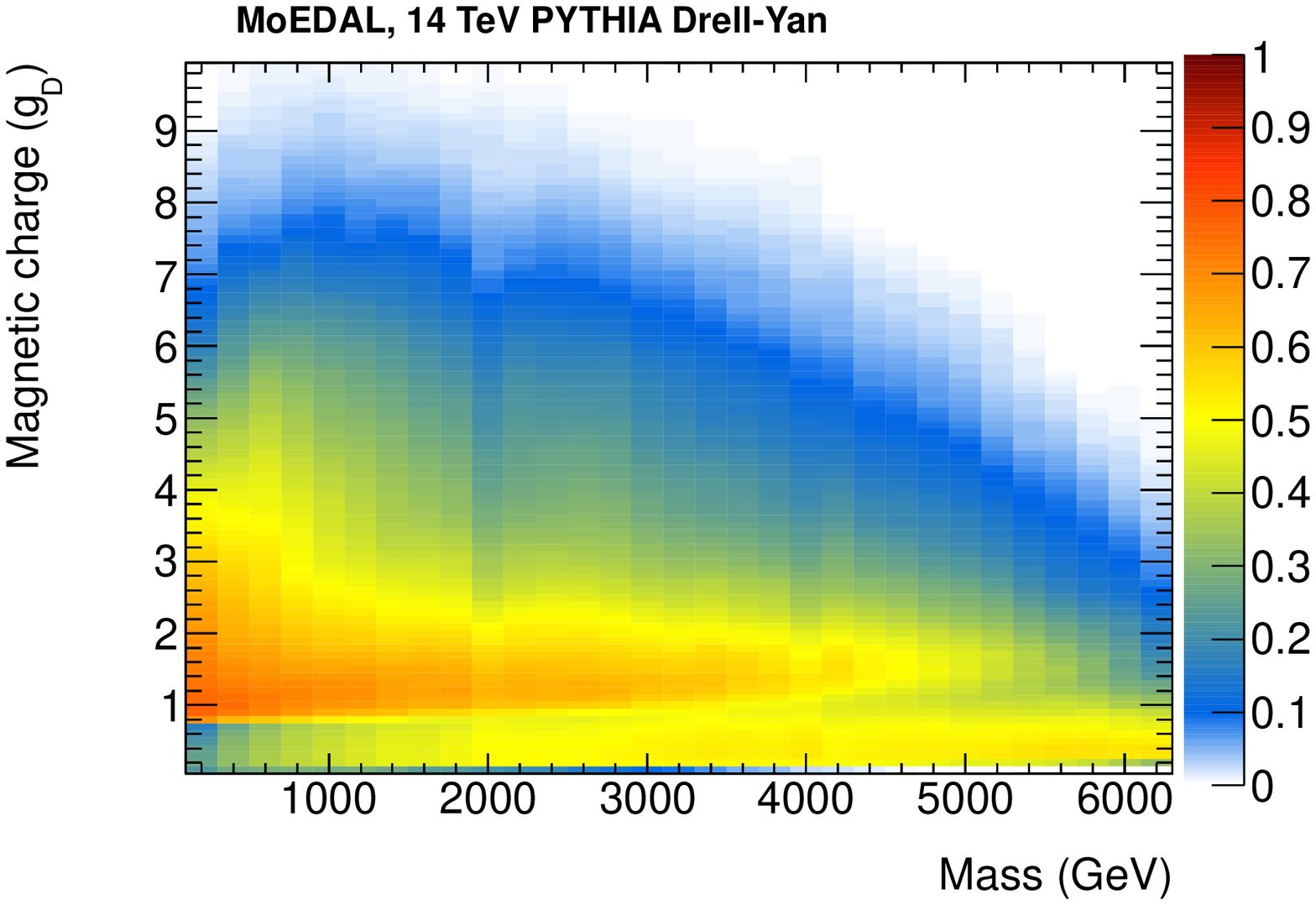}  
  \end{center}
  \caption{Acceptances as functions of HIP mass and charge, for electric (left) and magnetic (right) charges, for the MoEDAL detector, assuming a Drell-Yan pair production mechanism with 14 TeV $pp$ collisions. The two regions of high acceptance observed in these plots correspond to the two different track-etch modules to be used in MoEDAL: one is most sensitive to low charges and high mass, while the other (the VHCC) is sensitive to high charges. The relative binwise uncertainties in the acceptance $a$ are 15\% (systematics) and $(1000\cdot a)^{-1/2}$ (statistics).}
  \label{fig:HIPsensitivity_14TeV_MoEDAL}
\end{figure}

The quantity defined as acceptance is the probability (per event) that a HIP with given mass and charge would enter a sensitive region of the detector. Given that a HIP is inside the detector acceptance, the quantity called efficiency is defined as the probability that the event would further satisfy an offline selection. The efficiency depends on the selection criteria and on specific details of detector hardware and software performance such as visible energy, saturation, timing, noise, and reconstruction. It can also depend on the bending of the HIP in the solenoid magnetic field, which depends on the nature of the HIP charge. As these issues would best be addressed by the concerned detector collaborations, only acceptance is considered here. For ATLAS, CMS and LHCb, the acceptance is the fraction of events that produce a candidate for a first level trigger, for which a minimum requirement is that at least one HIP reaches the EM calorimeter where it must deposit 30 GeV or more. Note that in this definition the HIP can stop anywhere beyond that point (by contrast, in the ATLAS search~\cite{QballATLAS10}, an implicit veto on the hadronic energy deposition at higher level trigger forces the HIP to stop in the EM calorimeter, which lowers the effective energy spectrum of HIP candidates). An additional constraint is that the calorimeter signal should not be delayed by more than the window around the proton bunch crossing time (of length 25 ns) with respect to a $\beta=1$ particle~\cite{ATLASCaloTrig}. To model this loss, a linear decrease of acceptance from one to zero is assumed for delays of the time of arrival at the EM calorimeter between 10 and 20 ns. In the cases of MoEDAL and ALICE, there is no such timing constraint, and the acceptance is assumed to be the probability that at least one HIP traverses one of the MoEDAL track-etch detectors (with sufficient $z/\beta$ to produce an etch-pit cone) and the ALICE TPC, respectively. 


The acceptance depends on the HIP energy and pseudorapidity distributions (kinematics). The Drell-Yan process $q\bar{q}\rightarrow\gamma^*\rightarrow X\bar{X}$ is traditionally used as a benchmark model of kinematics of heavy spin-1/2 HIP pair production. To model such production in $pp$ collisions, we use the PYTHIA Monte-Carlo generator~\cite{Sjostrand:2006za}. We conservatively use a center-of-mass energy of 7 TeV to model LHC $pp$ collision data prior to 2013, even though 8 TeV are planned for the 2012 runs. Pseudorapidity, kinetic energy and velocity distributions using 50000 events for a center-of-mass energy of 14 TeV are shown in Fig. \ref{fig:DYkin}. It must be stressed, however, that HIP kinematics in nature may turn out to be different (see discussion in Section \ref{challenges}). To test the effect of the arbitrary choice of kinematics, a model of isotropic production of a HIP pair and two scattered protons in the center-of-mass frame is also considered. Such events were generated with the Genbod Monte-Carlo generator~\cite{CERNLIB}. These events are represented by the dashed lines in Fig.~\ref{fig:DYkin}. The energy distribution in the isotropic model peaks at $\sim\sqrt{s}/4$ and extends up to $\sim\sqrt{s}/2$. Model dependence is further discussed in Section \ref{dependence}.  

For a Drell-Yan fermion-pair production process at $pp$ collision center-of-mass energies of 7 and 14 TeV, acceptances as functions of mass and charge are shown in Figs.~\ref{fig:HIPsensitivity_7TeV} and \ref{fig:HIPsensitivity_14TeV} respectively, for electrically (left) and magnetically (right) charged HIPs, for the ATLAS (top), CMS (middle, top), LHCb (middle, bottom) and ALICE (bottom) detectors. Acceptances for the MoEDAL detector for the Drell-Yan process at 14 TeV center-of-mass energy are shown in Fig.~\ref{fig:HIPsensitivity_14TeV_MoEDAL}. In general-purpose detectors, an increase of acceptance with increasing mass is expected due to an increasing tendency for the particles to be produced in a more central region in the chosen production model. This is counterbalanced by a decrease at high mass due to a shift of the kinetic energy spectrum towards lower energies, which reduces the HIP range and thus the acceptance. In the case of electrically charged particles, a lower velocity also causes the energy loss to increase and reduces the HIP range (while the reverse is true for magnetically charged particles). In ATLAS, CMS and LHCb the timing requirement for a first level calorimeter trigger affects the high ($m\apprge 1000$ GeV) masses. This constraint is especially limiting for LHCb due to the large distance (12 meters) to the EM calorimeter. CMS features an acceptance $\sim 10\%$ higher than ATLAS thanks to the absence of solenoid magnet in front of the EM calorimeter barrel. The charge corresponding to $>1\%$ acceptance in ATLAS, CMS and LHCb lies around $z<200$ or $g<4g_D$. For ALICE and MoEDAL, it is around $z<500$ or $g<10g_D$. 

There are four dominant contributions to the systematic uncertainty in the calculated acceptances: detector material budget, computation of d$E$/d$x$ (expressions \ref{Bethe} and \ref{Bethe_mag}), bending in the magnetic field, and limited statistics of the simulated events. Varying the detector thicknesses within their uncertainties of $20\%$ produces changes in the HIP ranges, which in turn affect the acceptance by up to 10\%. Neglecting higher-order terms in the computation of d$E$/d$x$ also results in up to 10\% uncertainty in the acceptance. Monopole bending can affect the acceptance mostly in two ways: it can change the path length to the EM calorimeter and thus the amount of traversed material, and it can cause a monopole to cross the edge of the detector geometrical acceptance. As it can impact the acceptance in both directions depending on the sign of the monopole charge, the overall effect of bending is less than 5\%. When added in quadrature, the three contributions result in a 15\% relative uncertainty. A large number of events $N$ (usually $N=1000$) is simulated for each bin of mass and charge, among which $n$ events have at least one HIP inside the detector acceptance. The relative uncertainty in the acceptance $a=n/N$ due to simulation statistics (significant only for bins with low acceptance) is $n^{-1/2}=(Na)^{-1/2}$.

\section{Monopole stopping acceptances}
\label{trapped}

The ATLAS and CMS central beam-pipe sections are very likely to be replaced during the LHC shutdown scheduled for 2013~\cite{IBLTDR,CMSupgrades,ATLASPixelUpgrades,LS1plans2012}. The reason of the replacement is to make room to allow for new layers of pixel detectors to be inserted. It also offers the possibility for an early search for monopoles trapped in the obsolete beam pipes with the induction technique. A replacement of the whole ATLAS and CMS pixel detectors may occur around 2018~\cite{CMSupgrades,ATLASPixelUpgrades,CMSPixelUpgrades}, which would also allow to search for stopped monopoles in the pixel detectors. Here we investigate the potential stopping of monopoles inside the ATLAS and CMS beam pipes. The pipe prior to the replacement (for both ATLAS and CMS) is a beryllium cylinder of length 4~m, inner radius 29~mm and thickness 0.8~mm (or 0.148 g$\cdot$cm$^{-2}$) \cite{ATLAS08}. Due to the poisoning hasard of beryllium and its possible activation, the handling of the beam pipes to produce samples suitable for analysis with a SQUID apparatus may be expensive. There is a precedent of similar handling of the D0 beryllium beam pipe at the Tevatron~\cite{TEVATRONSQUID2000,TEVATRONSQUID2004}. 

\begin{figure}[tb]
  \begin{center}
    \includegraphics[width=0.49\linewidth]{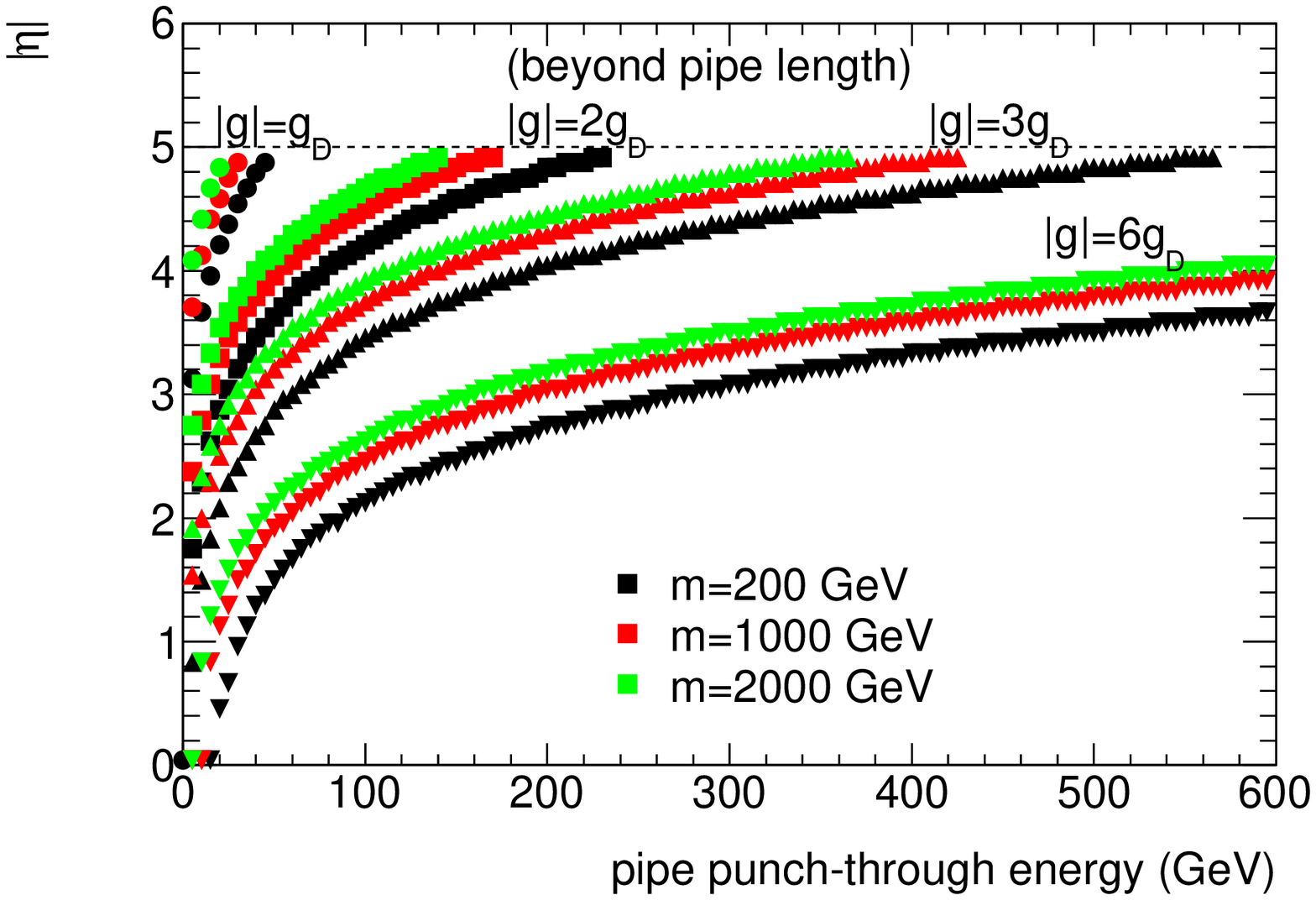}
    \includegraphics[width=0.49\linewidth]{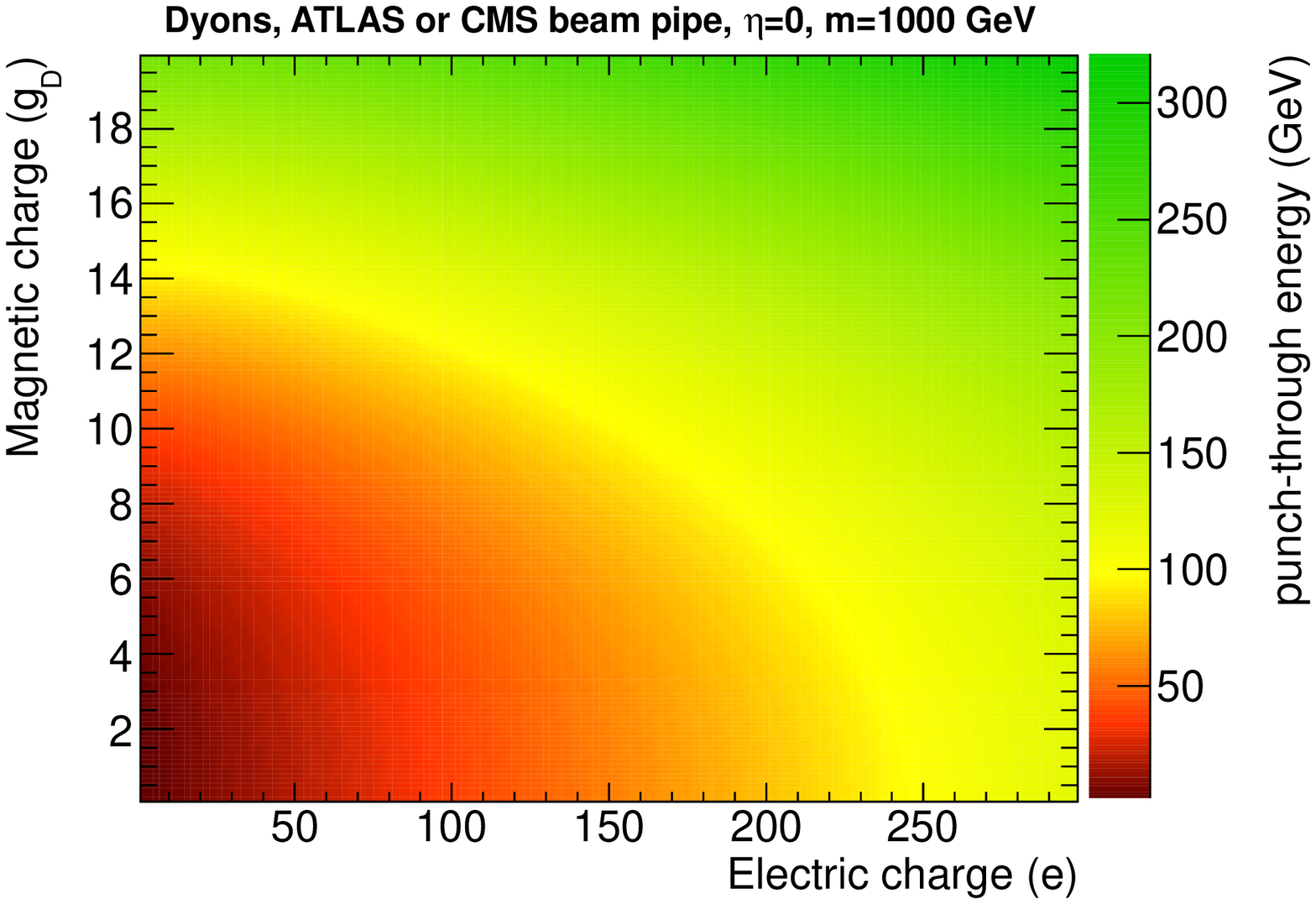}    
  \end{center}
  \caption{Energy below which magnetically charged particles stop in the ATLAS or CMS beam pipe, as a function of pseudorapidity for various magnetic charges and masses (left) and as a function of magnetic and electric charge for dyons with $m=1000$ GeV at $\eta=0$ (right).}
  \label{fig:thresh_pipe}
\end{figure}

The propagation of magnetically charged particles is simulated in beam-pipe material using the same machinery as described in the previous sections. The kinetic energy beyond which a HIP of given mass and charge punches through the beam pipe depends strongly on its angle of incidence. In the left plot of Fig.~\ref{fig:thresh_pipe} it can be seen that monopoles produced at low $|\eta|$ need to possess low energies in order not to punch through. The upper bound in $|\eta|$ is limited to 5 by the length and radius of the pipe. The right plot of Fig.~\ref{fig:thresh_pipe} shows the same punch-through energy, with $\eta=0$ and $m=1000$ GeV, but where the HIP is allowed to be a dyon, i.e., possess an electric charge in addition to its magnetic charge. 

\begin{figure}[tb]
  \begin{center}
    \includegraphics[width=0.49\linewidth]{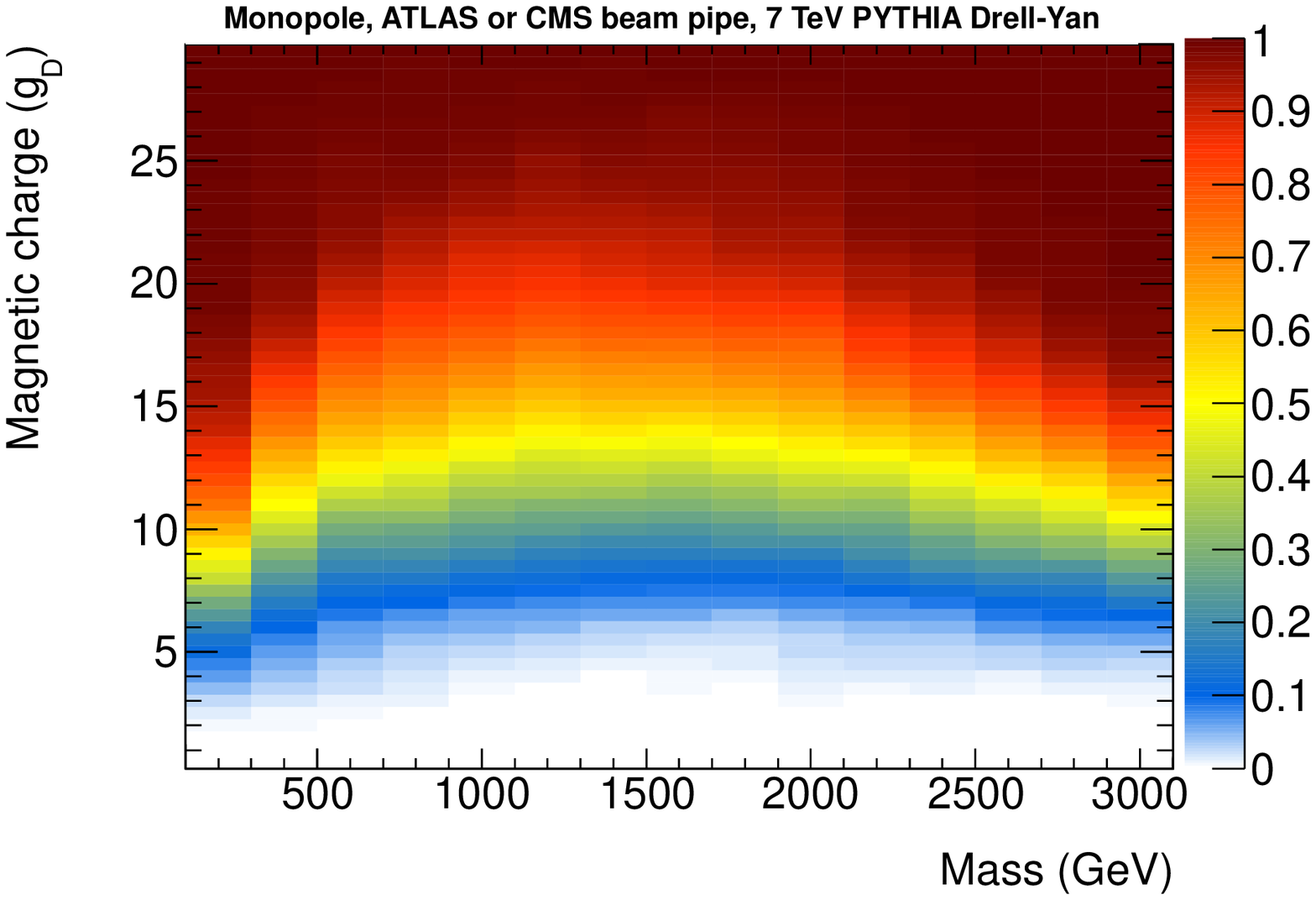}
    \includegraphics[width=0.49\linewidth]{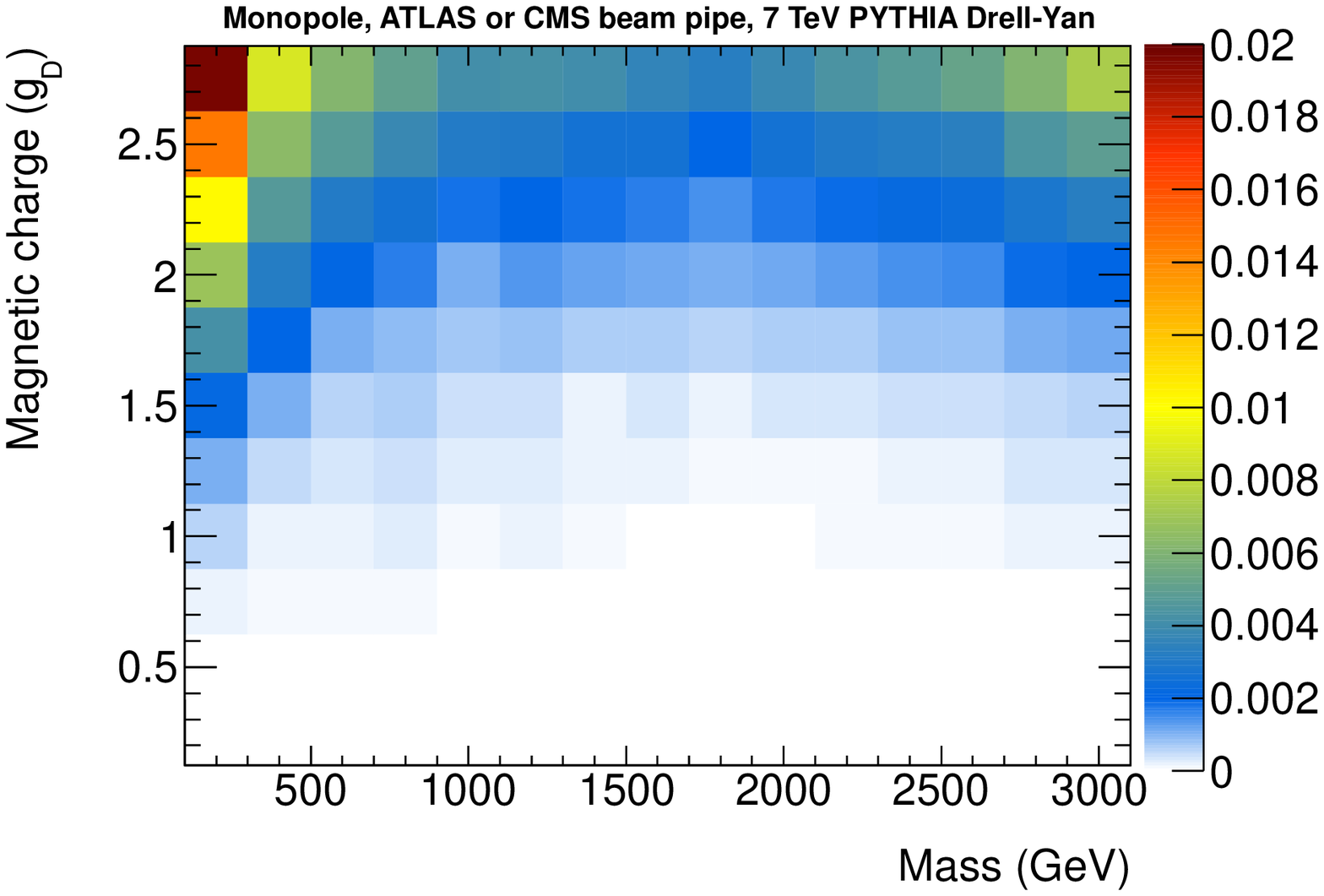}
    \includegraphics[width=0.49\linewidth]{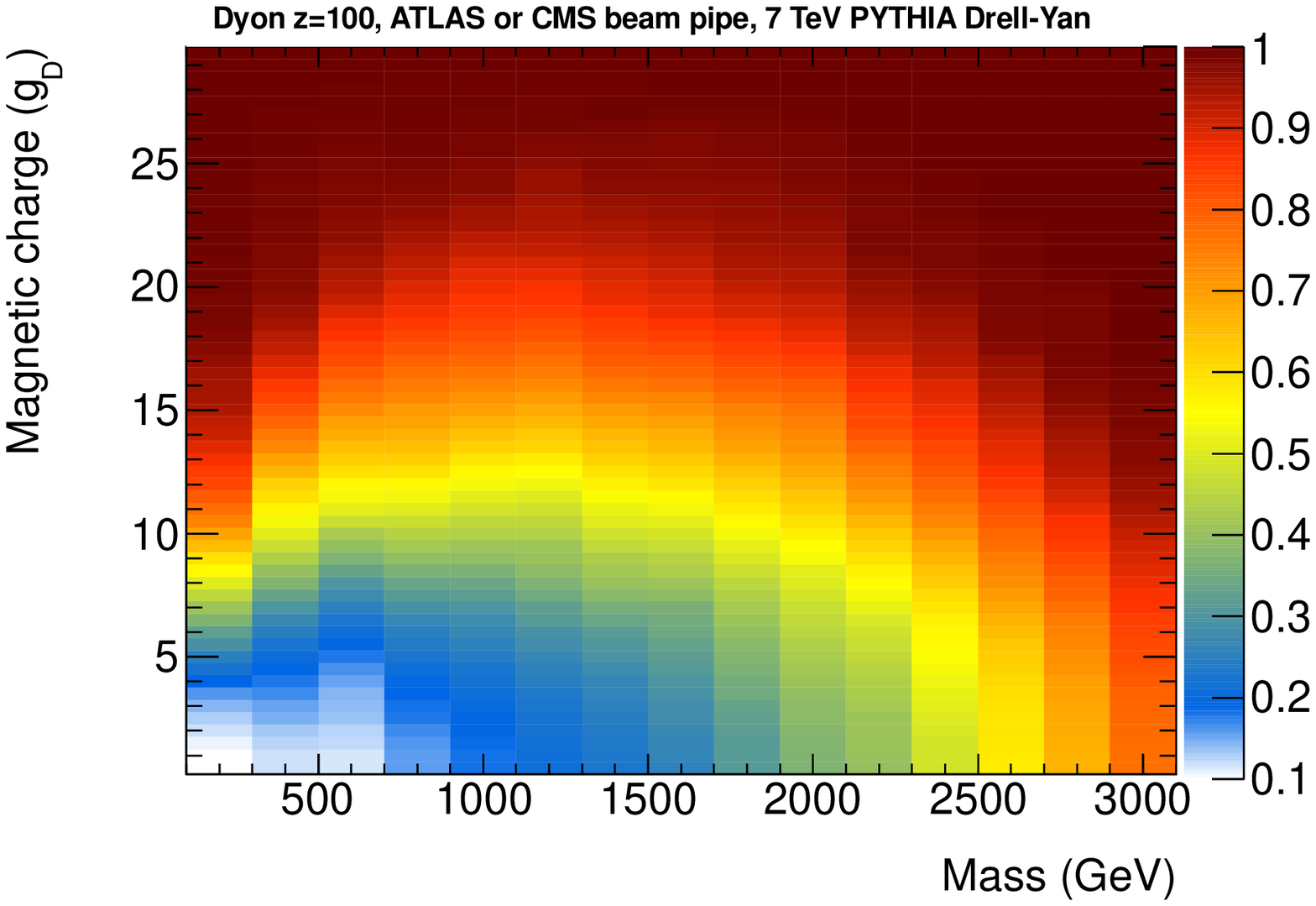}
    \includegraphics[width=0.49\linewidth]{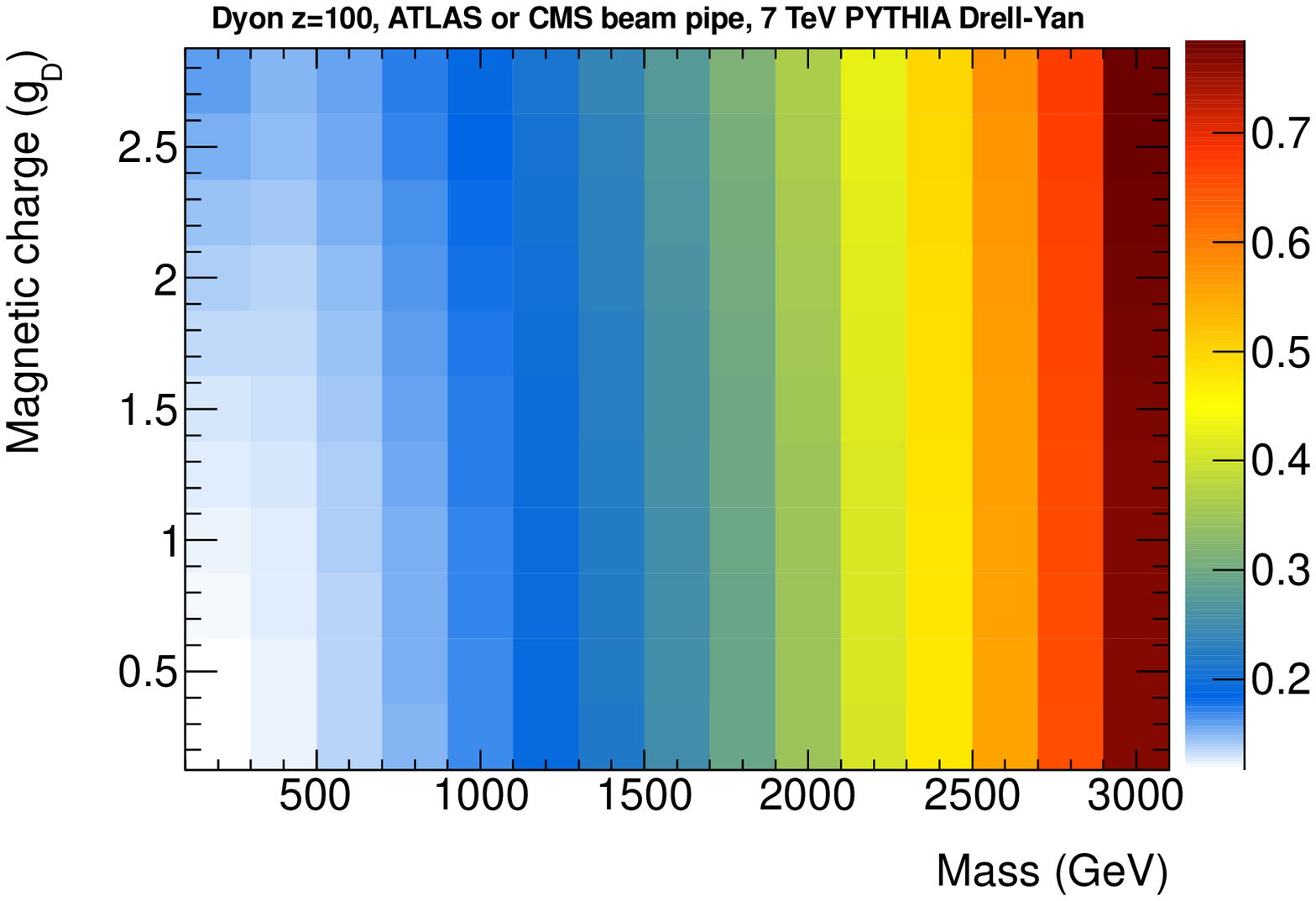}
  \end{center}
  \caption{Acceptance as a function of mass and charge, for magnetic monopoles (top) and dyons with electric charge $z=100$ (bottom), stopping in the ATLAS or CMS beam pipe, assuming a Drell-Yan pair production mechanism with 7 TeV $pp$ collisions. The right plots show the low-$g$ region with high statistics per bin (note that the acceptance scales are different). The relative binwise systematic uncertainty in the acceptance $a$ is 15\% in all plots. The relative binwise uncertainty from statistics is $(1000\cdot a)^{-1/2}$ for the left plots and $(50000\cdot a)^{-1/2}$ for the right plots.}
  \label{fig:HIPsensitivity_pipe_7TeV}
\end{figure}

\begin{figure}[tb]
  \begin{center}
    \includegraphics[width=0.49\linewidth]{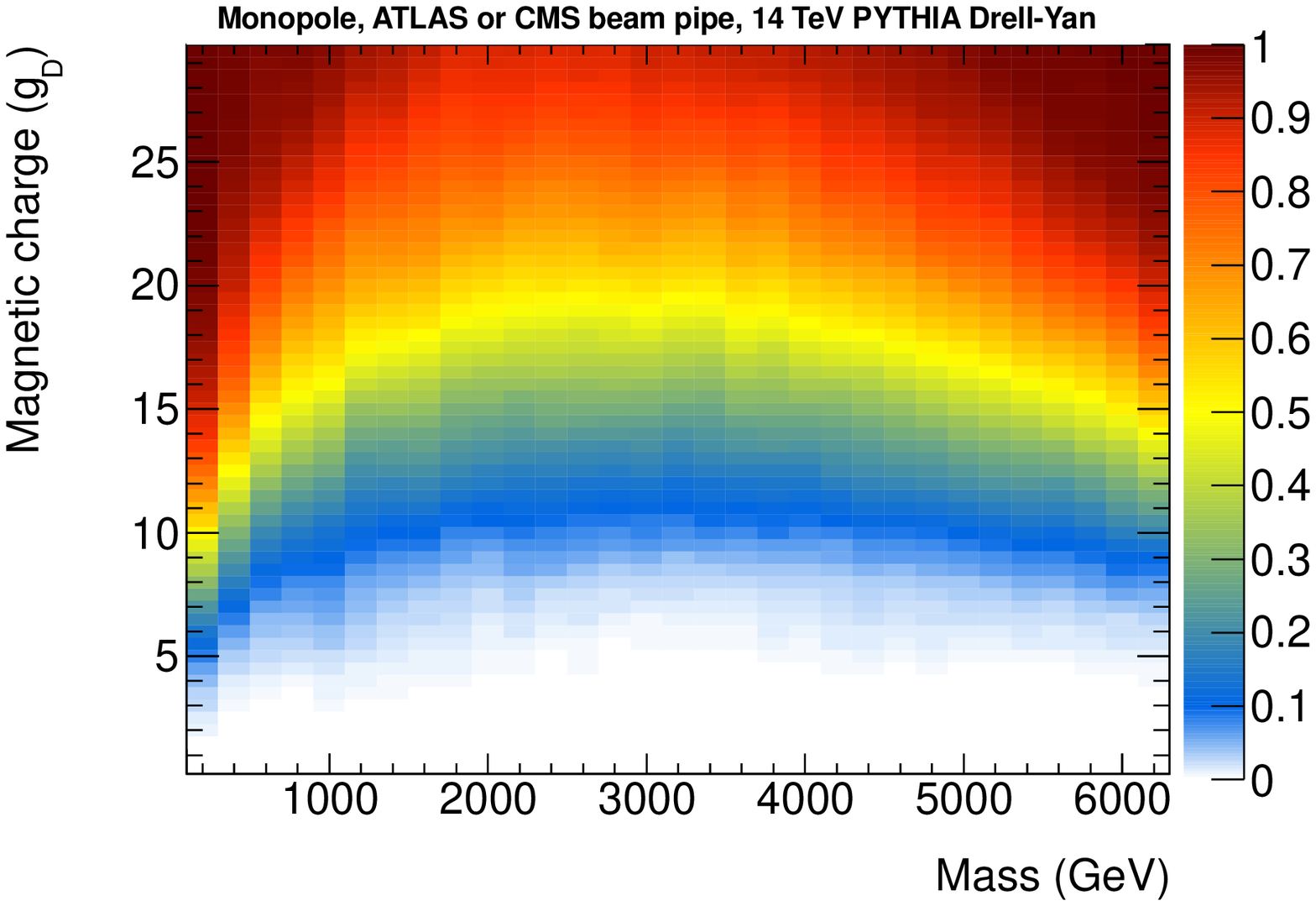}
    \includegraphics[width=0.49\linewidth]{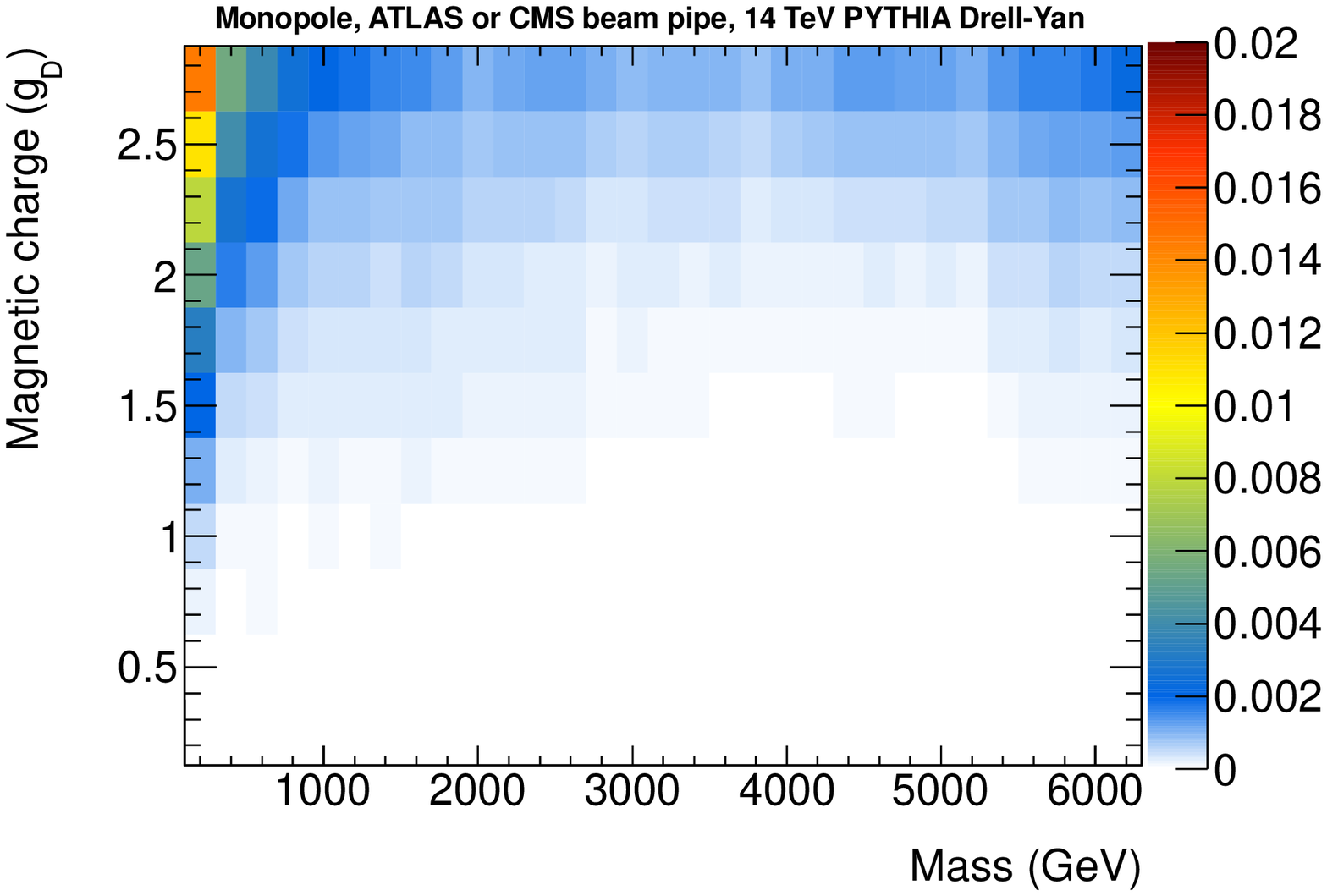}
  \end{center}
  \caption{Acceptance as a function of HIP mass and charge, for magnetic monopoles stopping in the ATLAS or CMS beam pipe, assuming a Drell-Yan pair production mechanism with 14 TeV $pp$ collisions, using the same pipe thickness as for 7 TeV collisions. The right plot shows the low-$g$ region with high statistics per bin (note that the acceptance scale is different). The relative binwise systematic uncertainty in the acceptance $a$ is 15\% in both plots. The relative binwise uncertainty from statistics is $(1000\cdot a)^{-1/2}$ for the left plot and $(50000\cdot a)^{-1/2}$ for the right plot.}
  \label{fig:HIPsensitivity_pipe_14TeV}
\end{figure}

The acceptance of an analysis of the ATLAS or CMS beam pipes with the induction method is defined as the probability (per event) that a magnetically charged particle with given mass and charge would stop inside the beam pipe. Acceptances as functions of mass and magnetic charge are shown in Figs. \ref{fig:HIPsensitivity_pipe_7TeV} and \ref{fig:HIPsensitivity_pipe_14TeV} for Drell-Yan kinematics in 7 TeV and 14 TeV $pp$ collisions, respectively. The same material thickness is used for 14 TeV collisions as for 7 TeV collisions. Above $|g|\simeq 20g_D$, all monopoles stop in the beam pipe and the acceptance is unity. The decrease of acceptance with mass at low mass is explained both by the fact that slower monopoles undergo lower d$E$/d$x$ (see Fig.~\ref{fig:dedx}) and by the shift of the $\eta$ spectrum towards more central directions (see top left plot in Fig.~\ref{fig:DYkin}), and the increase of acceptance with mass at large mass comes from the shift of the Drell-Yan energy spectrum towards lower energies (see top right plot in Fig.~\ref{fig:DYkin}). For $|g|=g_D$, less than one event in a thousand is expected to produce a monopole which stops in the beam pipe. However, dyons with $|g|=g_D$ and a high electric charge can still be easily stopped, as illustrated by the bottom plots in Fig.~\ref{fig:HIPsensitivity_pipe_7TeV} with $|z|=100$ for the dyon electric charge. The large acceptance increase at large dyon mass is due to the high d$E$/d$x$ at low velocity for an electric charge. 

To conclude this section about trapped monopoles, we investigate what are the highest and lowest magnetic charges one could possibly discover. The upper bound may come either from the monopole acceleration along the beam axis before reaching the beam pipe, or from saturation of the SQUID. A calculation with Equation~\ref{traj} shows that drifting in the CMS experiment's magnetic field causes monopoles to be lost in the beam pipe vacuum for extremely high charges, $|g|>10000g_D$. Concerning the SQUID dynamic range, in principle, there is no upper limit to the current which can be measured, but in practice there are limitations depending on the specifications of the apparatus. Currents corresponding to magnetic charges in the range $g_D \leq |g| \leq 100g_D$ can be observed with SQUID models which are typically used in geomagnetic studies~\cite{HERASQUID}. Very high magnetic charges $|g|\apprge 1000g_D$ might saturate such a SQUID at every pass. Such a behaviour would make the sample interesting enough to motivate a dedicated experiment with a less sensitive coil which would withstand higher currents. Finally, as discussed above, the lowest magnetic charge a  search can hope to detect is limited by the acceptance for monopoles, but not for dyons. In principle, for a dyon with a high electric charge, one could detect low magnetic charges by passing a given sample through the SQUID multiple times: backgrounds from magnetic dipoles would cancel out, while the permanent current induced by a tiny magnetic charge would build up, so that particles with $|g|=0.1g_D$ would be seen with this method~\cite{Moon71}.

\section{Model dependence} 
\label{dependence}

The acceptances shown above rely on the assumption of Drell-Yan kinematics. If a model of isotopic HIP production is chosen instead (see Fig.~\ref{fig:DYkin}), the LHC detector acceptance becomes higher due the higher HIP energy, which leads to a higher probability for at least one HIP to punch through the inner parts of the detectors. The acceptance still dies out sharply for HIP charges above a certain threshold. In fact, at $\eta\sim 0$, above the charges $6g_D$ in ATLAS and LHCb, $10g_D$ in CMS, $16g_D$ in MoEDAL, and $20g_D$ in ALICE, the acceptance is always zero regardless of the model, as energies higher than the collision energy minus the HIP pair mass are required to reach the sensitive parts of the detectors. Conversely, the probability for stopping in a thin part of the detector such as the beam pipe is lower in the case of the isotropic model than for Drell-Yan. 

It must also be stressed that, in addition to the energy spectrum, the HIP angular distribution has a large impact on the acceptance. Drell-Yan kinematics predict that the HIPs are produced centrally (see top left plot in Fig.~\ref{fig:DYkin}), which is an advantage for ATLAS, CMS and ALICE. In the isotropic model, the centrality is even greater, and the angular distribution does not depend on the HIP mass, which reduces the mass dependence of the acceptance. Conversely, a model where the HIPs are mostly produced in the forward regions would favour using LHCb and looking for trapped monopoles in the ATLAS and CMS beam pipes.

\section{Sensitivity and complementarity} 
\label{reach}

\begin{figure}[tb]
  \begin{center}
    \includegraphics[width=0.49\linewidth]{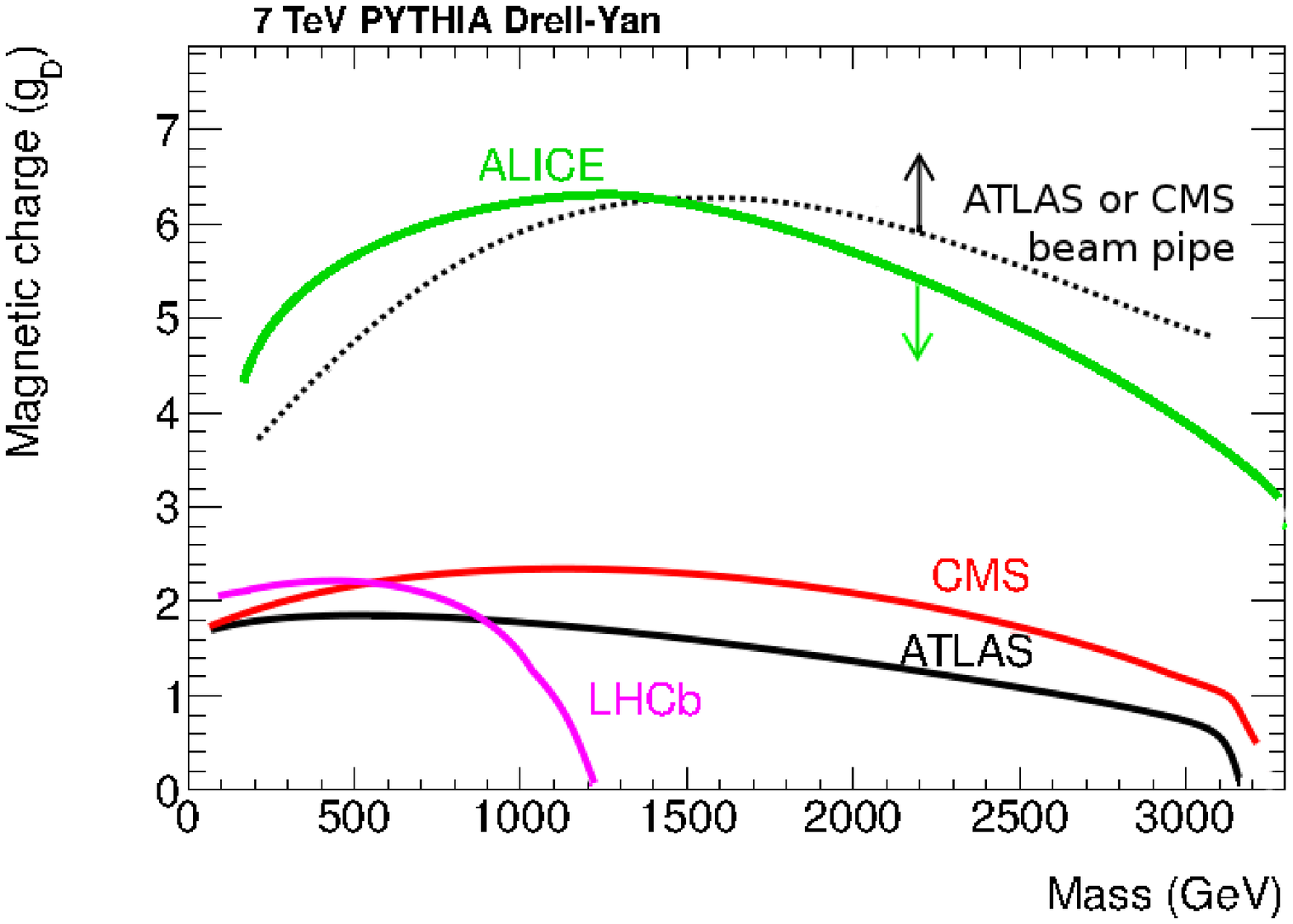}
    \includegraphics[width=0.49\linewidth]{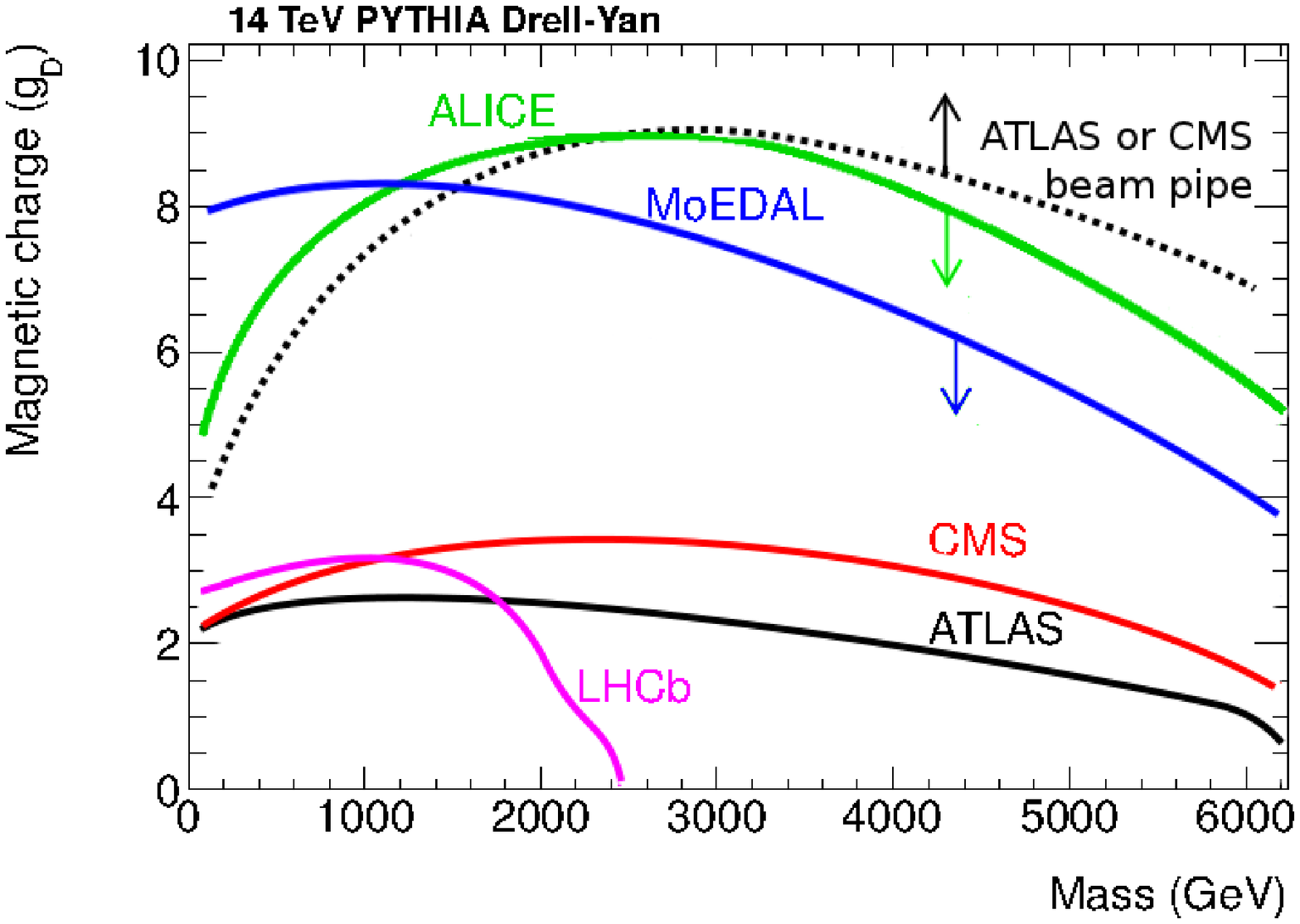}
  \end{center}
  \caption{Contours corresponding to 5\% acceptance in various detectors, as functions of monopole mass and charge, assuming a Drell-Yan pair production mechanism with 7 TeV (left) and 14 TeV (right) $pp$ collisions. The analysis of ATLAS or CMS beam pipe material with the induction technique (dashed line) is sensitive to charges above the curve (particles with high charges stop in the beam pipe), while all other detectors are sensitive to charges below the corresponding curves (particles with high charges stop before they can reach the detectors).}
  \label{fig:summary}
\end{figure}

\begin{figure}[tb]
  \begin{center}
    \includegraphics[width=0.49\linewidth]{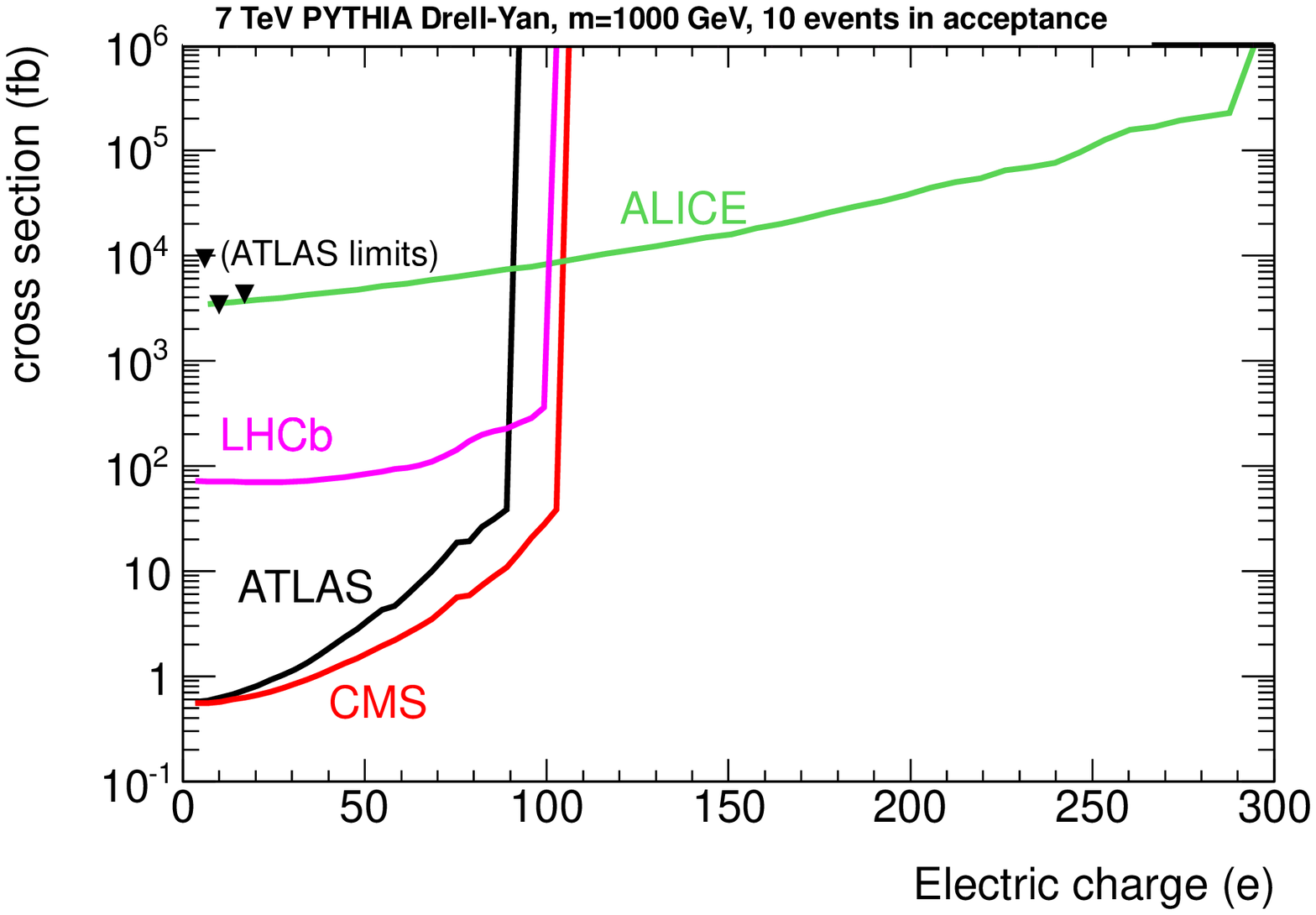}
    \includegraphics[width=0.49\linewidth]{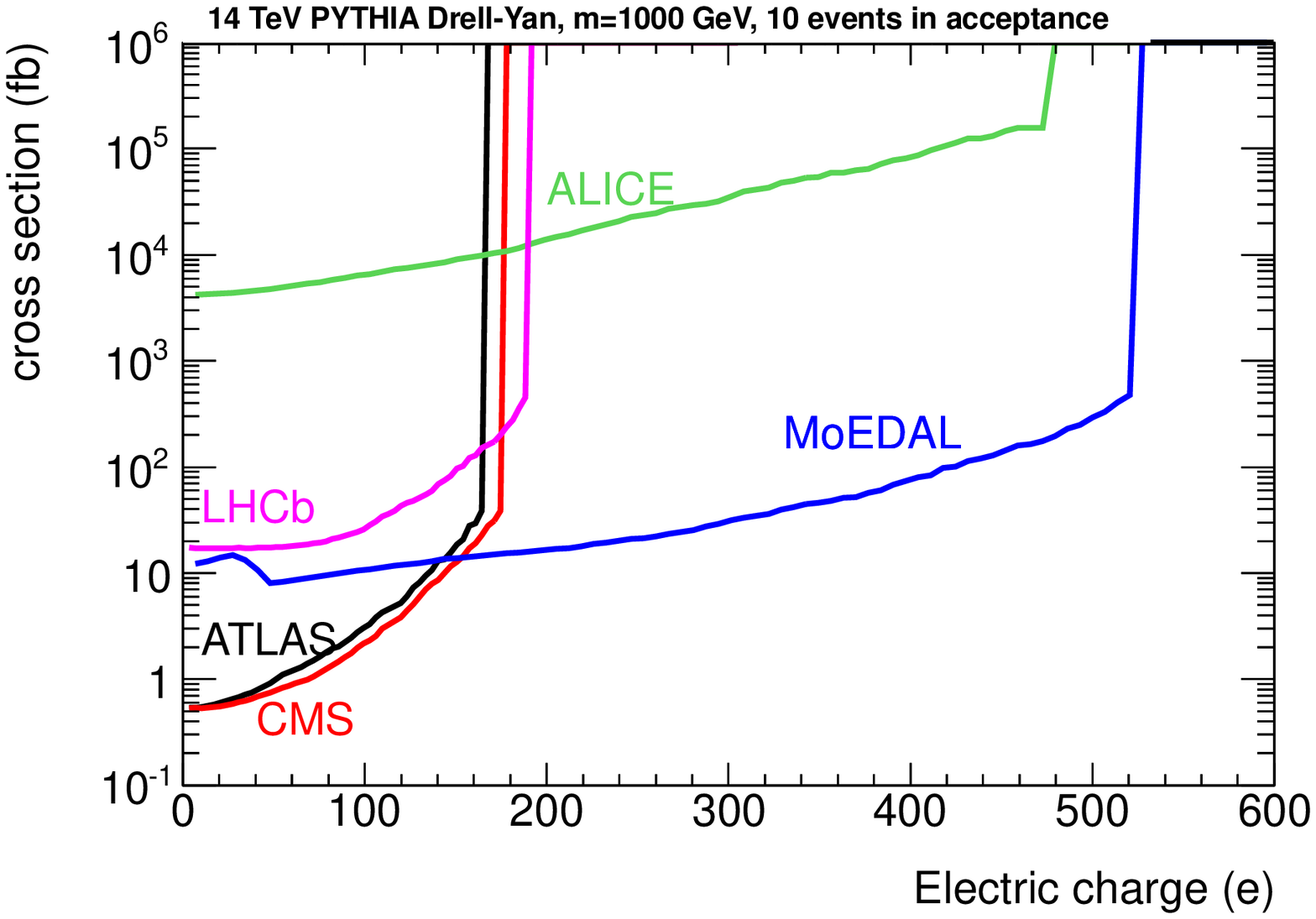}
    \includegraphics[width=0.49\linewidth]{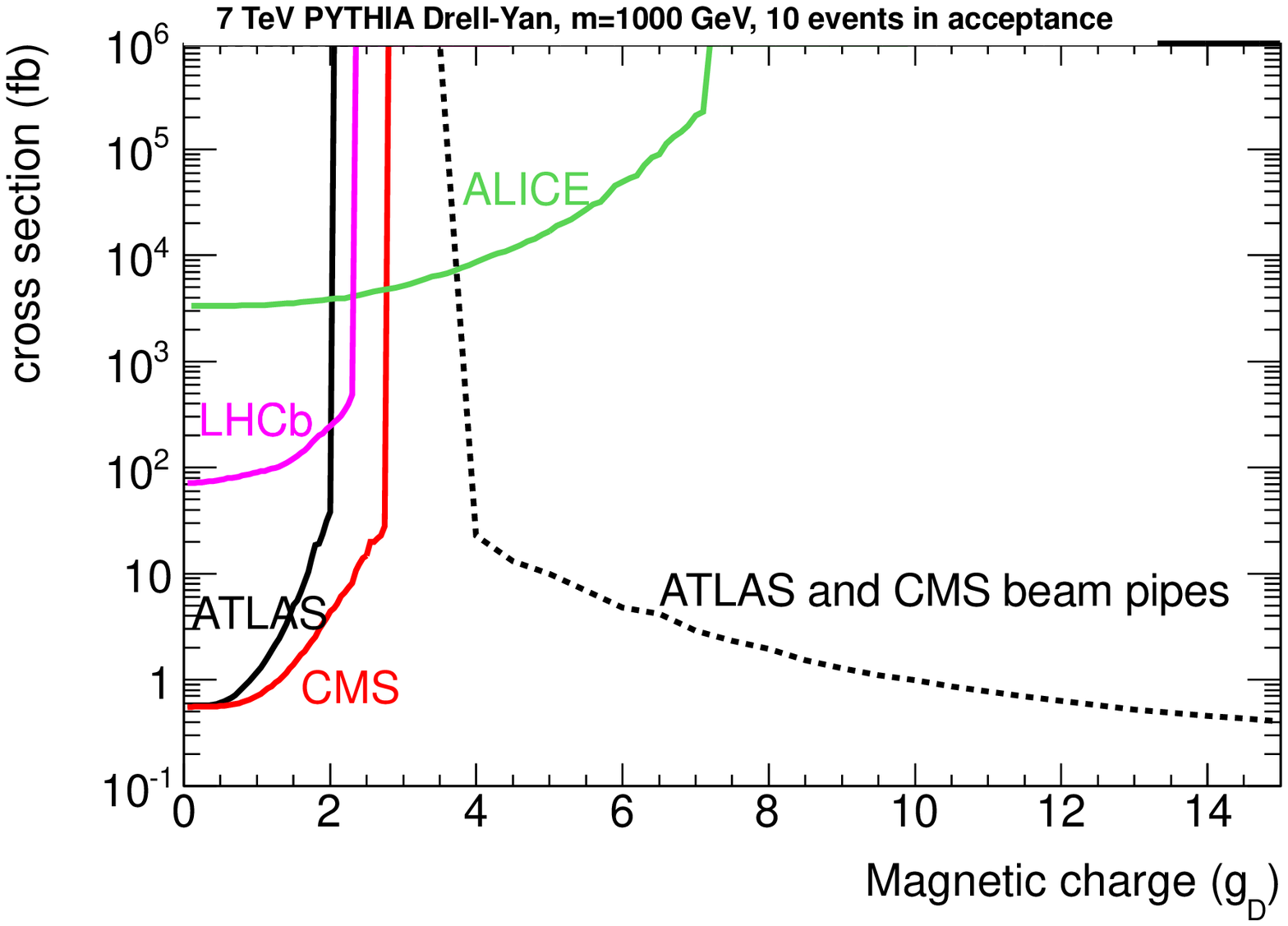}
    \includegraphics[width=0.49\linewidth]{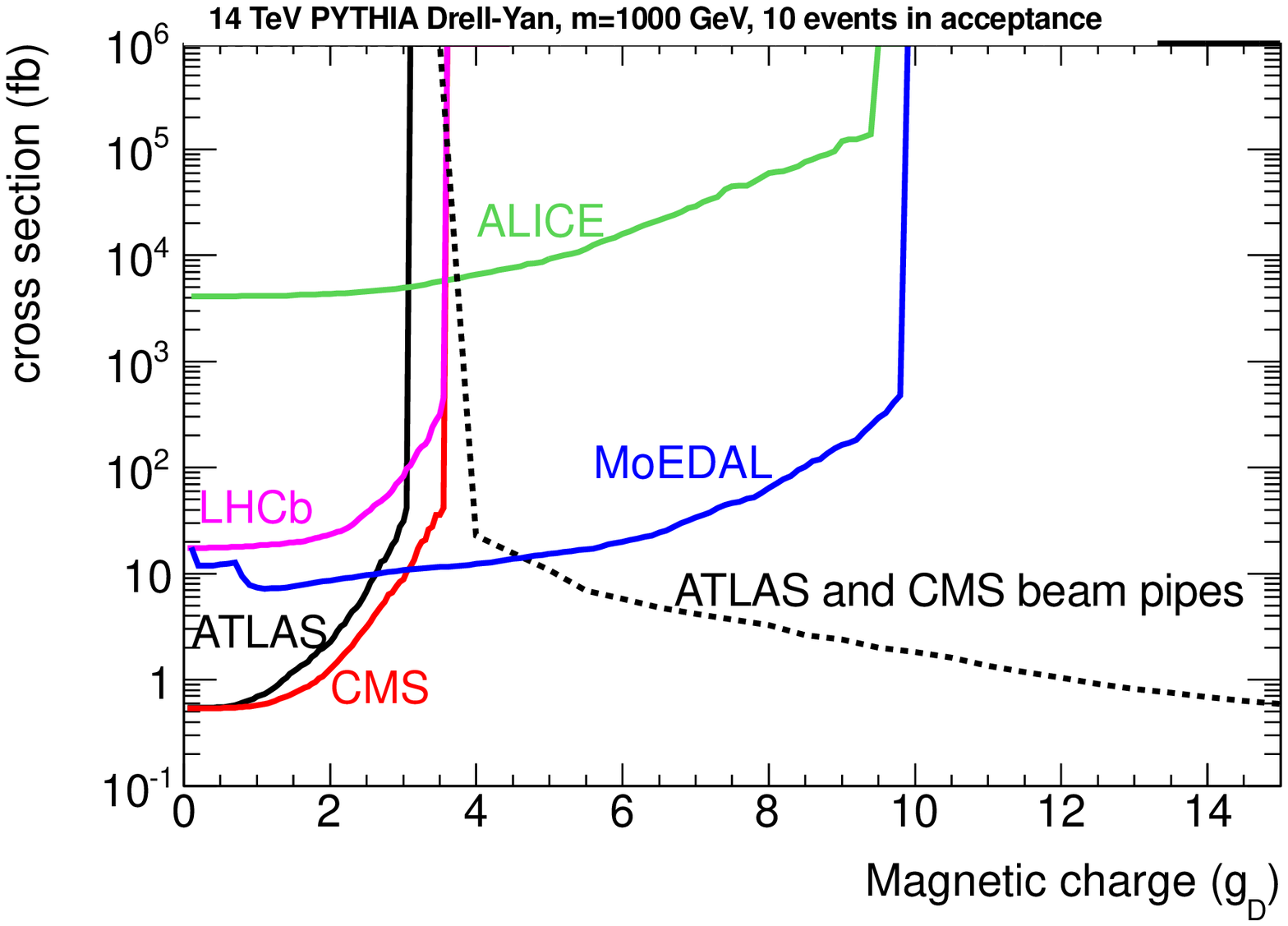}
  \end{center}
  \caption{HIP pair production cross sections for obtaining 10 events with at least one HIP inside the detector acceptance, as a function of HIP electric (top) and magnetic (bottom) charge, assuming $m=1000$ GeV and a Drell-Yan pair production mechanism with 7 TeV (left) and 14 TeV (right) $pp$ collisions. Integrated luminosities are from Table~\ref{tab:lumi}, corresponding to the full expected dataset at the end of the 2012 runs or about two years of 14 TeV runs. Cross section limits from ATLAS~\cite{QballATLAS10} are also indicated as triangles in the top left plot.}
  \label{fig:xs}
\end{figure}

Contours corresponding to regions of more than 5\% monopole detection acceptance for all detectors considered in this work are shown in Fig.~\ref{fig:summary} for the Drell-Yan production model with 7 TeV (left) and 14 TeV (right) collisions. 
The sensitivity in charge depends on the amount of material to be traversed before reaching the sensitive parts of the detector and is highest for ALICE, followed by MoEDAL, CMS and ATLAS. The sensitivity in mass is best for ALICE and for the induction method with the ATLAS and CMS beam pipes, and worst for LHCb due to the timing constraint. However, these curves are not weighted with luminosity, nor with the number of events needed for an unambiguous discovery. 

In order to get a benchmark of the discovery potential of each experiment, the HIP pair production cross section needed to have 10 events with at least one HIP inside the detector acceptance is estimated. These estimates are based on the integrated luminosities indicated in Table~\ref{tab:lumi}, corresponding to ATLAS and CMS each collecting 20 fb$^{-1}$ of both $7-8$ TeV and 14 TeV $pp$ collisions. The results are shown in Fig.~\ref{fig:xs} as a function of HIP charge in the case of $m=1000$ GeV and assuming a Drell-Yan production mechanism. Although the efficiency is not included, the ATLAS results~\cite{QballATLAS10} suggest that it can be between 50\% and 90\% for a well-designed search. ATLAS and CMS would have the best sensitivity for charges in the range $|z|<150$ or $|g|<3g_D$ thanks to a comparatively higher luminosity in $pp$ collisions. To probe high electric charges ($150<|z|<500$) and moderately high magnetic charges ($3g_D<|g|<4.5g_D$), MoEDAL or ALICE would be needed, with MoEDAL being sensitive to cross sections lower by more than two orders of magnitude. The induction method applied to the ATLAS and CMS beam pipes would offer the best performance for very high magnetic charges ($|g|>4.5g_D$). The LHCb experiment, despite its poor suitability for detecting very massive object, is complementary to ATLAS, CMS and ALICE in that it can probe the forward regions up to $\eta=4.9$ (the ATLAS and CMS trackers are limited to $|\eta|<2.5$) -- this is important as the event kinematics are unknown. 

\section{Conclusions} 
\label{conclusions}

Since spring 2010, the LHC has been colliding protons with the unprecedented center-of-mass energy of 7 TeV. HIPs carrying a high electric charge or a magnetic charge with masses around or above 1 TeV (beyond the reach of previous experiments) can potentially be produced and observed by the LHC experiments. The only dedicated HIP search carried out so far at the LHC~\cite{QballATLAS10} is limited in scope and cannot be interpreted for monopoles. In this article, sensitivities for HIP detection at LHC experiments are quantified and it is shown how best to exploit this potential by combining various techniques, which allow to explore complementary regions of phase space. 

The acceptance of the general purpose detectors for the observation of HIPs has been studied, and their strengths and weaknesses have been discussed. Centrally produced HIPs with modest charge ($|g|\apprle 2g_D$) can be probed by direct detection in the ATLAS and CMS detectors, which are exposed to the highest luminosity. This is assuming that a non-negligible fraction ($>5\%$) of the HIPs are produced with sufficient energy to reach the calorimeters, which is the case with Drell-Yan like kinematics. ATLAS has a lower sensitivity than CMS owing to its solenoid coil absorbing high-charge or low-energy HIPs before they can penetrate the calorimeters. Centrally produced HIPs with high charges ($|g|\apprle 9g_D$) or low energies could be probed in ALICE thanks to a lower material budget, but at the cost of a luminosity more than three orders of magnitude lower than in ATLAS and CMS for $pp$ collisions. MoEDAL and ALICE have no constraints on the time of arrival of the HIPs and can thus probe masses up to half the center-of-mass energy. HIPs produced in the forward regions (up to $\eta=4.9$) can potentially be searched for in MoEDAL and LHCb. MoEDAL, which will operate with 14 TeV collisions at the LHCb luminosity (ten times lower than ATLAS and CMS), has the unique advantages of a full angular coverage and a well-understood detection technique. A promising complementary method for probing the forward regions and very high charges in the whole angular range is the analysis of beam pipe material with a SQUID apparatus, to look for trapped magnetic monopoles or dyons. Monopoles with charge up to $|g|=10000g_D$ would be accessible with such a technique.

\section*{Acknowledgments} 
\label{Acknowledgments}
 
We are very grateful to J. Pinfold for supplying additional information about the layout of the MoEDAL experiment, and for useful discussions. This work was supported by two grants from the Swiss National Science Foundation.

%

\bibliographystyle{mystylem}
\bibliography{SMPpheno11}

\end{document}